# Static and Dynamic irreversible magnetic properties of high temperature superconductors

*Ruslan Prozorov*

**Department of Physics**

# Ph. D. Thesis

Submitted to the Senate of Bar-Ilan University
Ramat-Gan, Israel                    July 1998

*(with Highest Distinction)*

This work was carried out under the supervision of

*Professor Yosef Yeshurun*

Department of Physics, Bar-Ilan University.

# Acknowledgements

I extend my grateful thanks to Professor *Yosef Yeshurun* for his guidance during this research. Due to his help and his concern I have not only completed, but also enjoyed doing this work.

This work could not have been fulfilled without the active participation of my wife, *Tanya*. I appreciate her understanding and patience.

I am grateful to the *Clore Foundations* for awarding me a scholarship.

I thank all the people I worked with at Bar-Ilan. I am indebted in particular to *Avner Shaulov* and *Leonid Burlachkov* for their unobstrusive support and care. I also thank *Boris Shapiro, Yossi Abulafia, Shuki Wolfus, Lior Klein, Eran Sheriff, Dima Giller* and *Yael Radzyner* for their constructive criticism, suggestions and interest. I thank *Menahem Katz* who has shared my computer interests. I am obliged to *Yael Radzyner* and *Aviva Oberman* for the help in writing this thesis.

It is very difficult to mention everyone I would like to thank. I am obliged to many people I have met during the years of my Ph.D. project. Special thanks are due to *Marcin Konczykowskii* for help in heavy-ion irradiation. I gained a great deal from fruitful discussions with *Mike McElfresh, Mikhail Indenbom, Vitallii Vlasko-Vlasov, Anatolii Polyanskii, Lior Klein, Eli Zeldov, Lev Dorosinskii, Valerii Vinokur, Edouard Sonin, Vadim Geshkenbein, Ernst Helmut Brandt, Marcin Konczykowski, Kees van der Beek, Peter Kes, Lia Krusin-Elbaum, Leonid Gurevich, Maksim Marchevski, Richard Doyle* and *John Clemm*.

My cat, *Physia*, has helped me in understanding the fishtail effect.

To the memory of my brother *Serge*

# Table of Contents

# List of Figures

# List of Tables

# *Abstract*

This thesis describes experimental and theoretical study of static and dynamic aspects of the irreversible magnetic behavior of high-$T_c$ superconductors. Experimentally, conventional magnetometry and novel Hall-probe array techniques are employed. Using both techniques extends significantly the experimental possibilities and yields a wealth of new experimental results. These results stimulated the development of new theoretical analyses advancing our understanding of the irreversible magnetic behavior of type-II superconductors. The topics studied in this thesis include effects of sample geometry and magnetic field orientation, influence of heavy-ion irradiation and flux creep mechanisms.

Effects of sample geometry are addressed in a study of the thickness dependence of the irreversible magnetic properties of HTS thin films. Our experimental results reveal peculiar dependence of the current density and magnetic relaxation rate on the film thickness. In order to explain these data we investigate theoretically the distribution of the current density $j$ throughout the film thickness, and develop a new critical state model that takes into account this distribution. The results of our analysis are compared with measurements of the persistent current and the relaxation rate in a series of $Y_1Ba_2Cu_3O_{7-\delta}$ films of various thicknesses.

Effects of the geometrical and intrinsic anisotropy are investigated in thin $Y_1Ba_2Cu_3O_{7-\delta}$ films in an external field applied at an angle $\theta$ with respect to the film plane and films rotated in a constant magnetic field. We show that in practice one can always neglect the in-plane component of the magnetic moment and obtain the rotation curve *M vs. $\theta$* simply by mapping the *c*-component of the magnetic moment versus the effective magnetic field $H\cos(\theta)$. We also investigate the angular dependence of the magnetic properties of $Y_1Ba_2Cu_3O_{7-\delta}$ thin films irradiated with high-energy ions, and report on the first magnetic observation of a unidirectional anisotropy in Pb irradiated films.

Effects of field orientation are further examined in measurements of the angular dependence of the irreversibility line in $Y_1Ba_2Cu_3O_{7-\delta}$ films. The results of this study shed new light on the origin of the irreversibility line in these films. It is demonstrated that the irreversibility line in unirradiated films is above the melting line, due to pinning in a viscous vortex liquid. In heavy-ion irradiated films, the magnetic irreversibility is governed by the enhanced pinning on the columnar defects. Due to the one-dimensional nature of these defects, the irreversibility line exhibits strong angular anisotropy.

The novel local magnetic measurement technique, using an array of miniature Hall sensors, is addressed in detail both experimentally and theoretically. The critical state model is employed to describe the local DC and AC magnetic response. We further extend this model to account for relaxation effects in order to explain our observation of frequency dependent AC response.

Local magnetic measurements are utilized in the study of flux creep mechanisms. We develop a unique approach to obtain the parameters of the flux creep process from direct measurements of the time evolution of the field profiles across the sample. This method is also applied for determination of the electric field *vs.* current density ($E$-$j$ characteristics) in the sample. We employ these techniques in studying the dynamic behavior of $Y_1Ba_2Cu_3O_{7-\delta}$ and $Nd_{1.85}Ce_{0.15}CuO_{4-\delta}$ single crystals in the field region of the anomalous "fishtail" in their magnetization curves. We identify two different flux creep mechanisms on two sides of the "fishtail" peak: elastic (collective) creep below the peak and plastic creep at fields above it.

The phenomenon of magnetic relaxation is also discussed in light of the new concept of self-organization in the vortex matter. The original concept of the self-organized criticality is extended to include sub-critical behavior. We show that measurements of the magnetic noise spectra may yield new information about the particular flux creep mechanism.

# Chapter I. Introduction

Irreversible magnetic properties of type-II superconductors have attracted much interest, [1-3], especially since the 1986 discovery of high-temperature superconductors (HTS) [4]. Physical parameters of HTS differ significantly from those of conventional type-II superconductors, giving rise to a rich field-temperature (B-T) magnetic phase diagram. In particular, HTS are characterized by smaller coherence length $\xi$, larger London penetration depth $\lambda$, and high anisotropy of the electron mobility [2, 3]. The small $\xi$ leads to a weak vortex pinning, as the energy gained by a pinned vortex is proportional to $\xi^2$. Large values of $\lambda$ imply a soft vortex lattice, as all the elastic moduli are proportional to $1/\lambda^2$ [2, 3]. Both small pinning and soft lattice enable relatively easy thermal depinning of fluxons from pinning sites, thus creating a large domain of reversible magnetic phase within the mixed state. The irreversibility line, which separates irreversible and reversible regions, is thus sensitive to the time window of the experiment, as it is mostly determined by the vortex dynamics. Due to the large anisotropy, the magnetic phase diagram of HTS not only depends on the magnitude of the magnetic field but also on its orientation with respect to the crystal lattice [2, 3].

The experimental techniques for characterizing the magnetic properties of superconductors can be classified into three major categories: global, local and microscopic. Global techniques characterize the sample as a whole, yielding information on physical quantities averaged over the sample volume. A Vibrating Sample Magnetometer (VSM) and a Superconducting Quantum Interference Device (SQUID) are commonly used in global magnetic measurements. Local magnetic techniques employ, among others, magneto-optics [5-11] and miniature Hall-probe arrays [12-18]. These techniques map the magnetic induction on a specimen surface with spatial resolution of order of microns. Although the local information has proven extremely useful (see Chapter V), it is important to note that the magnetic induction is mapped only on the sample surface where the induction may differ significantly from that of the bulk [3, 19, 20]. Microscopic measurements, such as scanning tunneling microscopy or magnetic force magnetometry, are used for the study of individual vortices.

In this thesis we employ both global and local techniques to study various aspects of the irreversible magnetic behavior of HTS. (Microscopic measurements are beyond the scope of this thesis, which concerns the collective behavior of vortices rather than the behavior of individual vortices.) These include effects of sample geometry [ PR15, RP25, RP28, RP35 ] and field orientation [ RP10, RP15, RP17 - RP19, RP23 ], influence of heavy-ion irradiation [ PR10, RP17 – RP19, RP23 ], and flux creep mechanisms [ RP14, RP16, RP21, RP22, RP27, RP29, RP32, RP33 ]. These topics are discussed in five chapters of this thesis; each chapter includes new experimental data and theoretical analysis.

Chapter II presents a thorough investigation of the thickness dependence of the irreversible magnetic properties of HTS thin films. We begin with a detailed theoretical analysis of the distribution of the current density $j$ in superconducting films along the direction of an external field applied perpendicular to the film plane. This analysis was motivated by our experimental observations of a peculiar dependence of the current density and magnetic relaxation rate on the film thickness. While previous works assumed constant current density throughout the film thickness, our analysis reveals that in the presence of bulk pinning, $j$ is inhomogeneous on the length scale of order of the inter-vortex distance. This inhomogeneity is significantly enhanced in the presence of surface pinning. On the basis of these results we develop a new critical state model that takes into account current density variations throughout the film thickness, and shows how these variations give rise to thickness dependence of $j$ and the magnetic relaxation rate. The experimental part of this chapter describes our results of measurements of the persistent current and relaxation rates in a series of $Y_1Ba_2Cu_3O_{7-\delta}$ films of various thickness, and compares these results with our theoretical predictions.

Chapter III presents a theoretical and experimental study of the effects of field orientation and angular anisotropy of vortex pinning. We begin with an analysis of the irreversible magnetization in thin $Y_1Ba_2Cu_3O_{7-\delta}$ films in an external field applied at an angle $\theta$ with respect to the film plane. The results indicate that one can always neglect the in-plane component of the magnetic moment, despite the fact that the current produced by the in-plane component of the field may attain values close to the depairing current density. However, it is shown that this large current may induce anisotropy in the flux penetration into a thin film in inclined field. The second part of this chapter presents experimental results of the magnetization in thin $Y_1Ba_2Cu_3O_{7-\delta}$ films rotated in an external magnetic field. In a certain field range, $M(\theta)$ is not symmetric with respect to $\theta=\pi$ and the magnetization curves for forward and backward rotations do not coincide. Previous interpretations of similar results invoked complex models, such as lag of vortices behind the magnetic moment. We show that our results can be explained within the framework of the critical state model, and that one can obtain the rotation curve *M vs. θ* simply by mapping the *c*-component of the magnetic moment versus the effective magnetic field $H\cos(\theta)$. The third part of this chapter describes the angular dependence of the magnetic properties of $Y_1Ba_2Cu_3O_{7-\delta}$ thin films irradiated with high-energy Pb and Xe ions. We study the different types of defects produced by these ions, and describe the first magnetic observation of unidirectional anisotropy in Pb irradiated films.

The study of the effects of field orientation is extended in Chapter IV to include the angular dependence of the irreversibility line in $Y_1Ba_2Cu_3O_{7-\delta}$ films. The results of this investigation are utilized to obtain an insight into the origin of the irreversibility line in these films. It is demonstrated that the onset of the magnetic irreversibility in unirradiated films occurs well above the melting line due to pinning in a viscous vortex liquid. In heavy-ion irradiated films, the physics of the magnetic irreversibility is determined by the enhanced pinning by the columnar defects. Due to the one-dimensional nature of these defects, the irreversibility line exhibits strong angular anisotropy. A surprising result is that columnar defects not only enhance the critical current density, but also result in a faster magnetic relaxation. The latter may be suppressed by using crossed columnar defects. These observations are explained by considering easy directions for vortex motion in the presence of columnar defects.

In Chapter V we describe the novel local magnetic measurement technique using an array of miniature Hall sensors. This is followed by a description of our theoretical analysis of the local AC magnetic response, based on the critical state model. We further develop this model to account for relaxation effects in order to explain our observation of frequency dependent *AC* response. This Chapter also describes a unique approach to investigate the flux creep process using miniature

Hall-probe arrays. In this approach we utilize time-resolved measurements of the magnetic induction in HTS crystals to determine the electric field *vs.* current density (*E-j*) characteristics. In particular, we study the dynamic behavior of $YBa_2Cu_3O_{7-\delta}$ and $Nd_{1.85}Ce_{0.15}CuO_{4-\delta}$ compounds in the field range of the anomalous second peak ("fishtail") in their magnetization curves. From the functional dependence of the *E-j* curves we demonstrate that the fishtail is shaped by a competition between elastic and plastic flux creep mechanisms.

The problem of flux creep is also discussed in Chapter VI where we consider numeric analysis of the flux creep equation and discuss the issue of self-organization of the vortex matter during magnetic relaxation. We show that measurements of the magnetic noise spectra may yield new information about the particular creep mechanism. The original concept of the self-organized criticality is extended to include sub-critical behavior.

Chapter VII summarizes the main findings of this thesis, emphasizing the original contributions of this work to the understanding of the static and dynamic irreversible magnetic behavior of HTS.

# Chapter II. Thickness dependence of persistent current and creep rate in YBa$_2$Cu$_3$O$_{7-\delta}$ films

## A. *Analysis of the mixed state in thin films*

One of the strongest assumptions made in calculating the field distribution in thin films in perpendicular field is that the current density is constant throughout the sample thickness [21-23]. In such an approach, magnetic induction is proportional to the sample thickness, and sample magnetization *m* is independent of it. Correspondingly, current density extracted from magnetization measurements should not depend on film thickness. Experiments, however, show that current density [24-29] and magnetic relaxation rate [27-29] decrease with the increase of the film thickness. To explain this observed behavior one has to analyze field and current density distributions in thin films with pinning throughout the film thickness.

In this Chapter we summarize our theoretical and experimental studies of the thickness dependence of the persistent current density and magnetic relaxation rate in thin films [RP25, RP35].

Explanation of the observed decrease of *j* with the increase of the film thickness *d* is usually based on the idea that pinning on surfaces perpendicular to the direction of vortices is important, and must be taken into account [24-26]. However, this is not sufficient for understanding the thickness dependence of the magnetic relaxation rate, which decreases with the increase of the film thickness [27-29].

Another explanation of the observed thickness dependence of the current density may be based on collective pinning in a 2D regime, i. e., for longitudinal collective correlation length *L* larger than the film thickness. This case is carefully considered in [30, 31]. The pinning is then *stronger* and as a result, the critical-current density and the creep barrier are larger for thinner samples. This leads to a *slower* relaxation in thinner samples, contrary to our results. Moreover, this scenario is probably not suitable for the explanation of our experimental data discussed below, because the thickness of our films $d$>800 Å is larger than $L$≈40-100 Å in $YBa_2Cu_3O_{7-\delta}$ films.

In order to understand the experimental results we calculate the current density and magnetic induction distribution by using the "two-mode electrodynamics" theory suggested earlier to explain the AC response in bulk materials [32, 33]. The essence of this theory is that two length scales govern the penetration of fields and currents. The longer scale is of electrodynamics origin and, therefore, is universal: it exists, for example, in a superconductor in the Meissner state (the London penetration depth) or, in a normal conductor (the skin depth). The shorter scale is related to the vortex-line tension, so it is unique for a type-II superconductor in the mixed state. This scale was introduced into the continuous theory of type-II superconductors by Mathieu and Simon [34] (see also [35, 36]). When applying the two-mode electrodynamics to the critical state one may ignore the time variation, i.e., the two-mode electrodynamics becomes a *two-mode electrostatics* theory.

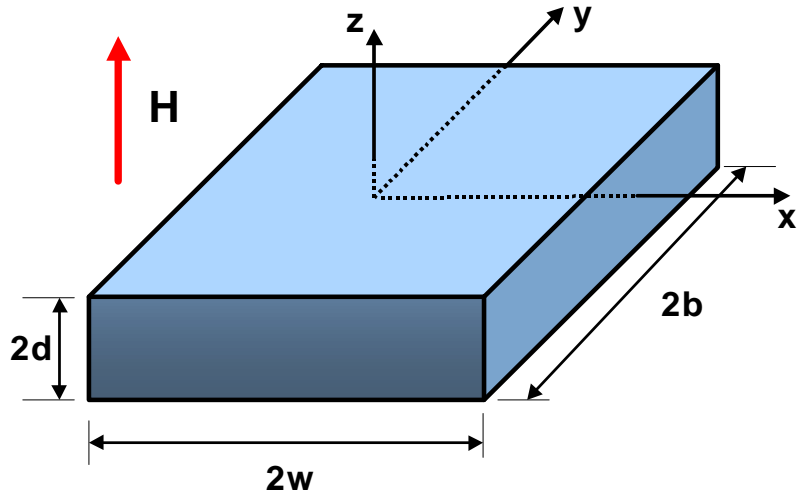

Figure 1. Geometry under study

Our analysis of a type-II thin superconducting film within the two-mode electrostatics theory leads to the conclusion that for strong enough bulk pinning and for film thickness larger than the Campbell penetration depth $\lambda_C$, inhomogeneity of the current density becomes important, *even in the absence of surface pinning*. Thus, inhomogeneity of the current distribution throughout the film thickness is a *distinctive* and inevitable feature of the perpendicular film geometry like, for example, the geometrical barrier [37]. Inhomogeneity of the current distribution is enhanced if the surface pinning supports the critical state. In this case, most of the current is confined to a layer of a depth of the order of the inter-vortex distance, which is usually much smaller than the London penetration depth $\lambda$. Because of this inhomogeneity, the *measured* average critical current density becomes thickness dependent. This current inhomogeneity also causes a thickness dependence of the magnetic relaxation rate. In the following, we present detailed calculations of the distribution of the current density $j$ and induction field $B$ in thin type-II superconducting film, resulting from surface and/or bulk pinning. We then introduce the first critical state model that takes into account the variation in *j* throughout the film thickness. Calculations based on this critical state model lead to thickness dependence in *j* and in the magnetic relaxation rate. These predictions are compared with the experimental data.

## 1. Basic electrodynamics equations

We consider the sample geometry depicted in Figure 1 and assume that our sample is a thin film, infinite in the *y*-direction, so that $d \ll w$ and $b \to \infty$. The vortex density *n* is determined by the *z*-component $B_z$ of the average magnetic field (magnetic induction) **B** in the film: $n = B_z/\phi_0$. Super-currents of density $j(x,z) = j_y(x,z)$ flow along the *y*-axis resulting in a Lorenz force in the *x*-direction, and a vortex displacement *u* along the *x*-axis.

We begin with equations of the electrodynamics describing the mixed state of type-II superconductors in such geometry. They include the London equation for the *x*-component of the magnetic field:

$$B_x - \lambda^2 \frac{\partial^2 B_x}{\partial z^2} = B_z \frac{\partial u}{\partial z} \qquad (1)$$

the Maxwell equation:

$$\frac{4\pi}{c} j_y = \frac{\partial B_x}{\partial z} - \frac{\partial B_z}{\partial x} \qquad (2)$$

and the equation of vortex motion:

$$\eta \frac{\partial u}{\partial t} + ku = \frac{\Phi_0}{c} j_y + \frac{\Phi_0}{4\pi} H^\otimes \frac{\partial^2 u}{\partial z^2} \qquad (3)$$

where

$$H^{\otimes} = \frac{\Phi_0}{4\pi\lambda^2} \ln\left(\frac{a_0}{r_c}\right) \tag{4}$$

is a field of order of the first critical field $H_{c1}$, $a_0 \approx \sqrt{\Phi_0/B_z}$ is the inter-vortex distance, and $r_c \approx \xi$ is an effective vortex core radius.

The equation of the vortex motion arises from the balance among four terms: (i) the friction force proportional to the friction coefficient $\eta$; (ii) the homogeneous, *linear* elastic pinning force $\propto k$ (i. e., assuming small displacements $u$); (iii) the Lorentz force proportional to the current density $j$; and (iv) the vortex-line tension force (the last term on the right-hand side of Eq.(3) is discussed in detail in Refs. [32, 33]).

In the bulk of an infinite in $z$-direction slab ("parallel geometry"), vortices move without bending so that the $x$-component $B_x$ is absent, and the Maxwell equation becomes: $4\pi j_y/c = -\partial B_z/\partial x$. Since $B_z$ is proportional to the vortex density, this current may be called a "diffusion current".

The case of thin film $d \ll w$ in perpendicular field ("perpendicular geometry") is essentially different: the diffusion current is small compared to the "bending current" $4\pi j_y/c \approx \partial B_x/\partial z$ (see the estimation below) and may be neglected for calculation of the distribution throughout the film thickness (along the $z$-axis). As a result, Eq.(3) becomes

$$\eta \frac{\partial u}{\partial t} + ku = \frac{\Phi_0}{4\pi}\frac{\partial B_x}{\partial z} + \frac{\Phi_0}{4\pi} H^{\otimes} \frac{\partial^2 u}{\partial z^2} \tag{5}$$

Equations Eq.(1) and Eq.(5) determine the distribution of the displacement $u(z)$ and of the in-plane magnetic induction $B_x(z)$. This also yields a distribution of the current density $4\pi j_y/c \approx \partial B_x/\partial z$. But these equations are still not closed, since the two components of the magnetic induction, $B_x$ and $B_z$, and current density $j_y(z)$ are connected by the Biot-Savart law. However, neglecting the diffusion current in the Maxwell equation we separate the problem into two parts: (1) determination of the distribution of fields and currents along the $z$-axis, taking the total current $I_y = cB_x^s/2\pi$ (here $B_x^s \equiv B_x(z=d)$ is the in-plane field on the sample surface) and the perpendicular magnetic-induction component $B_z$ as free external parameters; (2) determination of the external parameters $I_y$ and $B_z$ using the Biot-Savart law. The latter part of the problem (solution of the integral equation given by the Biot-Savart law) has already been studied carefully in previous works [20, 22, 23, 38, 39]. In this Section we concentrate on the analysis of the distribution of fields and currents throughout the film thickness.

The accuracy of our approach is determined by the ratio of the diffusion current to the bending current, since we neglect the diffusion current contribution to the total current. The diffusion current density is roughly $(c/4\pi)\partial B_z/\partial x \approx (c/4\pi)(B_z^w - H)/w$, whereas the bending current density is $(c/4\pi)\partial B_x/\partial z \approx (c/4\pi)B_x^s/d$, where $B_z^w$ is $B_z$ on the sample edge and $B_x^s$ is the planar component of magnetic induction on the sample surface [7, 22, 38]. Suppose, as a rough estimation, that $B_x^s \approx (B_z^w - H)$ (see also [40]). Then, the ratio between the diffusion and the bending current is approximately $d/w \approx 10^{-3} - 10^{-4}$ for typical thin films. Note that this condition does not depend on the magnitude of the critical current and is well satisfied also in typical single crystals, where $d/w \approx 0.1$. Therefore, the results we obtain below hold for a wide range of typical samples used in the experiment.

## 2. Two-mode electrostatics: two length scales

Let us consider the static case when vortices do not move and there is no friction. Then, Eq.(5) becomes

$$ku = \frac{\Phi_0}{4\pi}\frac{\partial B_x}{\partial z} + \frac{\Phi_0}{4\pi}H^{\otimes}\frac{\partial^2 u}{\partial z^2} \qquad (6)$$

Excluding the $B_x$ component of the magnetic induction from Eqs. (1) and (6) yields the equation for the vortex displacement:

$$-\frac{4\pi k}{\Phi_0}\left(u - \lambda^2\frac{\partial^2 u}{\partial z^2}\right) + \left(H^{\otimes} + B_z\right)\frac{\partial^2 u}{\partial z^2} - \lambda^2 H^{\otimes}\frac{\partial^4 u}{\partial z^4} = 0 \qquad (7)$$

The two length scales which govern distributions over the $z$-axis become evident if one tries to find a general solution of Eq.(7) in the form $B_x \sim u \sim \exp(ipz)$. Then, the dispersion equation for $p$ is bi-quadratic and yields two negative values for $p^2$. In the limit $k \ll 4\pi\lambda^2/\Phi_0(H^{\otimes}+B_z)$ (weak bulk pinning):

$$p_1^2 = -\frac{1}{\tilde{\lambda}^2} = -\frac{1}{\lambda^2}\left(1 + \frac{B_z}{H^{\otimes}}\right) \qquad (8)$$

$$p_2^2 = -\frac{1}{\lambda_C^2} = -\frac{4\pi k}{\Phi_0\left(H^{\otimes} + B_z\right)} \qquad (9)$$

Thus, the distribution along the $z$-axis is characterized by the two length scales: the Campbell length $\lambda_C$, which is the electrodynamics length, and length $\tilde{\lambda}$, given by Eq.(8), which is related to $\lambda$ and the vortex-line tension.

# 3. Current density and field distribution - effects of surface and bulk pinning

In order to determine distribution of currents and fields throughout the film thickness, one must add the proper boundary conditions to the general solution of Eq.(7). We look for a solution that is a superposition of two modes. In particular, for the vortex displacement we can write:

$$u(z) = u_0 \cosh\left(\frac{z}{\lambda_C}\right) + u_1 \cosh\left(\frac{z}{\tilde{\tilde{\lambda}}}\right) \tag{10}$$

Using Eq.(6) one has for the current density:

$$\frac{4\pi}{c} j_y = \frac{\partial B_x}{\partial x} \approx B_z \frac{u_0}{\lambda_C^2} \cosh\left(\frac{z}{\lambda_C}\right) - H^\otimes \frac{u_1}{\tilde{\tilde{\lambda}}^2} \cosh\left(\frac{z}{\tilde{\tilde{\lambda}}}\right) \tag{11}$$

The total current is

$$\frac{4\pi}{c} I_y = 2 B_x(d) = 2 B_z \frac{u_0}{\lambda_C} \cosh\left(\frac{d}{\lambda_C}\right) - 2 H^\otimes \frac{u_1}{\tilde{\tilde{\lambda}}} \cosh\left(\frac{d}{\tilde{\tilde{\lambda}}}\right) \tag{12}$$

Equation (10) is in fact a boundary condition imposed on the amplitudes of two modes, $u_0$ and $u_1$. The second boundary condition is determined by the strength of the surface pinning. If displacements are small, the general form of this boundary condition is

$$\alpha u(\pm d) \pm \left.\frac{\partial u}{\partial z}\right|_{z=\pm d} = 0 \tag{13}$$

where $\alpha = 0$ in the absence of surface pinning and $\alpha \to \infty$ in the limit of strong surface pinning. In the following parts of the section we consider these two limits.

## (a) Surface pinning

Let us consider the case of surface pinning in the absence of bulk pinning $k=0$, when the Campbell length $\lambda_C \to \infty$. By "surface pinning" we understand pinning due to surface roughness on the surfaces *perpendicular* to the vortex direction. The surface roughness is assumed to be much smaller than the film thickness $d$. Substituting $\lambda_C \to \infty$ in the general solution Eq.(10), we derive the displacement for surface pinning:

$$u(z) = u_0 + u_1 \cosh\left(\frac{z}{\tilde{\lambda}}\right) \tag{14}$$

where $u_0$ and $u_1$ are constants, which can be determined from the boundary conditions Eqs.(12) and (13). Note, however, that $u_0$ is irrelevant in the case of surface pinning, because the constant $u_0$ does not affect distributions of currents and fields.

The magnetic field $B_x$ is obtained from Eq.(6):

$$B_x(z) = -H^\otimes \frac{u_1}{\tilde{\lambda}} \sinh\left(\frac{z}{\tilde{\lambda}}\right) \tag{15}$$

and the current is determined from the Maxwell equation (2) neglecting the diffusion current:

$$\frac{4\pi}{c} j_y = -H^\otimes \frac{u_1}{\tilde{\lambda}^2} \cosh\left(\frac{z}{\tilde{\lambda}}\right) \tag{16}$$

It is important to note that the characteristic length $\tilde{\lambda}$, which varies between the London penetration length $\lambda$ and the inter-vortex distance $a_0 \approx \sqrt{\Phi_0 / B_z}$, is much smaller than $\lambda$ for a dense vortex array, $B_z \gg H^\otimes$. Taking into account that usually thin films have thickness less or equal to $2\lambda$, the effect of the vortex bending due to surface pinning may be very important: most of the current is confined to a thin surface layer of width $\tilde{\lambda}$.

The current density just near the film surface is $j_s \equiv j_y(z=d) = -(c/4\pi)H^{\otimes}(u_1/\tilde{\lambda}^2)\cosh(d/\tilde{\lambda})$. Thus,

$$u_1 = -\frac{4\pi}{c}\frac{\tilde{\lambda}^2 j_s}{H^{\otimes}\cosh\left(\dfrac{d}{\tilde{\lambda}}\right)} \tag{17}$$

The total current integrated over the film thickness $2d$ is:

$$I_y = \int_{-d}^{d} j_y(z)dz = -\frac{c}{2\pi}H^{\otimes}\frac{u_1}{\tilde{\lambda}}\sinh\left(\frac{d}{\tilde{\lambda}}\right) = 2\tilde{\lambda}\tanh\left(\frac{d}{\tilde{\lambda}}\right) \tag{18}$$

Thus, the *average* current density $j_a \equiv I_y/2d$ - the quantity derived in the experiment - decreases with thickness as

$$j_a = j_s\frac{\tilde{\lambda}}{d}\tanh\left(\frac{d}{\tilde{\lambda}}\right) \tag{19}$$

yielding $j_a = j_s(\tilde{\lambda}/d)$ for $\tilde{\lambda}/d \ll 1$ as found experimentally [24, 25, 27, 28].

The field and the current distribution over the film thickness are:

$$j_y(z) = \frac{I_y}{2\tilde{\lambda}}\frac{\cosh(z/\tilde{\lambda})}{\sinh(d/\tilde{\lambda})} = j_s\frac{\cosh(z/\tilde{\lambda})}{\cosh(d/\tilde{\lambda})} \tag{20}$$

$$B_x(z) = \frac{2\pi}{c}I_y\frac{\sinh(z/\tilde{\lambda})}{\sinh(d/\tilde{\lambda})} = \frac{4\pi}{c}j_s\tilde{\lambda}\frac{\sinh(z/\tilde{\lambda})}{\cosh(d/\tilde{\lambda})} \tag{21}$$

Thus, the current penetrates into a small depth $\tilde{\lambda}$ and is exponentially small in the bulk beyond this length.

## (b) Bulk pinning

A remarkable feature of the perpendicular geometry is that, even in the absence of surface pinning, vortices are bent. This is in striking contrast with the parallel geometry where the diffusion current distribution is homogeneous along the direction of vortices and, therefore, does not bend them. Absence of surface pinning means that at the surface $\partial u / \partial z = 0$ (a vortex is perpendicular to an ideal surface). This yields the relation between $u_0$ and $u_1$ [see Eq.(10)]:

$$u_1 = -u_0 \frac{\tilde{\lambda}}{\lambda_C} \frac{\sinh(z/\lambda_C)}{\sinh(z/\tilde{\lambda})}$$

Then, Eq.(10) becomes

$$\frac{4\pi}{c} I_y = 2(B_z + H^\otimes) \frac{u_0}{\lambda_C} \sinh\left(\frac{d}{\lambda_C}\right) \tag{22}$$

The current distribution is

$$j_y(z) = I_y \left( \frac{1}{2\lambda_C} \frac{B_z}{H^\otimes + B_z} \frac{\cosh\left(\frac{z}{\lambda_C}\right)}{\sinh\left(\frac{d}{\lambda_C}\right)} + \frac{1}{2\tilde{\lambda}} \frac{H^\otimes}{H^\otimes + B_z} \frac{\cosh\left(\frac{z}{\tilde{\lambda}}\right)}{\sinh\left(\frac{d}{\tilde{\lambda}}\right)} \right) \tag{23}$$

In the limit $d<<\lambda_C$ Eq.(23) yields

$$j_y(z) = I_y \left( \frac{1}{2d} \frac{B_z}{H^\otimes + B_z} + \frac{1}{2\tilde{\lambda}} \frac{H^\otimes}{H^\otimes + B_z} \frac{\cosh\left(\frac{z}{\tilde{\lambda}}\right)}{\sinh\left(\frac{d}{\tilde{\lambda}}\right)} \right) \tag{24}$$

Another interesting case is that of the dense vortex array, $B_z >> H^\otimes$:

$$j_y(z) = \frac{I_y}{2\lambda_C} \frac{\cosh\left(\frac{z}{\lambda_C}\right)}{\sinh\left(\frac{d}{\lambda_C}\right)} = j_s \frac{\cosh\left(\frac{z}{\lambda_C}\right)}{\sinh\left(\frac{d}{\lambda_C}\right)} \tag{25}$$

where again $j_s$ is the current density near the film surface. Remarkably, current density is inhomogeneous even in the absence of surface pinning. We illustrate this in Figure 2, where we plot $j_y(z)/j_b$ vs. $z/d$ at different ratios $d/\lambda_C$. "Uniform" bulk current density $j_b=I_y/2d$ corresponds to the limit $d/\lambda_C =0$. Physically, such current profiles reflect Meissner screening of the in-plane component $B_x$ of the self-field.

For the average current density we have

$$j_a = j_s \frac{\lambda_C}{d} \tanh\left(\frac{d}{\lambda_C}\right) \qquad (26)$$

which is similar to the case of the surface pinning, Eq.(19), with $\tilde{\lambda}$ replaced by $\lambda_C$.

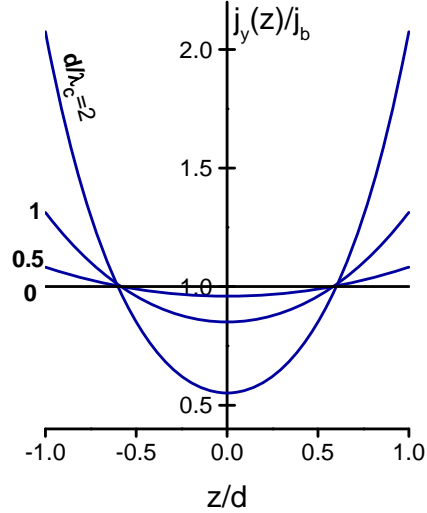

Figure 2. Normalized current density distributions $j_y(z)/j_b$ vs. $z/d$ calculated from Eq.(25) at different ratios $d/\lambda_C$.

Thus, in the perpendicular geometry, the current distribution is strongly inhomogeneous: the whole current is confined to a narrow surface layer of width $\tilde{\lambda}$ (surface pinning), or $\lambda_C$ (bulk pinning).

## 4. Critical state

In the theory given in the previous sections we have assumed that currents and vortex displacements are small. In this section we deal with the critical state when the current density equals its critical value $j_c$. Let us consider how it can affect our picture, derived in the previous sections for small currents.

### (a) Surface pinning

If vortices are pinned only at the surface, the value of the critical current depends on the profile of the surface, and one may not use the linear boundary condition imposed on the vortex displacement, Eq.(13). However, the $z$-independent vortex displacement $u_0$ does not influence the current density and field distribution in the bulk as shown in Chapter II.A.3(a) (see Eqs.(15) and (16)). Therefore the bulk current density and field distribution derived from our linear analysis can be used even for the critical state.

### (b) Bulk pinning

In this case our theory must be modified for the critical state. In particular, for large currents the bulk pinning force becomes nonlinear and, as a result, the current and field penetration is not described by simple exponential modes. Formally, this non-linearity may be incorporated into our theory assuming a $u$ - dependent pinning constant $k$ and, therefore, $k$ varies along the vortex line.

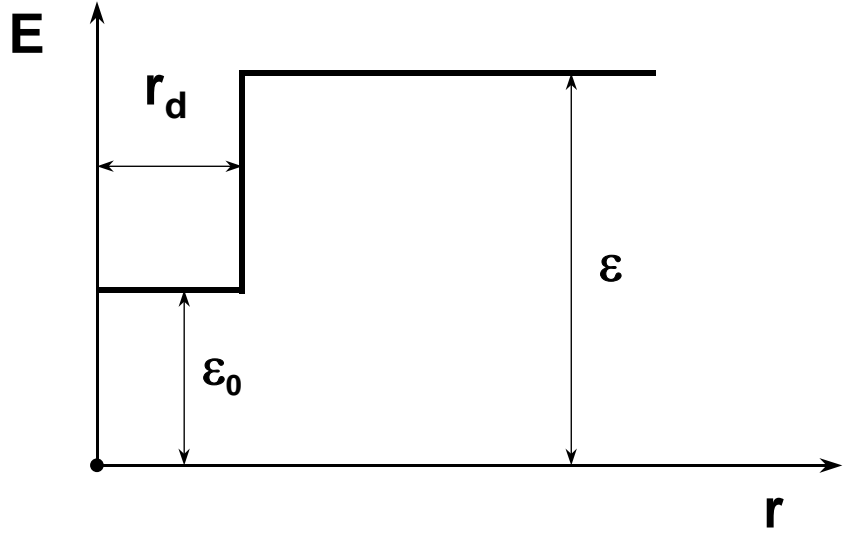

Figure 3. Vortex energy (per unit length) *vs.* vortex displacement in the vicinity of the pinning center of radius $r_d$.

As an example, let us consider the case of a strongly localized pinning force when the vortex is pinned by a potential well of small radius $r_d$ like that sketched in Figure 3: the vortex energy per unit length (vortex-line tension) is given by $\varepsilon$ for vortex line segments outside the potential well and by $\varepsilon_0$ for segments inside the well. Thus, the pinning energy per unit length is $\varepsilon$-$\varepsilon_0$. In fact, such a potential well model may describe pinning of vortices by, for example, columnar defects or planar defects, such as twin or grain boundaries [41, 42]. The latter is very relevant in laser ablated thin films. Therefore, we also use such strong pinning potential as a rough qualitative model for other types of pinning sites, in order to illustrate the effect of bulk pinning on current distribution and magnetic relaxation.

If the current distribution were uniform, such a potential well would keep the vortex pinned until the current density $j_y$ exceeds the critical value $c(\varepsilon - \varepsilon_0)/\Phi_0 r_d$. The escape of the trapped vortex line from the potential well occurs via formation of the un-trapped circular segment of the vortex line (see Figure 4). In this case, both the critical-current density and the energy barrier for vortex depinning do not depend on film thickness [41].

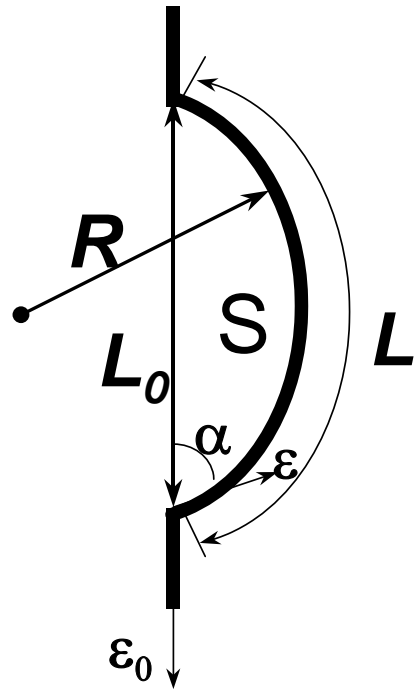

Figure 4. Model for the vortex line depinning out of the trapping potential

However, in perpendicular geometry the current distribution is not homogeneous. In order to find it for the critical state, we may use the following approach. The vortex line consists of the trapped and untrapped segments as shown in Figure 4. The un-trapped segment is beyond the potential well, therefore there is no bulk pinning force acting on it. This means that Eq.(6) describes the shape of this segment at $k=0$.

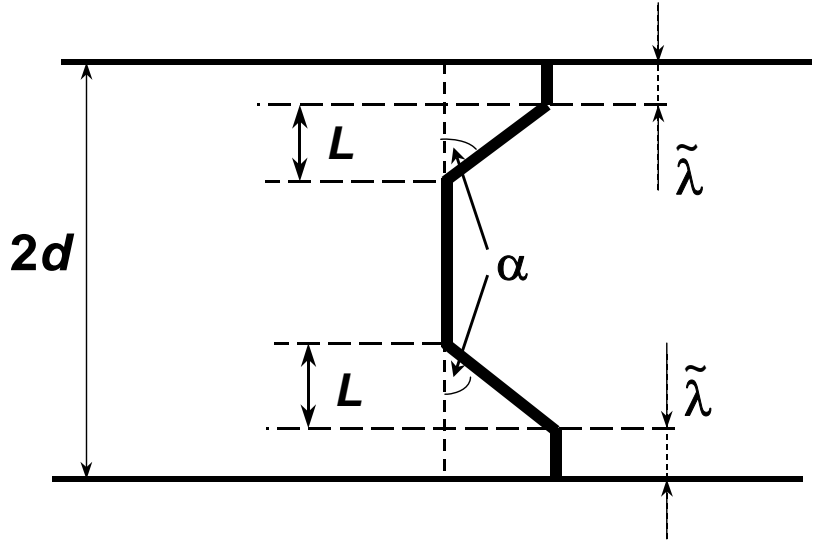

Figure 5. Schematic configuration of the vortex line in thin film

Applying the theory of Chapter II.A.3(a), one determines that the total current $I_y = \int_{-d}^{d} j_y(z) dz$ is concentrated near the film surfaces within a narrow surface layer of width $\tilde{\lambda}$. Inside the surface layer the vortex line is curved, but has a straight segment of length $L$ outside the layer, as illustrated in Figure 5. As for the vortex-line segment trapped by the potential well, we assume that it is straight and vertical, neglecting its possible displacements inside the potential well. Formally speaking, our approach introduces a non-homogeneous bulk-pinning constant $k$ assuming that $k=0$ for the un-trapped segment and $k=\infty$ for the trapped one. The energy of the vortex line in this state is determined by the line tensions ($\varepsilon$ and $\varepsilon_0$) and is given by

$$E = 2\varepsilon \frac{L}{\cos(\alpha)} - 2\varepsilon_0 L - 2\frac{\Phi_0}{c} I_y L \tan(\alpha) = 2L \tan(\alpha)\left( \varepsilon \sin(\alpha) - \frac{\Phi_0}{c} I_y \right) \quad (27)$$

where the contact angle $\alpha$ is determined by the balance of the line-tension forces at the point where the vortex line meets the line defect:

$$\cos(\alpha) = \frac{\varepsilon}{\varepsilon_0} \qquad (28)$$

## 5. Magnetic relaxation in thin films

We now discuss the effect of various current distributions on the thickness dependence of magnetic relaxation. We first show, in Chapter II.A.5(a), that uniform current density cannot explain the experimentally observed thickness dependence. We also show, in Chapter II.A.5(b), that inhomogeneous current density distribution, resulting from the surface pinning only, cannot explain the experimental data as well. We demonstrate that only the presence of bulk pinning and the resulting current inhomogeneity may lead to an accelerated relaxation in thinner films. We also discuss in Chapter II.A.5(b) the general case when both bulk and surface pinning are present.

### (a) Homogeneous current density distribution

As pointed out above, if the current distribution is uniform throughout the film thickness, a trapped vortex may escape from the potential well (Figure 3) via formation of a circular segment of the vortex line (Figure 4), with energy

$$E = \varepsilon L - \varepsilon_0 L_0 - \frac{\Phi_0}{c} j_y S \qquad (29)$$

where $L$ and $L_0$ are the lengths of the vortex line segment before and after formation of the loop, $S$ is the area of the loop [41]. If the loop is a circular arc of the radius $R$ and the angle $2\alpha$ (Figure 4), then $L_0 = 2R\sin(\alpha)$, $L = 2R\alpha$, and $S = R^2(2\alpha - \sin(2\alpha))/2$, where the contact angle $\alpha$ is given by Eq.(28). Then,

$$E = 2R(\varepsilon\alpha - \varepsilon_0 \sin(\alpha)) - \frac{\Phi_0}{2c} j_y R^2 (2\alpha - \sin(2\alpha)) = (2\alpha - \sin(2\alpha))\left(\varepsilon R - \frac{\Phi_0}{2c} j_y R^2\right) \qquad (30)$$

The height of the barrier is determined by the maximum energy at $R_c = \varepsilon c / \Phi_0 j_y$:

$$E_b = (2\alpha - \sin(2\alpha)) \frac{\varepsilon^2 c}{2\Phi_0 j_y} \qquad (31)$$

As one might expect, this barrier does not depend on the film thickness. We stress that this estimation is valid only for the 3D case when $d > R_c$. If $d < R_c$ the energy barrier is obtained from Eq.(62) by substituting $d = R_c$. This case of uniform current, however, leads to a thickness *independent* current density, thus cannot describe the experimental data.

## (b) Inhomogeneous current density distribution
### (i) Surface pinning

In this case, the whole current is confined to the surface layer of width $\tilde{\lambda}$. It is apparent from Eq.(9) that for typical experimental fields (~1 T) $\tilde{\lambda}$ is smaller than the film thickness. This means that current flows mostly in a thin surface layer. Thus, all creep parameters, including the creep barrier, are governed by the total current $I_y$, and not by the average current density $I_y/2d$. Then, apparently, the critical current density and the creep barrier are larger for thinner films, similar to the case of the collective-pinning effect mentioned above. Thus, also this scenario cannot explain the observed accelerated relaxation in the thinner films.

### (ii) Short-range bulk pinning

Let us consider the relaxation process for a critical state supported by the short-range pinning force discussed in Chapter II.A.3(b). The energy $E$ of the vortex line is given by Eq.(27). The average critical current density corresponds to $E=0$ and is inversely proportional to the film thickness [see also Eq.(26)]:

$$j_c = \frac{I_y}{2d} = \frac{c\varepsilon}{2d\Phi_0}\sin(\alpha) \qquad (32)$$

The energy barrier is given by the maximum energy at $d=L+\tilde{\lambda}\approx L$ when the whole vortex line has left the potential well (Figure 6):

$$E_b = \tan(\alpha)\left(2d\varepsilon\sin(\alpha) - 4d^2\frac{\Phi_0}{c}j_a\right) \qquad (33)$$

where $j_a=I_y/2d$ is the average current density. If $j_c>j_a>j_c/2$, then $\partial E_b/\partial d < 0$, i.e., the barrier is larger for thinner films. However, for $j_a<j_c/2$ the derivative $\partial E_b/\partial d > 0$, and the barrier *increases* with the increase of the film thickness. Thus, under this condition ($j_a<j_c/2$) the magnetic relaxation rate is larger in the thinner samples.

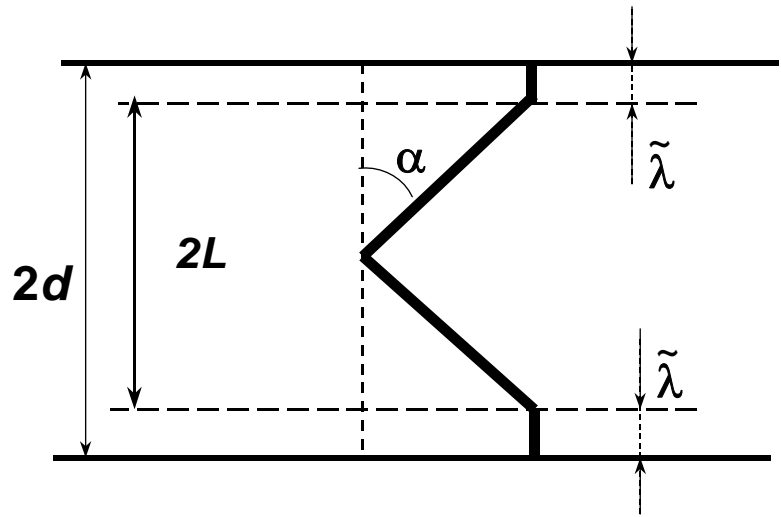

Figure 6. Maximum energy barrier for vortex depinning for dilute defect structure

The above analysis did not take into account the possibility of dense defects. By "dense" we mean that the distance $r_i$ from the neighbor potential well is less than $d\tan(\alpha)$. In this case, as is apparent from Figure 7, the maximal energy (the barrier peak) is smaller than the barrier calculated in Eq.(33). Then the barrier energy is given by

$$E_b = r_i\left(2\varepsilon\sin(\alpha) - 4d\frac{\Phi_0}{c}j_a\right) \tag{34}$$

In this case $\partial E_b/\partial d < 0$ and the energy barrier for thinner films is always larger. Therefore, one can see faster relaxation in thinner films only if the films are so thin that $d < r_i/\tan(\alpha)$ and the energy barrier is given by Eq.(33). From the experimental results shown below we infer that the average distance between effective defects $r_I \geq 1000$ Å in agreement with direct measurements using atomic force microscopy.

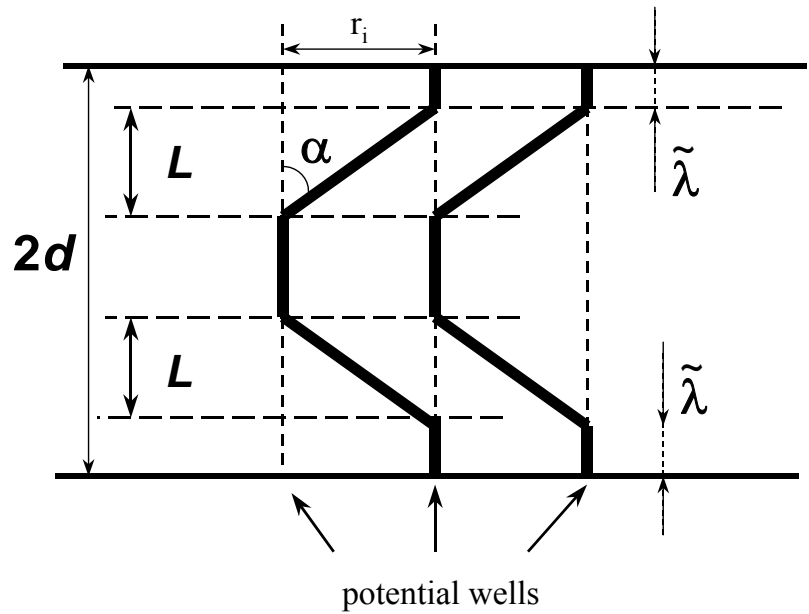

Figure 7. Maximum energy barrier for vortex depinning for a dense defect structure

To conclude, if the average current density in thin films becomes small enough compared to the original critical current density and if the films are thin enough, the relaxation *at the same average persistent current* is predicted to be faster for the thinner films.

### *(iii)* General case

In the simplified picture of the critical-state relaxation outlined in the previous subsection, the total current was concentrated within a very thin layer of the width $\tilde{\lambda}$. It was based on the assumption that the pinning force disappears when the vortex line leaves the small-size potential well, whereas inside the potential well the pinning force is very strong. As a result, outside the thin surface layers of the width $\tilde{\lambda}$ the vortex line consists of two straight segments (Figure 5). In the general case, the distribution of the pinning force may be smoother and the shape of a vortex line is more complicated, but the tendency must be the same: the current confined in a narrow surface layer drives the end of a vortex line away from the potential well to the regions where the pinning force is weaker and the vortex line is quite straight with the length proportional to thickness of the film if the latter is thin enough. Therefore, the barrier height for the vortex jump is smaller for smaller *d*.

We also note that we do not consider an anisotropic case and limit our discussion to isotropic samples. The effect of anisotropy on the barrier height was considered in details in [41]. In the presence of anisotropy the circular loop becomes elliptic and the vortex-line tension ε must be replaced by some combination of vortex-line tensions for different crystal directions. These quantitative modifications are not essential for our qualitative analysis.

Our scenario assumes that the current is concentrated near the film surfaces. In general, the width of the current layer may vary from $\tilde{\lambda}$ to the effective Campbell length $\lambda_C$. One may then expect a *non-monotonous* thickness dependence when $\lambda_C$ is comparable with *d*. As we see, the Campbell length is an important quantity in determining whether current density inhomogeneity must be taken into account or not (in the absence of the surface pinning). The length $\lambda_C$ can be estimated from the micro-wave experiments: according to Golosovskii et al. [43] $\lambda_C \approx 1000\sqrt{H}$, where the magnetic field *H* is measured in *Tesla*. For $H \leq 0.2\ T$ this results in $\lambda_C \approx 450$Å or $2\lambda_C \approx 900$ Å, which has to be compared with the film thickness.

## **B.** *Experiments in* $Y_1Ba_2Cu_3O_{7-\delta}$ *films*

A decrease of the measured current density with an increase of the film thickness is reported in numerous experimental works [25-28, 44-47]. This is consistent with the predictions given above for either surface or/and bulk pinning. Both pinning mechanisms predict similar ~1/*d* dependence of *j* and it is, therefore, impossible to distinguish between surface and bulk pinning in this type of measurements. Only the additional information from the thickness dependence of the relaxation rate allows the drawing of some conclusions about the pinning mechanisms. Measurements of magnetic relaxation in films of different thickness reported here were originally discussed in detail in [27, 28].

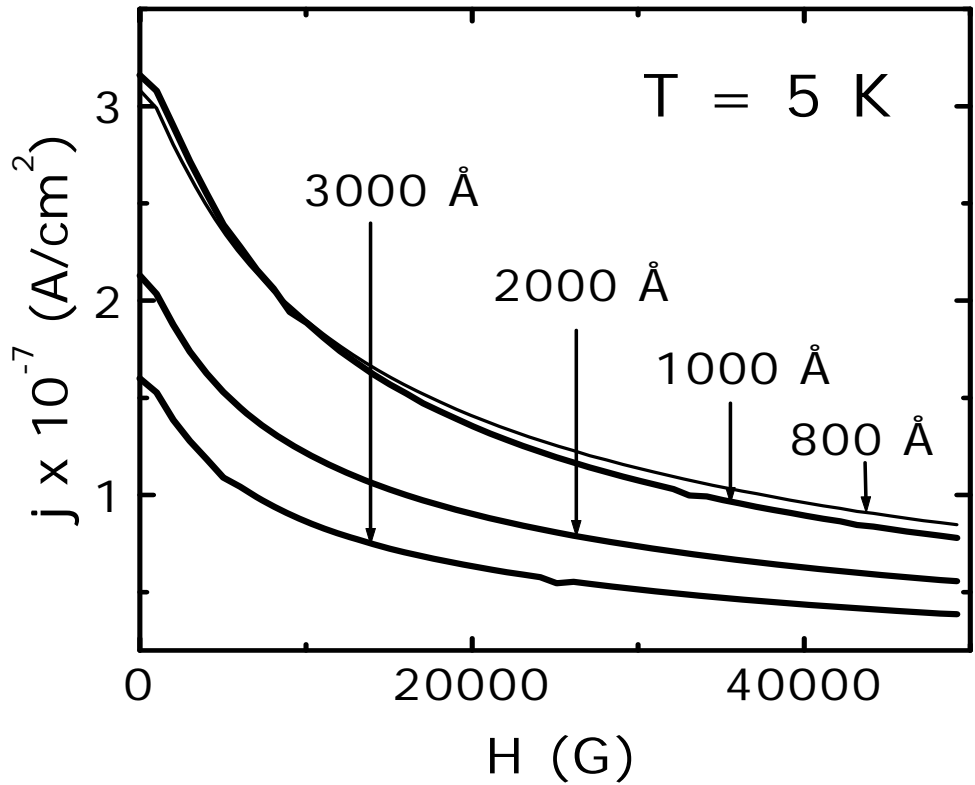

Figure 8. Average persistent current density as a function of magnetic field at *T*=5 *K* for films of different thickness.

Measurements were conducted on four 5×5 mm$^2$ *YBa$_2$Cu$_3$O$_{7-\delta}$* films of thickness 2d=800, 1000, 2000 and 3000 Å, prepared by the laser ablation technique on *SrTiO$_3$* substrates [48]. All samples had T$_c$≈89 *K*. The morphology of the samples was examined by atomic-force microscopy (AFM) technique and was found to be similar: the average grain size 100-5000 Å and the inter-grain distance 50 Å. The magnetic moment was measured as a function of field, temperature and time, using a *"Quantum Design"* SQUID magnetometer (see Chapter I.A).

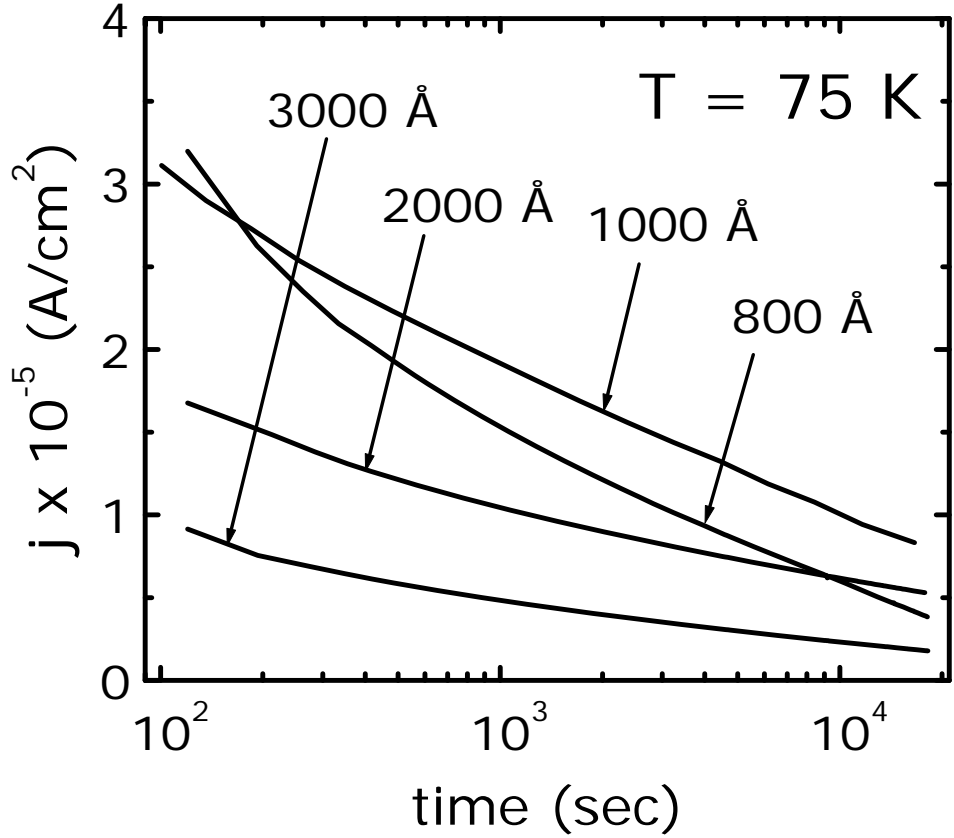

Figure 9.  Time evolution of the average persistent current density at *T*=75 *K* for films of different thickness.

The *average* persistent current density was extracted from the magnetic hysteresis loops using the Bean model, Eq.(104). Figure 8 shows the persistent current density *j* at *T*=5 *K* as a function of the applied magnetic field *H*. Apparently, *j* is larger in thinner films. The same trend is found at all temperatures. These observations are in good agreement with Eqs.(19) and (26). We note, however, that since the value of $j_s$ is not known, we cannot point to the dominance of pure surface, pure bulk or a mixed type of pinning.

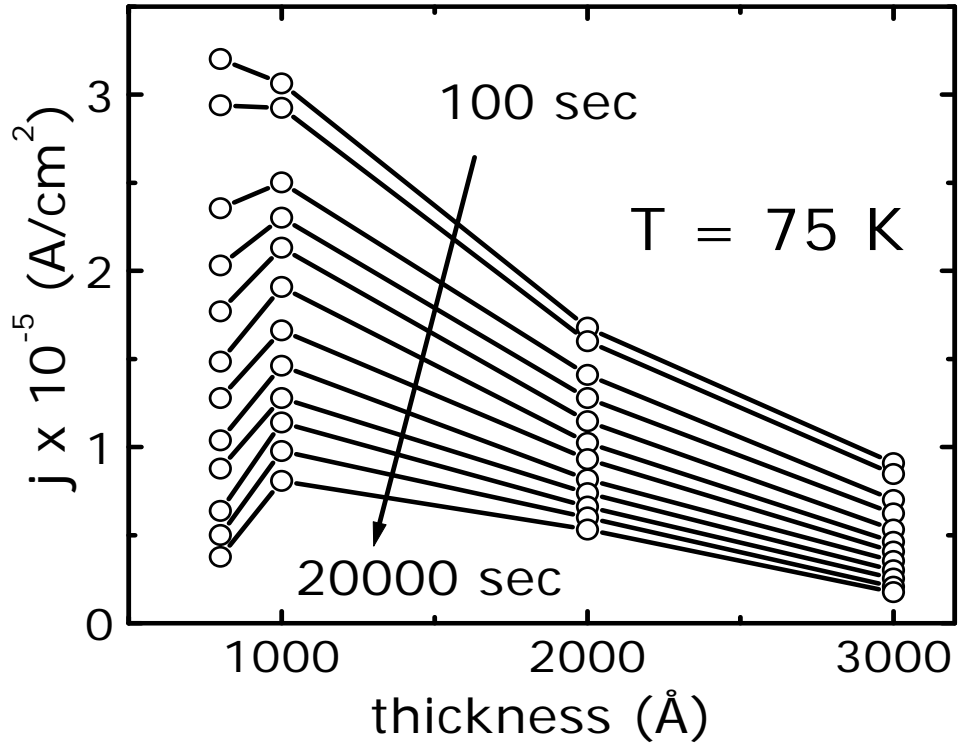

Figure 10. Thickness dependence of the average persistent current density at T=75 K taken at different times.

Figure 9 shows typical relaxation curves at $H=0.2\ T$ (ramped down from 1 $T$) measured in films of different thickness. The interesting and unexpected feature is that curves cross, i. e., the relaxation is faster in thinner films. This is further illustrated in Figure 10 where $j$ vs. $d$ is plotted at different times. At the beginning of the relaxation process, the average current density in the thinner films is larger. However, in the thinner films, the current density decreases much faster than in the thicker ones; as a result $j_a$ exhibits a non-monotonous dependence on thickness at later times, as shown in Figure 10. The faster relaxation in thinner films is in qualitative agreement with our results, discussed in Chapter II.A.5, in particular in subsections of Chapter II.A.5(b) discussing short-range bulk pinning and the general case. There, we find that such acceleration of the relaxation in thinner films may be understood only if we consider inhomogeneous bulk current density. In reality, it is very probable that *both* surface and bulk pinning mechanisms lead to inhomogeneous current density with a characteristic length scale in between the short (surface pinning) length $\tilde{\lambda}$ and the larger Campbell length.

## C. *Summary and conclusions*

Based on the two-mode electrostatics approach we built a consistent theory of the critical state in thin type-II superconducting films *throughout the film thickness*. We show that, irrespective of the pinning mechanism, current density is always larger near the surface, and decays over a characteristic length scale, which is in between $\tilde{\lambda}$ (of order of the inter-vortex distance $a_0$) and the Campbell length $\lambda_C$. The length scale $\tilde{\lambda}$ is determined by the (finite) vortex tension and by the boundary conditions which force vortices to be perpendicular to the surface of the superconductor, whereas the Campbell length $\lambda_C$ is determined by bulk pinning potential.

Following this novel physical picture we conclude that:

- Current density and magnetic induction in thin films in perpendicular field are highly inhomogeneous throughout the film thickness. Surface pinning significantly enhances these inhomogeneities.
- Averaged over the thickness current density decreases with the increase of film thickness approximately as 1/d.
- Magnetic relaxation is *slower* in thinner films in the following cases:
  - In the absence of bulk pinning, i.e., only surface pinning is effective.

> In the presence of bulk pinning, provided that the ratio between thickness and distance between neighboring defects is above a certain threshold $d/a \sim 1$.

- Magnetic relaxation is *faster* in thinner films only if bulk pinning is effective and the ratio $d/a$ is below this threshold.

In the experimental data presented here the measured average current $j_a$ decreases with the increase of film thickness as predicted, and the relaxation rate is larger for the thinner films, suggesting that $d/a \leq 1$, and the effective distance between defects $\geq 1000$ Å.

# Chapter III. $Y_1Ba_2Cu_3O_{7-\delta}$ thin films in inclined field

## A. Anisotropic thin film in inclined field

In the previous chapter we have shown that the important parameter characterizing a thin sample is the total current $I_y$, see e. g. Eq.(22). This current can be estimated from the measurements of magnetic moment, a technique that is not sensitive to a distribution of the current density along the $z$-axis. When the magnetic field is perpendicular to the sample plane one can use a model suggested by Gyorgy *et. al.* [49]. We discuss applicability of this model and summarize the conversion formulae between the measured magnetic moment and averaged current density in Appendix B. It is worth noting that results of Appendix B are not confined to the thin film geometry, but describe the sample of arbitrary thickness to width ratio.

In this chapter we extend the analyses of thin film in perpendicular field to the case of films in inclined field. Due to geometrical effects and magnetic anisotropy this situation is qualitatively different from the case when external magnetic field is applied perpendicular to the sample plane. In the case of an anisotropic sample, tilted magnetic field results in the appearance of persistent currents of different densities and the formulae derived in Appendix B should be modified. (Note, that in the case of in-plane anisotropy these formulae have to be modified as well, even if $\theta=0$. See discussion after Eq.(36)). In the following we consider the sample in an inclined magnetic field, as depicted in Figure 11. As we will see, in thick samples, all the components of the magnetic moment are significant and anisotropy must be explicitly taken into account. However, in thin films due to the extreme geometry, we can always consider only one component of the magnetic moment – along the film normal, usually *c*-direction.

Below we re-write the most often used formula, Eq.(103), for an arbitrary thick rectangular sample with anisotropic persistent current densities, and show experimentally that with a good accuracy in thin films magnetic moment is oriented along the film normal.

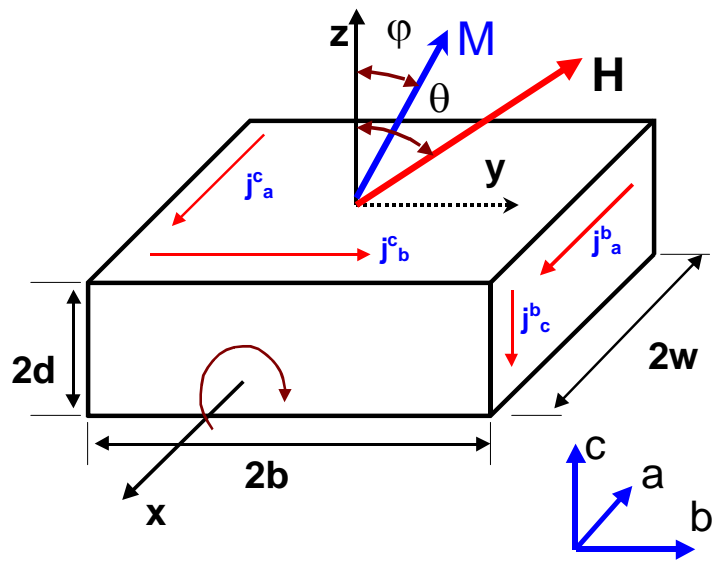

Figure 11    Geometrical arrangement considered for an anisotropic superconductor in an inclined field.

We will consider the case of a long bar, so that w<<b. This allows us to avoid complications due to U-turns [38] where out-of-plane currents mix with the in-plane currents. In this geometry the components of the magnetization (*m=M/V*) are:

$$m_c \approx \frac{j_b^c w}{2c} \tag{35}$$

and

$$m_b = \begin{cases} \dfrac{j_a^b d}{2c}\left(1 - \dfrac{j_a^b}{j_c^b}\dfrac{d}{3w}\right) & \dfrac{j_c^b}{j_a^b} \geq \dfrac{d}{w} \\ \dfrac{j_c^b w}{2c}\left(1 - \dfrac{j_c^b}{j_a^b}\dfrac{w}{3d}\right) & \dfrac{j_c^b}{j_a^b} \leq \dfrac{d}{w} \end{cases} \quad (36)$$

Here we have introduced anisotropic current densities $j_x^y$, where the subscript denotes the direction of the current flow (*a, b, c*, see Figure 11) and the superscript denotes the direction of the magnetic field. This terminology is necessary because current densities depend on the way the vortices move. For example, in highly anisotropic compounds $j_a^c \ll j_a^b$, because in the first case the current density is determined by the propagation of the vortices perpendicular to the film plane and current density is determined by the pinning on crystal structure imperfections, whereas the current $j_a^b$ is due to the strong pinning of vortices crossing the Cu-O planes.

Note that Eq.(36) is the anisotropic form of Eq.(103) for magnetization calculated along the *b*-direction.

The convenient quantity to look for is the ratio $R=m_b/m_c$. This ratio determines the angle $\varphi$ between the total magnetic moment and the c-axis, i. e., $\varphi=\arctan(R)$.

From the above formulae we get:

$$R = \begin{cases} \dfrac{j_a^b}{j_b^c}\dfrac{d}{w}\left(1 - \dfrac{j_a^b}{j_c^b}\dfrac{d}{3w}\right) & \dfrac{j_c^b}{j_a^b} \geq \dfrac{d}{w} \\ \dfrac{j_c^b}{j_b^c}\left(1 - \dfrac{j_c^b}{j_a^b}\dfrac{w}{3d}\right) & \dfrac{j_c^b}{j_a^b} \leq \dfrac{d}{w} \end{cases} \quad (37)$$

In thick samples, both cases can be realized and global measurements alone are not sufficient to determine the anisotropy. Here one can use magneto-optics for direct visualization of flux penetration [5-7, 10, 50]. It should be emphasized that the usual notion of magnetic anisotropy does not represent the ratio $\varepsilon = j_b^c / j_a^b \leq 1$, since the latter is the result of irreversible current densities and, as we will see below, may be very large. Another complication is that current densities depend on the magnetic field and it is not clear which component of the tilted magnetic field one must consider calculating the particular component of current density.

In the case of thin films the situation is greatly simplified, since for typical samples $d/w \approx 10^{-4} - 10^{-5}$ and one may safely write:

$$R_{film} \approx \frac{j_a^b}{j_b^c}\frac{d}{w} \tag{38}$$

Thus, by measuring experimentally the ratio $R_{film}(H,T)$ one can estimate the ratio of the current densities. Such measurements can be performed using a magnetometer sensitive to two components of the magnetic moment. We measured the vector magnetic moment using a Quantum Design MPMS SQUID magnetometer, see Appendix A for discussion. The MPMS system used in these experiments utilizes two orthogonal pickup coils to measure the components of the magnetic moment both parallel $M_L$ and perpendicular ($M_T$) to the field. As shown in Figure 11, the angle of sample rotation $\theta$ is defined as the angle between the film normal (crystallographic c-axis) and the field axis.

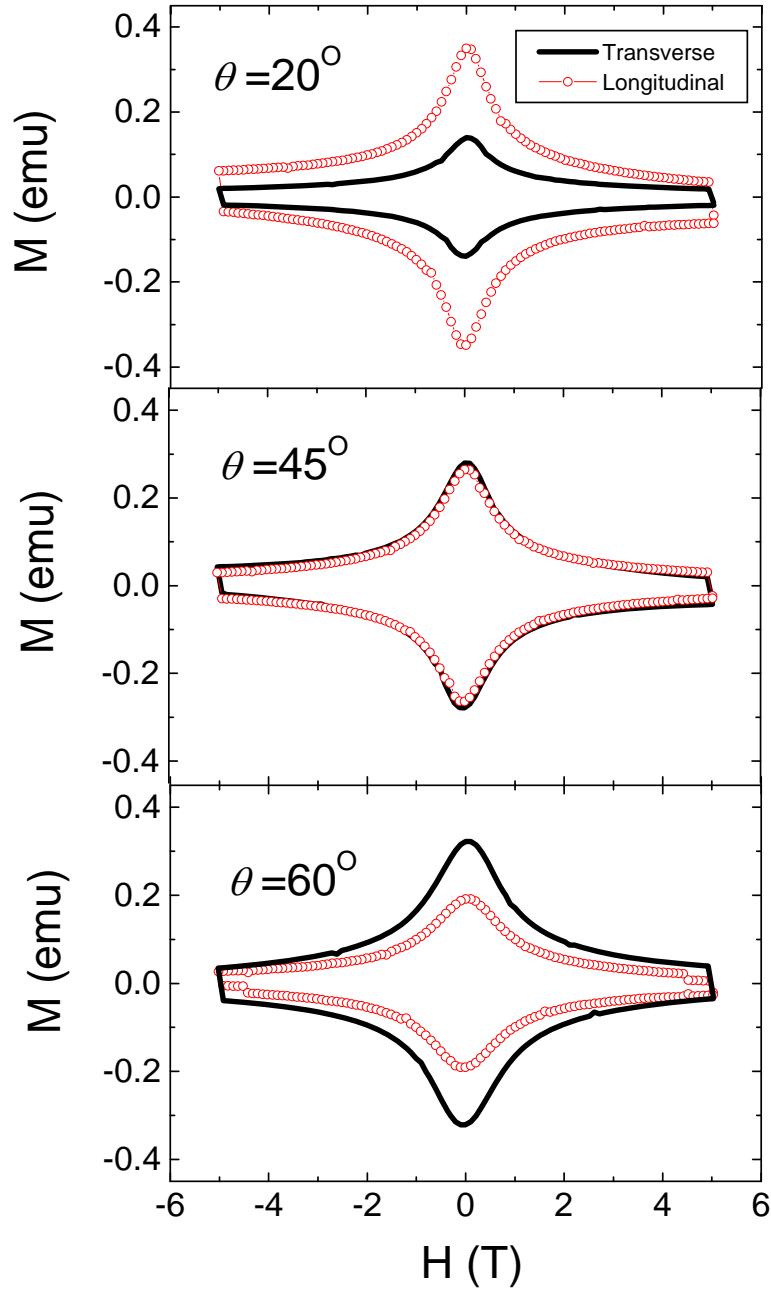

Figure 12. Magnetization loops measured in thin $YBa_2Cu_3O_{7-\delta}$ film at T=20 K at angles $\theta$=20°, 45° and 60°. Symbols are the longitudinal component and the solid line is the transverse component of $M$.

Figure 12 shows the raw data measured on $YBa_2Cu_3O_{7-\delta}$ thin film of $w$ = 0.1 cm and $d$=4·10$^{-5}$ cm at T=20 K at three different angles $\theta$=20°,45° and 60°. Note change in the magnitudes of the components of the magnetic moment. Using rotation of the coordinate system, measured $M_L$ and $M_T$ can be combined to reconstruct the components of the magnetic moment along the *ab*-plane ( $M_{ab} = \cos(\theta)M_T - \sin(\theta)M_L$ ) and the *c*-axis ( $M_c = \cos(\theta)M_T + \sin(\theta)M_L$ ) as well as the total magnetic moment **M**. The direction of **M** can be quantified by φ, defined in Figure 11 as the angle between **M** and the *c*-axis (i.e., the direction **c**).

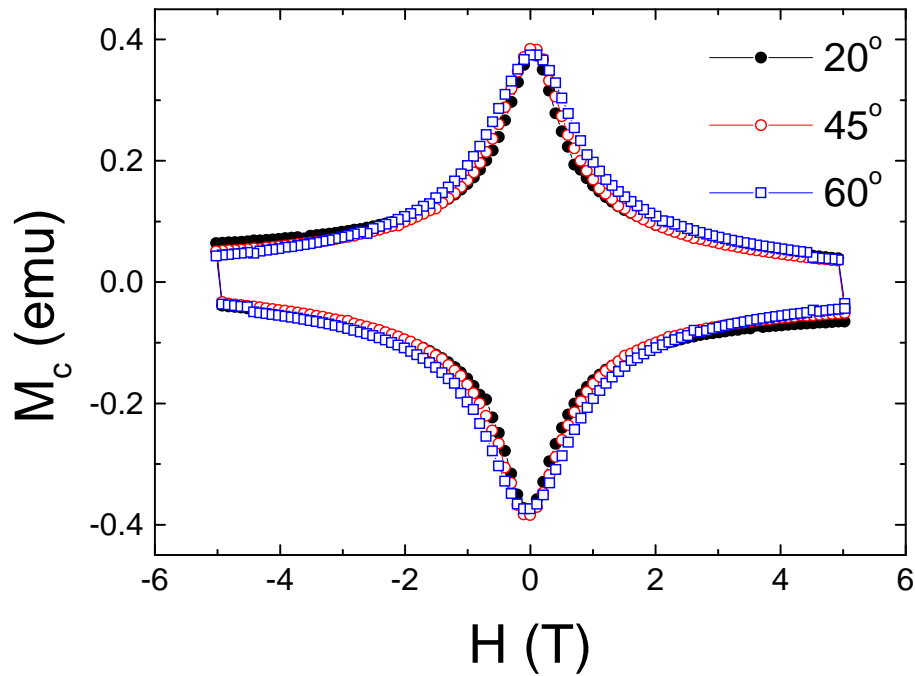

Figure 13   Normal-to-plane component of the magnetic moment calculated from the loops of Figure 12.

Figure 13 presents $M_c$ component of the magnetic moment (normal-to-plane) calculated using data of Figure 12 at different angles θ. Surprisingly, $M_c$ scales for these three angles θ, indicating very small sensitivity of the total magnetic moment to the in-plane component.

To identify the direction of the magnetic moment we calculate the angle $\varphi = \arctan(M_{ab}/M_c)$ from the data of Figure 12. The angle φ is plotted in Figure 14 as a function of field. We note, firstly, that in a worst case φ does not exceed 10° and for intermediate magnetic fields (< 1 Tesla) this angle is less than $2^o$. Thus the magnetic moment always points in the $c$-direction. Another observation is that angle φ consistently increases with the increase of magnetic field indicating increasing dominance of current $j_a^b$ at larger fields. For our film d/w~$10^{-4}$ and $R_{film}$~0.17 for φ=$10^o$ and $R_{film}$~0.035 for φ=$2^o$. Thus, using Eq.(38), we get: $j_a^b / j_b^c \approx 350-90$. These values are, indeed, much larger than the known anisotropy (ε~1/7) of $YBa_2Cu_3O_{7-\delta}$.

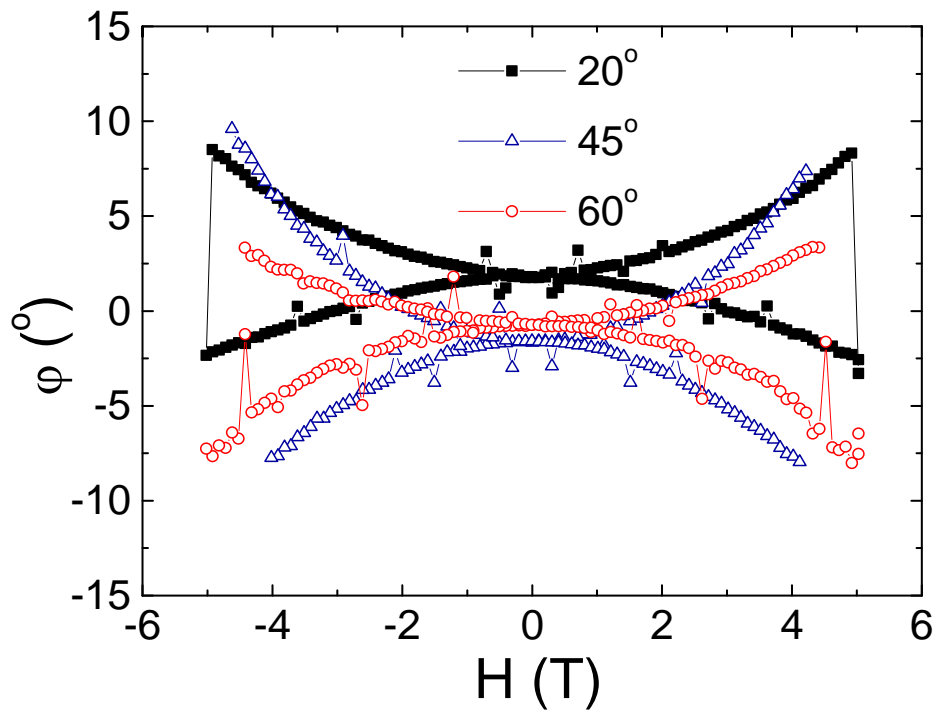

Figure 14. Angle φ between the magnetic moment and the c-axis as deduced from data of Figure 12.

These values indicate that current density $j_a^b$ dominates at high fields when the "usual" in-plane current $j_b^c$ dies out. Theoretically, the current density $j_a^b$ should represent the intrinsic pinning current, which may be close to the depairing one. If $j_b^c \sim 10^6 \, A/cm^2$ then $j_a^b \sim 10^8 - 10^7 \, A/cm^2$, then the upper limit is in a good agreement with theoretical values of the depairing for *YBa$_2$Cu$_3$O$_{7-\delta}$* compound [2].

In conclusion, in thin films one can almost always neglect the in-plane magnetic moment and approximate the total magnetic moment by its *c*-component. At the same time some peculiarities in the field dependence of the calculated persistent current density may be explained considering the second component of the magnetic moment. One of the examples is the observed *in-plane* anisotropy of the flux penetration in inclined magnetic field [8, 9, 51]. According to our scenario, current $j_a^b$ replaces current $j_a^c$, whereas current $j_b^c$ remains intact. Thus, penetration along the *b*-side will be inhibited by a large current $j_a^b$ inducing the apparent in-plane anisotropy.

## B. *Thin film rotated in external magnetic field*

Instead of measuring the magnetization loops at fixed angle $\theta$, one may fix the external magnetic field and rotate the sample. We have analyzed this situation in Ref. [ RP15 ] and summarize these results in this Section.

Experiments with rotating samples may supply ample information regarding the superconducting properties and they were used for studies of the 3D anisotropic Ginzburg-Landau theory in a single $YBa_2Cu_3O_{7-\delta}$ crystal [52-54], pinning strength distribution in polycrystalline materials [55-60], intrinsic anisotropy [61, 62] and interaction of vortices with pinning sites and external magnetic fields [56, 60, 63]. The conclusions in these works were drawn from the peculiarities observed in rotation curves at different ambient conditions. For example, it was found that in relatively high fields, well above the lower critical field $H_{c1}$, the $M(\theta)$ curves in forward and backward rotations do not coincide, but shift in relation to another on an angle $\Delta\theta$ [56, 61-63]. This was taken as evidence that the magnetic moment rotates frictionally, lagging behind the sample's *c*-axis, due to the interaction of vortices with the external field, thus exhibiting irreversible (hysteretic) behavior and an apparent phase shift between the two magnetization curves. Such a phase shift was not observed in lower fields, where the two curves were found to coincide and, it aappears to indicate that the sample is in a pure reversible (or Meissner) state. In this Section we show that all observed peculiarities may be explained using the modified critical state model [64].

Our conclusions are based on measurements of the angular dependence of the magnetization curves $M(\theta)$ of a thin *$YBa_2Cu_3O_{7-\delta}$* film. As discussed in Chapter III.A, in thin films we can safely neglect the in-plane component of a magnetic moment.

Therefore, in thin films, the results of rotational experiments are clear and much easier to interpret.

As we show below for fields below the Bean's full penetration field $H^*$ (and not only below $H_{c1}$) $M(\theta)$ is symmetric with respect to $\theta=\pi$ and the rotation curves for forward and backward rotation coincide (see also [61]). For $H>H^*$ the curves are asymmetric and they do not coincide. Moreover, the backward curve is a mirror image (with respect to $\theta=\pi$) of the forward one. Thus there is no "phase leg" between magnetic moment and field, but true magnetic irreversibility resulting from the sample remagnetization during rotation.

Many authors have pointed out the importance of the geometry for the proper analysis of the rotation experiments. In previous works this was limited to a consideration of the demagnetization in a reversible state [61, 65, 66], i.e., to the actual field on the sample edge which varies during rotation because of demagnetization. Maintaining this point, we further assert that demagnetization affects the rotation curves in an irreversible state. In particular, we find that the interval of angles for which the sample undergoes remagnetization shrinks dramatically as a result of a flat geometry. This results in sharp changes in $M(\theta)$.

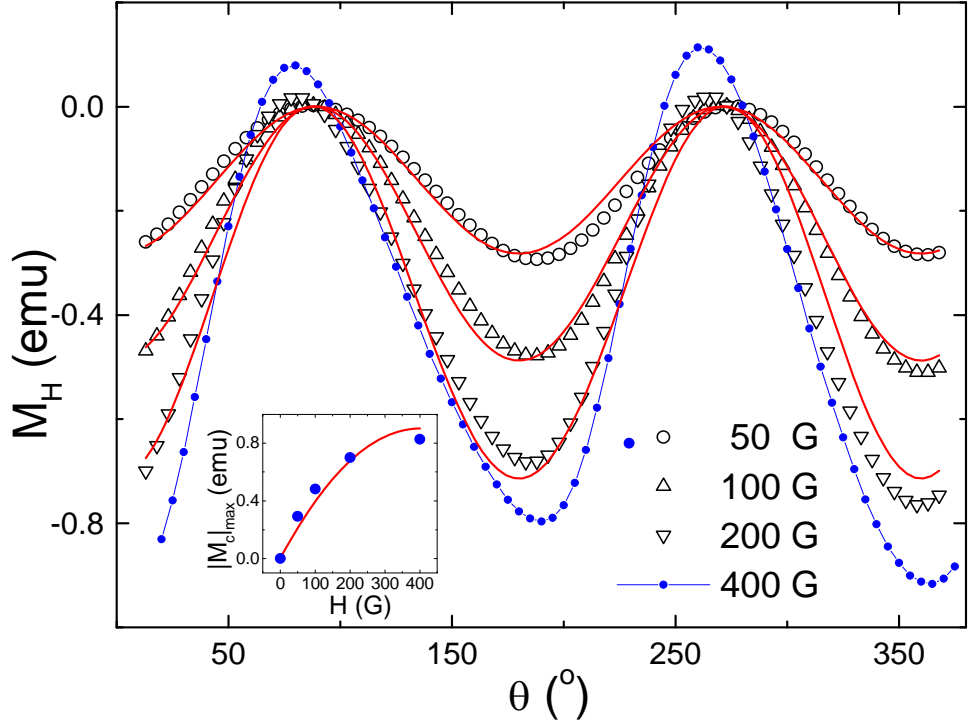

Figure 15. Forward rotation curves for $H$=50, 100, 200 and 400 $G$. *Inset*: maximal value of $M_H$ vs. $H$

A thin $YBa_2Cu_3O_{7-\delta}$ film of thickness $2d$=1000 Å and lateral dimensions of 5×5 mm$^2$ was laser ablated on a $SrTiO_3$ substrate [48]. The film was $c$-oriented, so that the $c$-axis points in a normal to a film surface direction with angular dispersion less than 2º and the $ab$-plane coincides with the film plane. For the rotation magnetic measurements we used an *"Oxford Instruments"* vibrating sample magnetometer (VSM) that enables sample rotation relative to an external magnetic field with a 1º precision (see Chapter I.B for discussion). The rotation axis is always perpendicular to the $c$-axis. Samples were zero field cooled down to the desired temperature when an external magnetic field $H$ was turned on. The component of the magnetic moment $M_H$ along the external field was then measured while the sample was rotated one full turn ("forward rotation") and then back ("backward rotation"). The initial field application was always along the $c$-axis. We find that turning on the external field at other angles does not yield any new information, since the rotation curve becomes independent of this angle after one full rotation - (see also [62]).

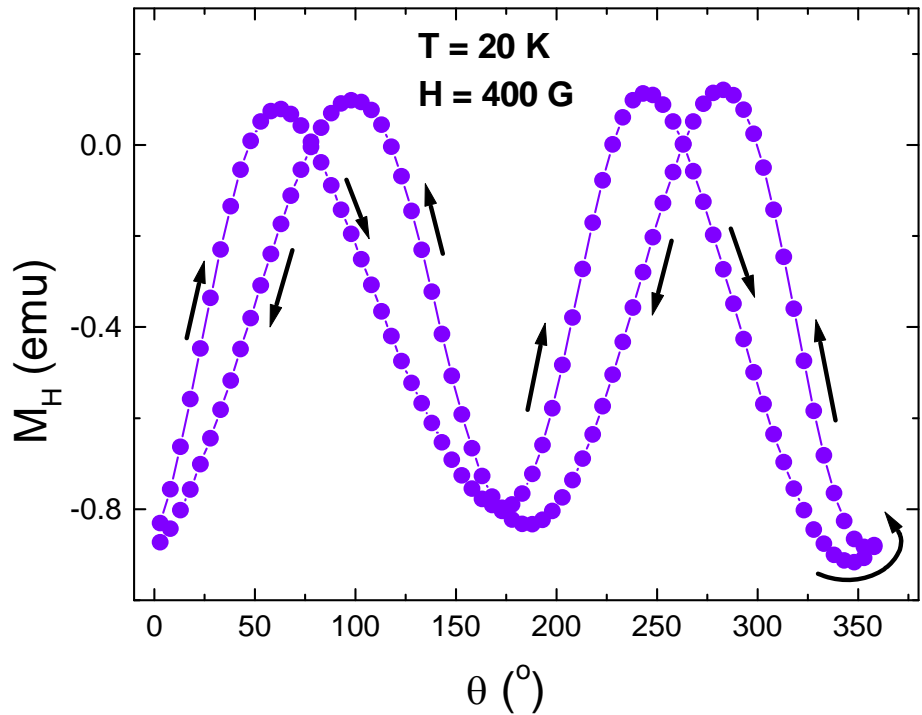

Figure 16. Forward and backward rotation curves at *H*=200 G

The symbols in Figure 15 present the measured angular dependence of the zero-field-cooled (ZFC) magnetization, $M_H(\theta)$, for the *YBa$_2$Cu$_3$O$_{7-\delta}$* film at different values of *H* at *T*=20 K. At this temperature, the apparent (reduced due to demagnetization effects) $H_{c1}\approx50$ *G*, and $H^*$=380 *G*, as determined from direct "static" *M(H)* measurements. The rotation curves are symmetric with respect to $\theta=\pi$, in spite of the fact that the applied fields are definitely larger than $H_{c1}$. This implies that such reversibility with respect to direction of rotation is not indicative of the *true* magnetic reversibility.

The apparently asymmetric curve for $H = 400\ G > H^*$ is shown for comparison. In Figure 16 this curve is shown for both forward and backward directions. It appears as if there is a simple phase shift between the curves. However, high - field measurements performed at *H*=1.5 *Tesla*, shown in Figure 17 demonstrate that the backward rotation curve is a mirror image of the forward one with respect to θ=π. Therefore, we conclude that the observed asymmetry does not imply a phase shift, but a true magnetic hysteresis with respect to rotation, which means a reverse of the magnetic moment when the direction of rotation is reversed.

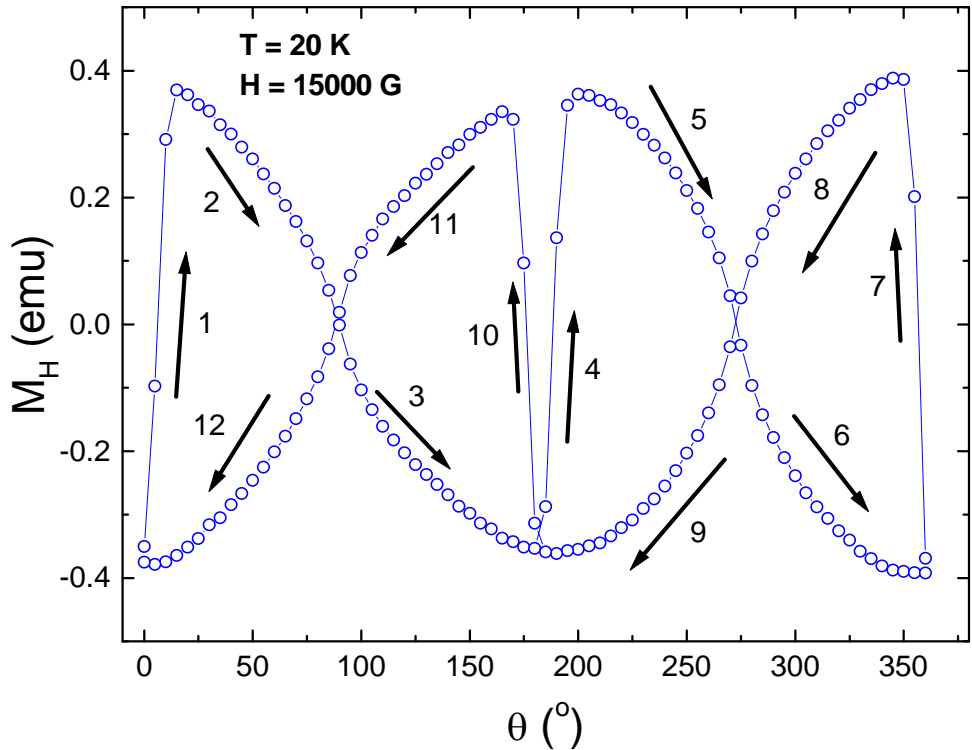

Figure 17. Forward and backward rotation curves at *H*=1.5 Tesla (mirror image)

In order to understand the variation of *M* during rotation in a system with isotropic pinning one has to consider separately the variation of its components along the c-axis ($M_c$) and the ab-plane ($M_{ab}$). This leads to a consideration of the projections of the applied field on the c-axis ($H_c$) and the ab-plane ($H_{ab}$), respectively. The *measured* magnetization is $M_H = M_c \cos(\theta) + M_{ab} \sin(\theta)$ and, therefore, as discussed in Chapter III we can safely write for thin films in the *irreversible* state $M_H \approx M_c \cos(\theta)$. The effective field $H_c$ cycles during the rotation between ±H and the calculation of *M(θ)* is thus analogous to the calculation of *M(H)*. In Figure 18 we demonstrate the validity of this concept in thin films by comparing direct *M vs. H* measured along the *c*-axis and $M_c = M_H/\cos(\theta)$ *vs.* $H\cos(\theta)$. We stress that this claim is valid only if a sample does not have any induced anisotropy of pinning, e.g., twin boundaries or columnar defects (see Chapter III.C).

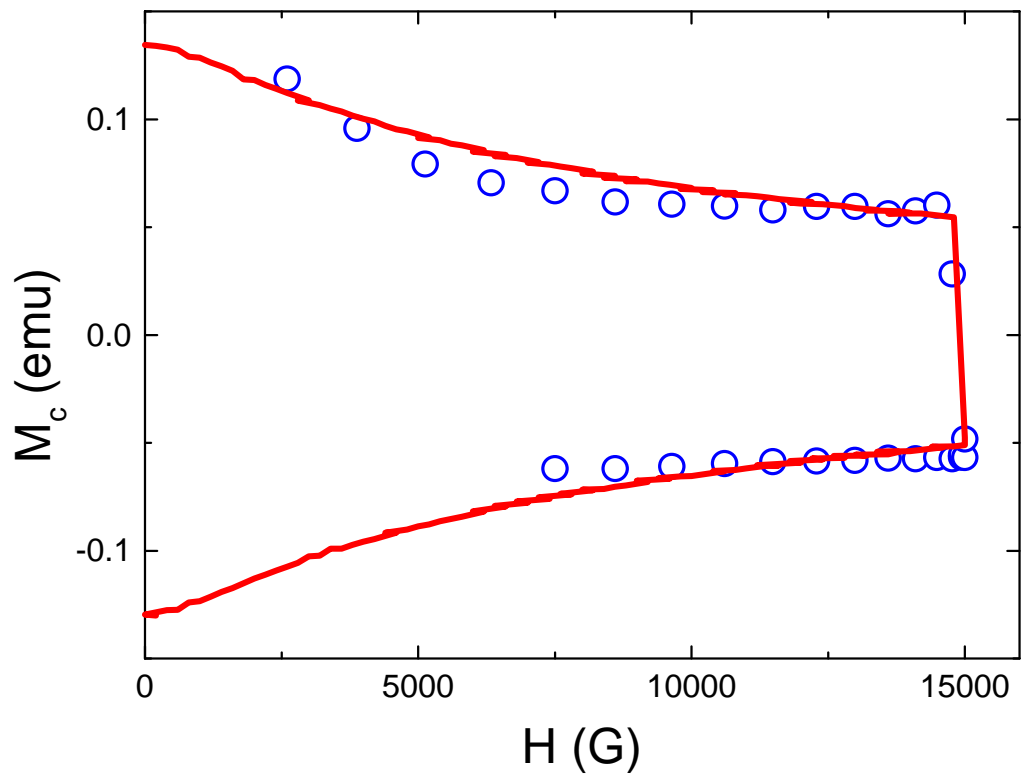

Figure 18. Comparison of the direct (solid line) *M(H)* measurement and reconstruction from the rotation as discussed in the text

This approach leads directly to the consideration of the two field regimes: (a) Moderate fields - $H \leq H^*$, where the sample is only partially occupied by vortices and (b) *High fields* – $H \geq H^*$, where vortices occupy the entire sample space. For fields below $H_{c1}$ the correct analyses of the data was given previously in a number of reports, e.g. [61, 66].

For the sake of clarity, the analysis below is based on the critical state model for an infinite slab. Whereas in a fully magnetized state the volume magnetization for a thin sample and for an infinite slab is the same and given by formulae of Chapter I.A, the remagnetization process is quite different [22, 38, 40]. Nevertheless, we use the critical state model, in order to demonstrate a general approach to the problem, avoiding unnecessary complications of the analysis. In fact even linear with field approximation for a remagnetization stage satisfactory describes the results.

### 1. Intermediate fields $H<H^*$

Utilizing the parameter $x=1-\cos(\theta)$ one may express the difference between the external magnetic field $H$ and its projection on the $c$-axis during rotation, as $\Delta H=Hx$. In the following we describe a ZFC experiment and consider forward rotation only. The backward rotation may be obtained from the formulae below by substituting $\theta_{back}=2\pi-\theta$. In this field regime, the curve obtained by such a substitution coincides with the forward curve.

In the framework of the Bean model we get for the projection of the total moment along the $c$-axis for fixed external field $H$:

$$M_c = -\frac{H^2}{8H^*}\left(x^2+2x-4\right)-H(1-x) \tag{39}$$

Note that for $x=0$ we recover the Bean result for partial magnetization [3]. The component $M_c$ varies continuously with $\theta$ in a whole interval of angles implying that the magnetic flux profile inside the sample changes for any change in $\theta$. Finally, Eq.(39), expressed in terms of $\theta$, yields for the measured component of the total moment along the direction of the external magnetic field:

$$M_H = H\cos(\theta)\left\{\frac{H}{8H^*}\sin^2(\theta)-\left(1-\frac{H}{2H^*}\right)\cos(\theta)\right\} \tag{40}$$

Apparently, the *forward* rotation curve described by Eq.(40) is symmetric with respect to $\theta=\pi$ and therefore it coincides with the backward rotation curve, obtained by substitution $\theta_{back}=2\pi-\theta$. In other words, reversibility with respect to the direction of rotation does not imply a "true" magnetic reversibility that is expected either in the Meissner state or in the unpinned state. The magnetic moment along the c-axis reaches a maximum value of

$$|M_c|_{max} = H\left(1 - \frac{H}{2H^*}\right) \qquad (41)$$

Figure 15 shows a good agreement between Eq.(40) and the experiment. In this figure the symbols represent the experimental data whereas the solid lines are fits to Eq.(40) with a single parameter $H^* \approx 380\ G$, for all curves. The value of $H^*$ has been determined from a fit of the maximum value of $M_c$ to Eq.(41) (inset to Figure 15) and was verified through independent measurements of standard magnetization loops in that sample.

Another implication of Eq.(40) is that as long as the applied field $H$ is smaller than or equal to $H^*$, the component $M_H$ of the total magnetic moment is less than (or equal to) zero in the whole angular range. We show below that for $H>H^*$, $M_H$ becomes positive at certain angles. This crossover from negative to positive values of $M_H$ may serve as a sensitive tool for experimental determination of $H^*$.

## 2. Full magnetization ($H>H^*$)

When the applied field is larger than $H^*$ magnetic flux penetrates the entire sample. In this case, the projection of the magnetic moment along the *c*-axis in the interval $x=(0,2)$ (i.e. $\theta=(0,\pi)$) according to the Bean model is:

$$M_c = \begin{cases} \dfrac{3}{4}Hx - \dfrac{H^2}{8H^*}x^2 - \dfrac{H^*}{2}, & x \leq 2\dfrac{H^*}{H} \\ \dfrac{H^*}{2}, & x \geq 2\dfrac{H^*}{H} \end{cases} \quad (42)$$

Again, we note that for $x=0$ we recover the Bean results for full penetration $M_c = H^*/2$. For $x \geq 2H^*/H$ the magnetization is constant as predicted by Bean for $H > H^*$. The remagnetization process occurs in the interval $x < 2H^*/H$ and it is completed when the moment reverses its sign (i.e., changes from $-H^*/2$ to $+H^*/2$).

The measured component of the magnetic moment along the direction of the external field may be determined from Eq.(42) by substituting $x = 1 - \cos(\theta)$:

$$M_H = \begin{cases} \dfrac{H}{8}\cos(\theta)\left\{6 + 2\left(\dfrac{H}{H^*}-3\right)\cos(\theta) - \dfrac{H}{H^*}\left(1+\cos^2(\theta)\right) - 4\dfrac{H^*}{H}\right\}, & \theta \leq \theta_r \\ \dfrac{H^*}{2}\cos(\theta) & \theta \geq \theta_r \end{cases} \quad (43)$$

where, $\theta_r = \arccos(1 - 2H^*/H)$, mod $\pi$ is the angle at which the remagnetization process is completed. An interesting implication of Eq.(43) is that for $H > H^*$ the resulting magnetization versus angle curves become *asymmetric* with respect to $\theta = \pi$.

In conclusion, we find that in the rotation experiments a *forward* and a *backward* magnetization versus angle curves coincide for magnetic fields less or equal to the field of full penetration $H < H^*$ and not, as previously believed at $H > H_{c1}$. On the contrary, for larger magnetic fields $H < H^*$, a *backward* rotation curve is a mirror image of a *forward* one with respect to $\theta = \pi$. In this field range the rotation curves may be highly distorted. The degree of un-harmonicity depends upon sample geometry and increases with decrease of the sample thickness (for given planar dimensions).

## C. *Unidirectional anisotropy of the pinning force in irradiated thin films*

The above description has to be modified if the pinning force is anisotropic and the anisotropy axis does not coincide with one of the sample symmetry axes. Such a situation is borne out in the experiment; for example, samples with planar and columnar defects. The former are usually presented in $YBa_2Cu_3O_{7-\delta}$ compounds in the form of twin boundaries, whereas the latter may be introduced deliberately by using high-energy heavy-ions [67-77]. The advantage of heavy-ion irradiation is in the ability to produce pinning centers of well-defined size, orientation and concentration. Indeed, the microstructure of the defects introduced by the irradiation depends on the kind of ions used for the irradiation, the energy of the ions and the dose of irradiation. In particular, high-energy (*GeV*-range) Pb irradiation introduces columnar defects, whereas Xe ions produce point-like and cluster-like defects [67-69]. Extensive studies of HTS crystals show that columnar defects produced by Pb irradiation serve as efficient pinning centers with unidirectional features [72, 73], namely, that the critical current, deduced from the width of the magnetization curves, reaches the maximum when the flux lines are aligned along the direction of the irradiation.

The study of superconductors with artificial pinning centers is of particular interest for thin films, which are more favorable than crystals for practical applications. In spite of an intensive experimental study of irradiation effects on thin films of copper-oxide superconductors (see e.g. [68]), the angular dependence of the critical current was hardly examined. The exception we are aware of is a nice work by Holzapfel et. al. [75], who have studied the angular dependence of the *transport* critical current in thin films irradiated with heavy-ions and found direct evidence for a unidirectional pinning in these films.

In this Section we report on the angular dependence of the *magnetic* properties in *YBa$_2$Cu$_3$O$_{7-\delta}$* thin films irradiated with Pb and Xe. The results show that in films irradiated with Pb ions, where columnar defects are manifested, a unidirectional enhancement of the critical current along the direction of irradiation is observed, similar to the effect which has been observed in crystals [72, 73]. In films irradiated with Xe ions, however, this effect was not found. This Section summarizes the results originally obtained in [ RP10 ].

## 1. Experimental

Thin *YBa$_2$Cu$_3$O$_{7-\delta}$* films were laser ablated on (100) *MgO* and *SrTiO$_3$* substrates, as described in detail in [48]. The films were irradiated by either 0.86 *GeV* Pb$^{+53}$ or 5 *GeV* Xe$^{+44}$ ions along the film normal or at 45° relative to it. The irradiation was carried out at the *Grand Acceleratoeur National d'Ions Lourdes* (GANIL), in Caen, France. The Pb$^{+53}$ ions produce continuous cylindrical amorphous tracks along their path with a diameter of about 7 nm. The Xe$^{+44}$ ions, on the other hand, do not produce columnar defects, but rather clusters of point-like defects, which are dispersed around the ions' trajectories [67-69]. All of the measurements reported in this Section were performed using an *"Oxford-Instruments"* vibrating sample magnetometer (VSM), see for description Chapter I.A. We denote the angle between the direction of the irradiation and film normal by $\theta_{irr}$ and the angle between the direction of the external magnetic field and the film normal by θ. As justified in Chapter III we will assume that vector magnetic moment points primarily along the film normal.

Figure 19 shows a typical set of magnetization curves, at 44 K, for a sample irradiated with Pb ions along $\theta_{irr}=+45^o$ measured at various θ. The width ΔM of the magnetization loops increases with the increase of the angle from $\theta=-45^o$ to $+45^o$. Apparently, the maximum width is reached for the loop measured at $\theta=+45^o$, namely when the external field is along the direction of the irradiation. The persistent current density j was derived from the width of such magnetization curves using Eq.(103).

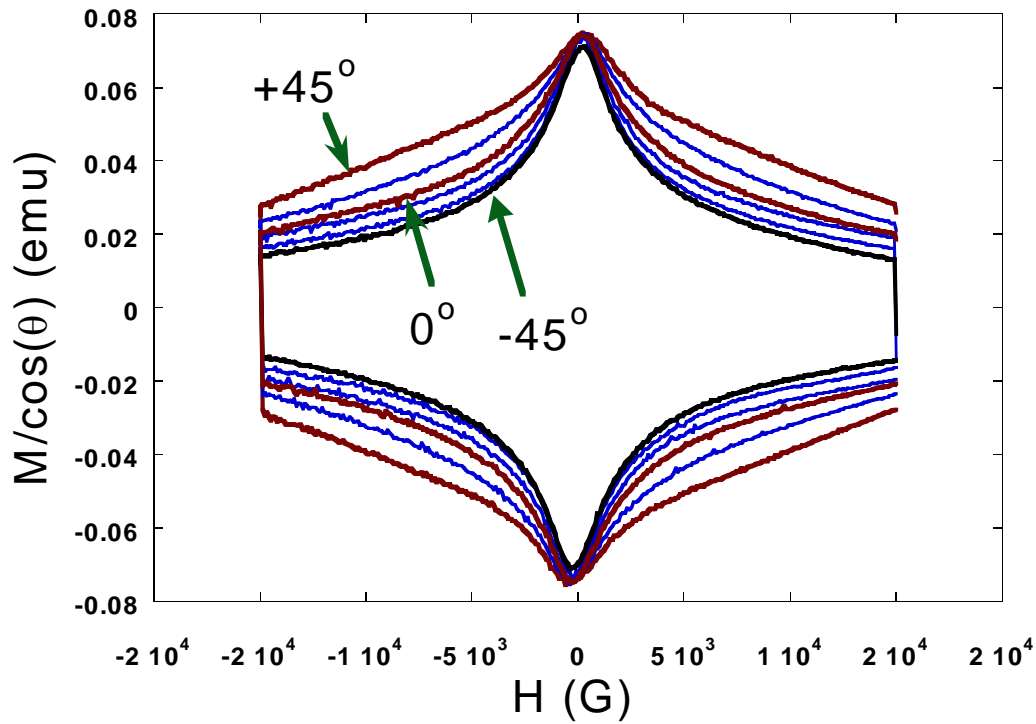

Figure 19. Magnetization along the c-axis measured at various angles $\theta$ in a sample irradiated with Pb at $\theta_{irr}=+45^o$

Figure 20 summarizes the derived current density plotted as a function of angle $\theta$ at T=64 K. The figure demonstrates a distinctive angular anisotropy of $j(\theta)$ with a sharp peak at $\theta=\theta_{irr}=45^o$, consistent with the transport data of Ref. [75]. It is quite clear that without columnar defects, the direction $\theta=45°$ is exactly equivalent to the direction $\theta=-45°$, and therefore any difference in the magnetic behavior between these directions may only be due to an interaction between the vortices and the columnar defects.

A similar pinning anisotropy was observed for films that were irradiated by Pb ions along $\theta_{irr}=0$. Figure 21 shows magnetization curves measured at angles $\theta=-45°$, $0°$ and $+45°$ for a sample irradiated with Pb along $\theta_{irr}=0$. The curves measured at $\pm 45°$ are almost identical, whereas the width of the curve for the field along the direction of irradiation exhibits a substantial increase for sufficiently large fields.

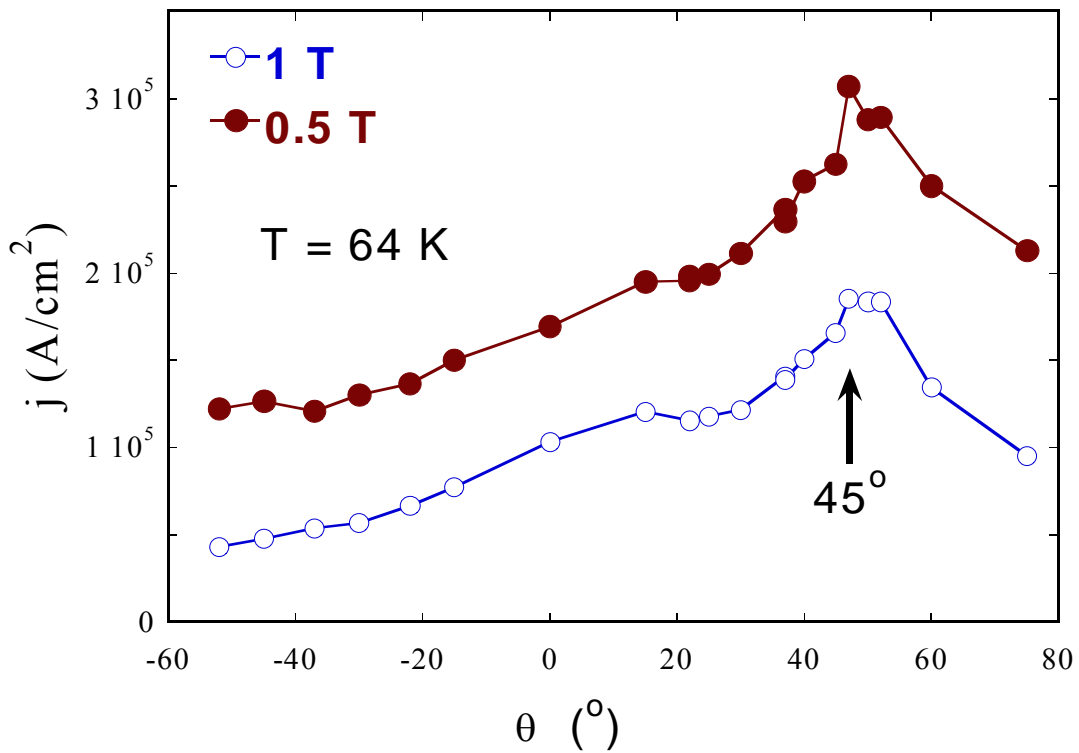

Figure 20. Persistent current density *vs.* $\theta$ evaluated from the magnetization loops in a sample irradiated with Pb ions along $\theta_{irr}=+45°$

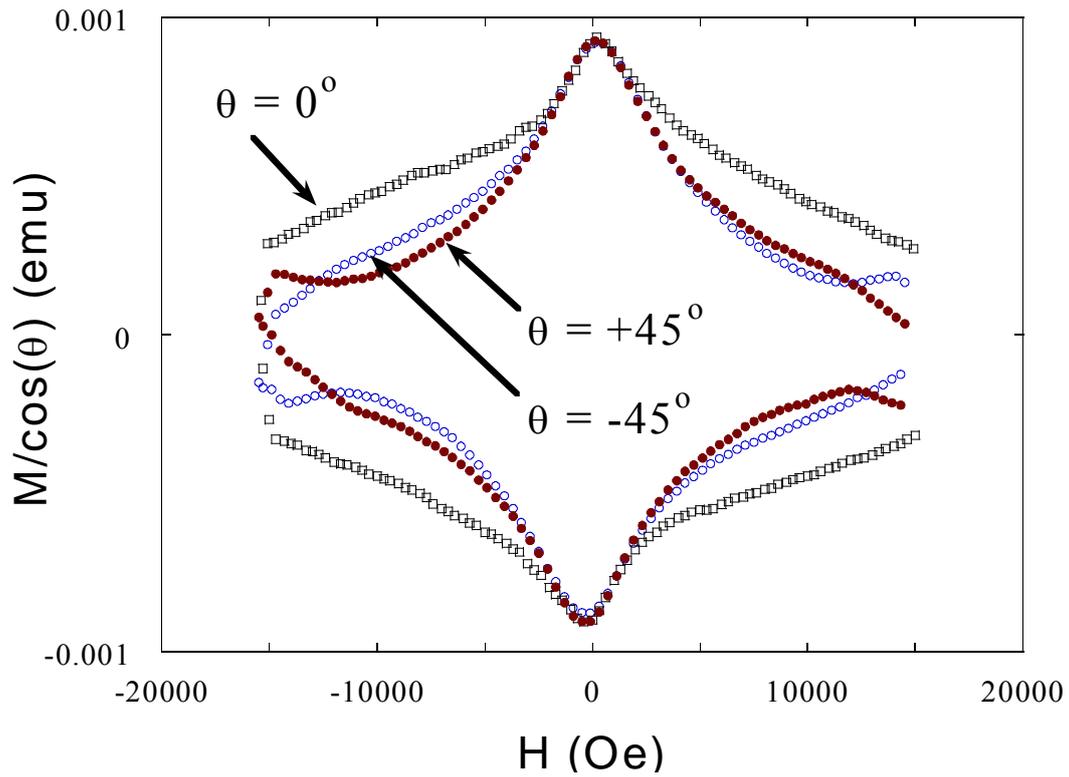

Figure 21. Magnetization loops in a sample irradiated with Pb along $\theta_{irr}=0$

Another demonstration of the anisotropic enhancement of the critical current is presented in Figure 22. In this figure the critical current, obtained from the widths of magnetization curves at $H=1$ $T$, is plotted versus temperature for three angles $\theta=-45^o$, $0^o$ and $+45^o$ for a sample irradiated along $\theta_{irr}=45^o$. It is apparent that the unidirectional pinning is more pronounced for low temperatures, and that it exists almost to the transition temperature.

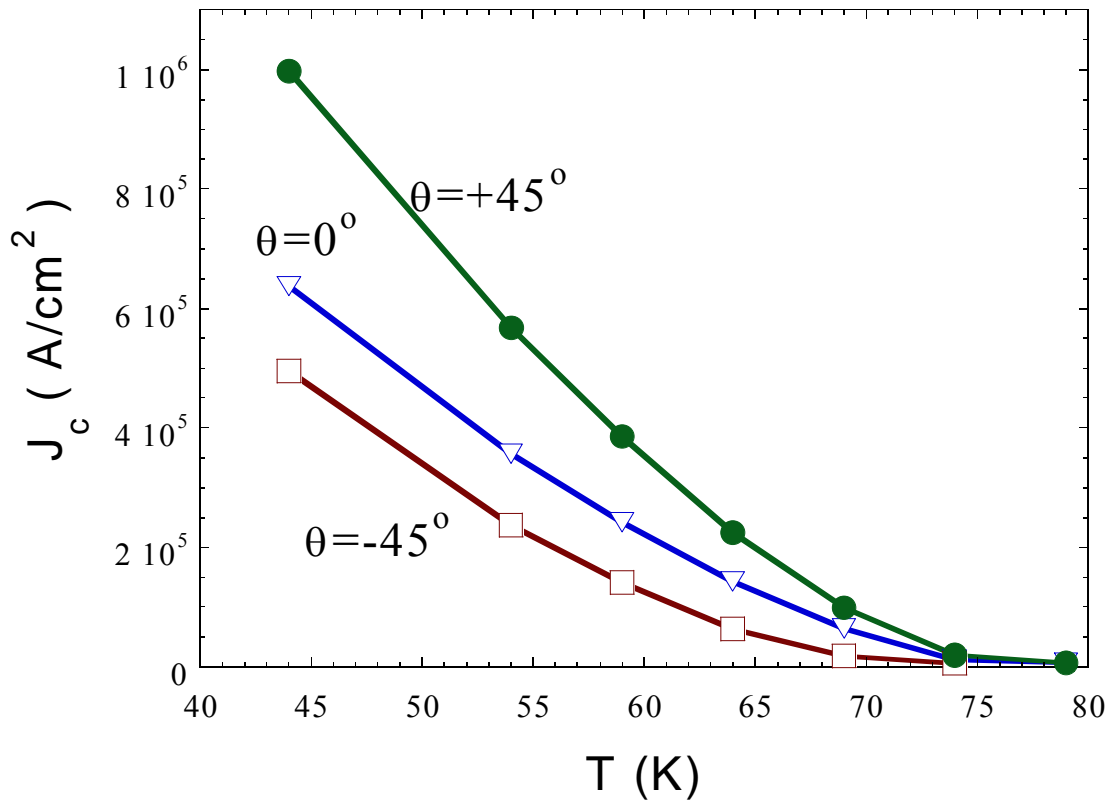

Figure 22. Persistent current density *vs.* temperature in a sample irradiated with Pd ions along $\theta_{irr}=+45°$ for three different angles $\theta$.

In contrast to the Pb irradiation, films irradiated with Xe ions have not exhibited unidirectional magnetic anisotropy. Figure 23 presents magnetization loops for a sample irradiated by Xe ions along $\theta_{irr}=+45°$, for $\theta=-45°$ and $+45°$, at T=64K. The magnetization loops are almost identical. Such $\theta$ independent magnetization was observed at all temperatures.

## 2. Scaling analysis of the pinning force

A powerful method to analyze the mechanism of pinning is the study of the volume pinning force density $F_p=|j \times B|/c$ [78]. At high enough fields, when variations of the magnetic induction in the sample (described in Chapter I.C) are smaller than the external field $H$ one can assume B=H and estimate pinning force as a function of temperature and magnetic field.

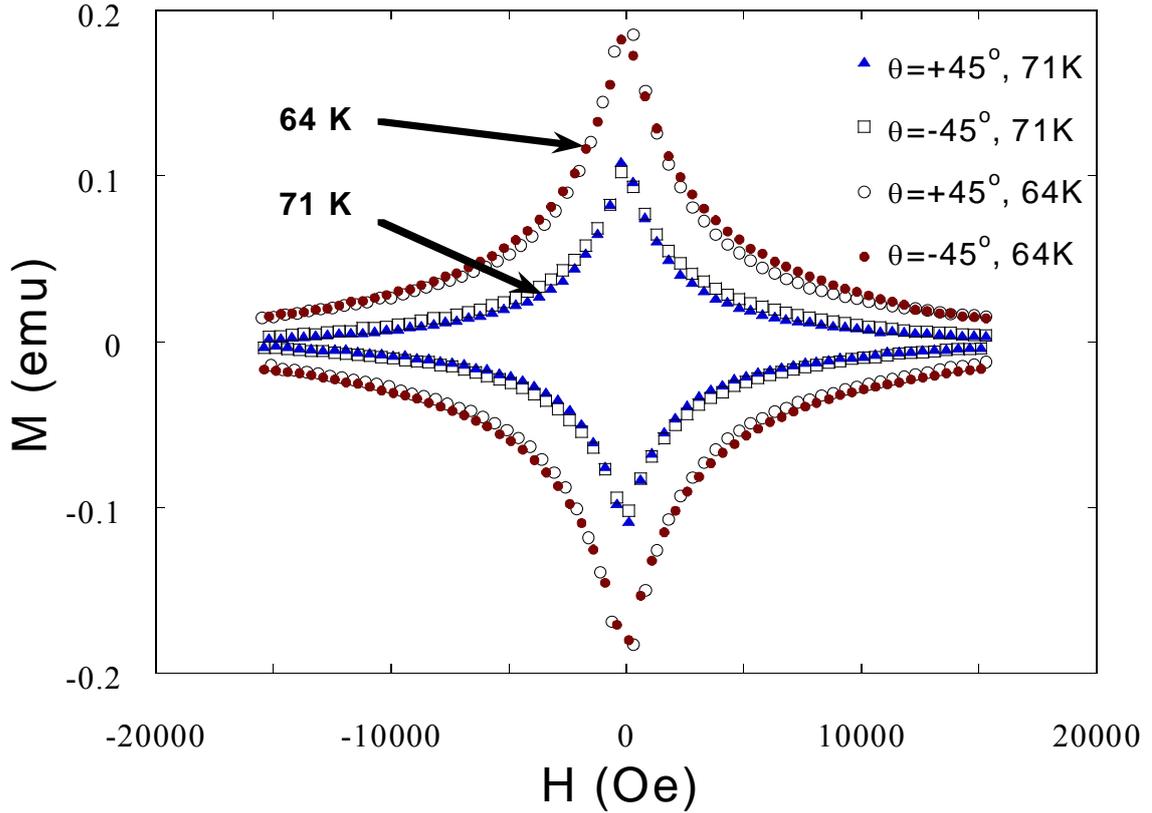

Figure 23. Magnetization loops in a sample irradiated with Xe ions along $\theta_{irr}=+45°$ for angles $\theta=\pm45°$.

According to its definition, $F_p$ is 0 at B=0 and at $H_{irr}$, the latter being the irreversibility field. These features were used to present the pinning force in a scaled form using variables $f=F_p/F_m$ and $b^*=H/H_{irr}$, where $F_m(T)$ is the maximal value of $F_p(T,H)$ and $H_{irr}$ is the irreversibility field. However, usually in YBa$_2$Cu$_3$O$_{7-\delta}$ thin films $H_{irr}$ is experimentally unreachable, so one can use another characteristic field, i. e., magnetic field value at the peak of the pinning force $F_p(T,H_m)=F_m$. If the temperature independent scaling form of $f_p$ exists using $H_{irr}$, it must exist using $H_m$ as well. Figure 24 summarizes scaled pinning forces obtained in different samples for different angles $\theta$.

As shown in numerous studies scaled pinning force may be represented in the form

$$f_p = Kb^p (1-ab)^q \tag{44}$$

where $a=H_m/H_{irr}$, $K=1/f_{p,max}$ and $p$ and $q$ are characteristic exponents usually used for analysis of the pinning mechanism [79-81].

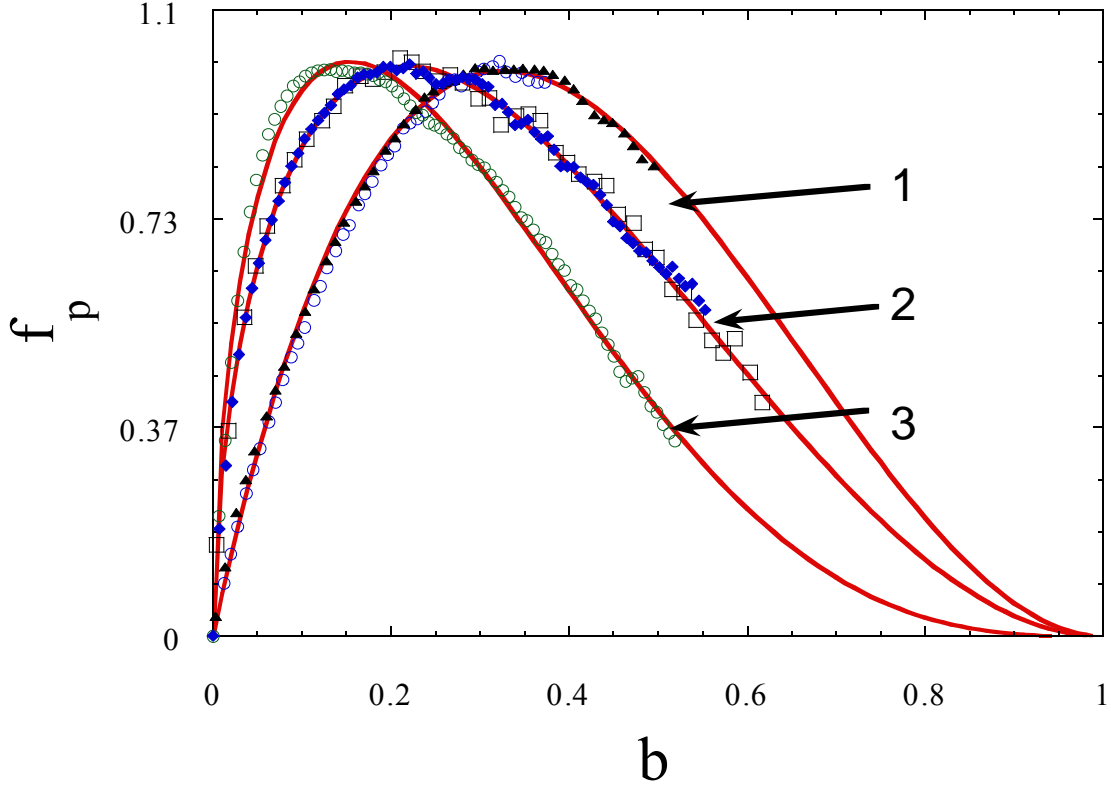

Figure 24. Scaled pinning force for (1) Pb irradiated samples when $\theta=\theta_{irr}$; (2) non-irradiated and Pb $\theta_{irr}=+45°$ irradiated measured at $\theta=-45°$; (3) Xe irradiated sample.

We find from our data of Figure 24 that there are three distinctive functional forms of $f_p$: **(i)** for the Pb irradiated samples with the direction of the external field along the columnar defects where $f_p=6.45b^{0.96}(1-b/3)^2$; **(ii)** for Pb $\theta_{irr}=+45°$ irradiated sample with the field perpendicular to the defects ($\theta=-45°$), and for the non-irradiated sample $f_p=3.82b^{0.55}(1-b/4.55)^2$; **(iii)** for the Xe irradiated sample $f_p=4.64b^{0.55}(1-b/6.67)^3$. The different functional form of the scaled pinning force, which is manifested in the exponents $p$ and $q$ of Eq.(44), reflects a different type of pinning mechanism, associated with different defects in our samples. The connection between *p* and *q* has been studied in several theoretical and experimental works. For example, taking into account the creep phenomena, Niel [81] has calculated the pinning force for different types of pinning centers and found that *p*=1/2, *q*=2 describe pinning by a surface. Quite similar exponents (*p*=0.6 and *q*=2.2) were found experimentally by Juang et. al. [82] for non-irradiated *Tl-Ba-Ca-Cu-O* thin films. They suggested that this functional form of $f_p$ manifests the surface core pinning. In our data, the pinning force, in the case of the Pb irradiation where the field direction is perpendicular to the direction of irradiation, exhibits the same pinning mechanism as that of that in non-irradiated sample, with the exponents *p*≈1/2, *q*=2. It is therefore plausible that this form of $f_p$ is due to pinning on planar defects, which always exist in laser-ablated films. The sample irradiated with Xe along $\theta_{irr}$=+45°) manifests pinning mechanism which differs from the virgin sample, with *q*=3 instead of 2. It was suggested that the exponent *q*=3 arises from a combined effect of the shear modulus $C_{66}$ and the tilt modulus $C_{44}$, in a contrast to a dominant $C_{66}$ pinning in the case of *q*=2 [79, 80].

## D. *Summary and conclusions*

In conclusion, we have analyzed the angular dependence of the magnetization of $Y_1Ba_2Cu_3O_{7-\delta}$ thin films irradiated with high-energy Pb and Xe ions. The Pb irradiated samples exhibit a unidirectional anisotropy, so that persistent current density reaches a maximum when the direction of the external magnetic field coincides with the direction of the irradiation. The Xe irradiated samples did not show such magnetic anisotropy. The differences in the pinning efficiency for the various samples and different mutual direction of the columnar tracks and external field are clearly reflected in the functional form of the pinning force scaled with the irreversibility field.

# Chapter IV. Origin of the irreversibility line in $Y_1Ba_2Cu_3O_{7-\delta}$ films

While previous chapters dealt with superconducting samples at low temperatures when the critical state is well established and persistent current is large, this chapter addresses the important issue of the onset of irreversibility. The origin of the irreversibility line (*IRL*) in the field-temperature (*H-T*) phase diagram of high-temperature superconductors is intriguing and is still a widely discussed topic [3, 17, 37, 83-90]. Experimentally, this line is defined as the borderline at which the magnetic response of the sample changes from irreversible to reversible In high-temperature superconductors, large fluctuations and relatively weak pinning lead to a rich *H-T* phase diagram with a variety of dynamic and static transitions which can be responsible for the appearance of magnetic reversibility [3, 85, 86, 91-93]. Thus, a thorough experimental investigation of the *IRL* is important for the understanding of the vortex-lattice behavior in superconductors in general and of the mechanisms responsible for the onset of irreversible magnetic response, in particular.

Several models, such as thermally activated depinning [84-88, 94, 95], vortex-lattice melting [96-103] and a transition from vortex glass to vortex fluid [104-109], were proposed to identify the origin of the *IRL* in high-temperature superconductors. Attention was also given to the possibility of pinning in the vortex-liquid phase [3, 91, 92, 110] and to different dissipation mechanisms above the melting line [3, 94, 111-113]. Irreversibility due to geometrical [17, 37, 89] or surface barriers [114] have also been proposed, but these mechanisms are less probable in thin YBCO films with strong pinning. The irreversibility line may be affected by sample-dependent properties such as the nature and density of pinning centers and by intrinsic or extrinsic anisotropy. For example, in superconductors with columnar defects, the irreversibility line may either be identified with the Bose-glass transition [3, 113, 115-123], or related to the concept of a trapping angle [124]. The configuration of the columnar defects is also very important, since it affects the possibility of different types of depinning mechanism. A splayed configuration, for example, inhibits creep from columnar defects [125, 126]. Similarly "crossed" defects (i.e., defects at angles $\pm\theta$) were shown to act collectively, i.e., they introduce unidirectional anisotropy such that the current density reaches its maximum for magnetic field directed in a mid angle between defects [11, 127].

Experimentally, the situation is even more complex, since different techniques (magnetization loops, field-cool *vs.* zero-field-cool DC magnetization, peak in the imaginary part of the first harmonic, etc.) yield different *IRL*s [107, 108, 128]. To a great extent, the reliability of the determination of *IRL* depends on the criterion for the onset of the irreversibility. We determine the irreversibility temperature at given DC field by the onset of third harmonic in the *ac* response, which, we believe, is one of the most reliable methods for contact-less determination of the *IRL* [76, 129, 130]. In most experiments $T_{irr}$ is measured as a function of the external field *H*. This information is insufficient to distinguish between different models for the origin of the irreversibility. Additional information, such as the frequency dependence of the *IRL* [76, 107, 108, 131] or its angular variation [99, 101-103, 118, 120, 122, 132], is needed.

In this Section we summarize results of the extensive study of the angular dependence of the irreversibility temperature $T_{irr}(\theta)$ in thin YBCO films before and after irradiation with Pb ions published in [130, 133].

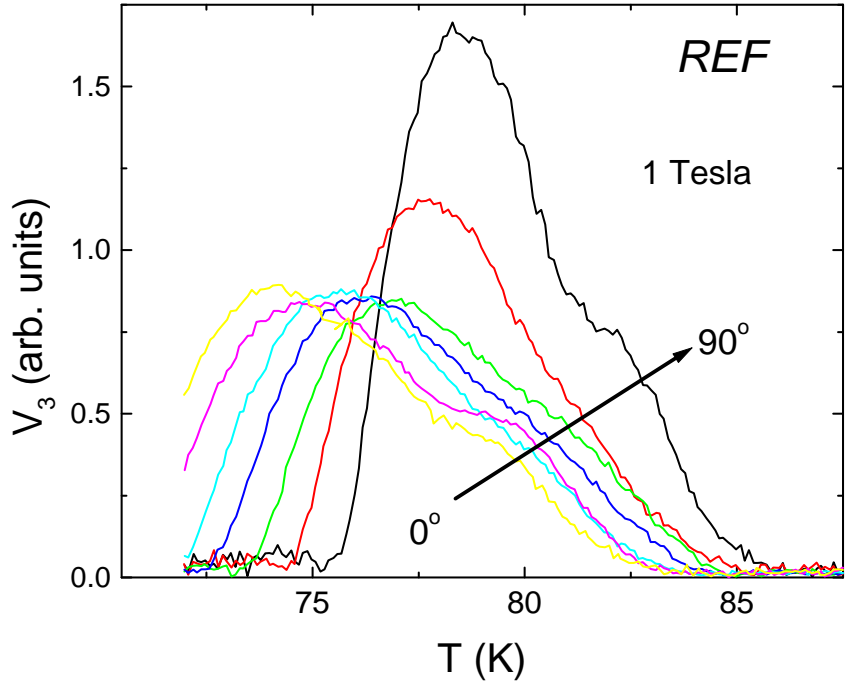

Figure 25. The third harmonic signal $V_3$ versus temperature during field-cooling at 1 *T* for sample *REF* at $\theta$ =0, 10°, 30°, 40°, 60°, 80°, 90°.

## A. *Experimental*

Three 1500 Å $YBa_2Cu_3O_{7-\delta}$ films were laser ablated on $SrTiO_3$ substrate [48]. All three samples have the same lateral dimensions of 100 x 500 μm². One film, denoted as *REF*, was used as a reference sample. The other two, *UIR* and *CIR*, were irradiated at GANIL with $2\times10^{11}\ ions/cm^2$ 5.8 *GeV* Pb ions along the *c*-axis and along $\theta = \pm 45°$, respectively. (*UIR* and *CIR* stand for "uniform irradiation" and "crossed irradiation", respectively). The superconducting transition temperatures, measured by a *Quantum Design SQUID* (see Chapter I.A) susceptometer and defined as the onset of the Meissner expulsion in a DC field of 5 G, are $T_c \approx 89\ K$ for the samples *REF* and *UIR* and 88 *K* for *CIR*.

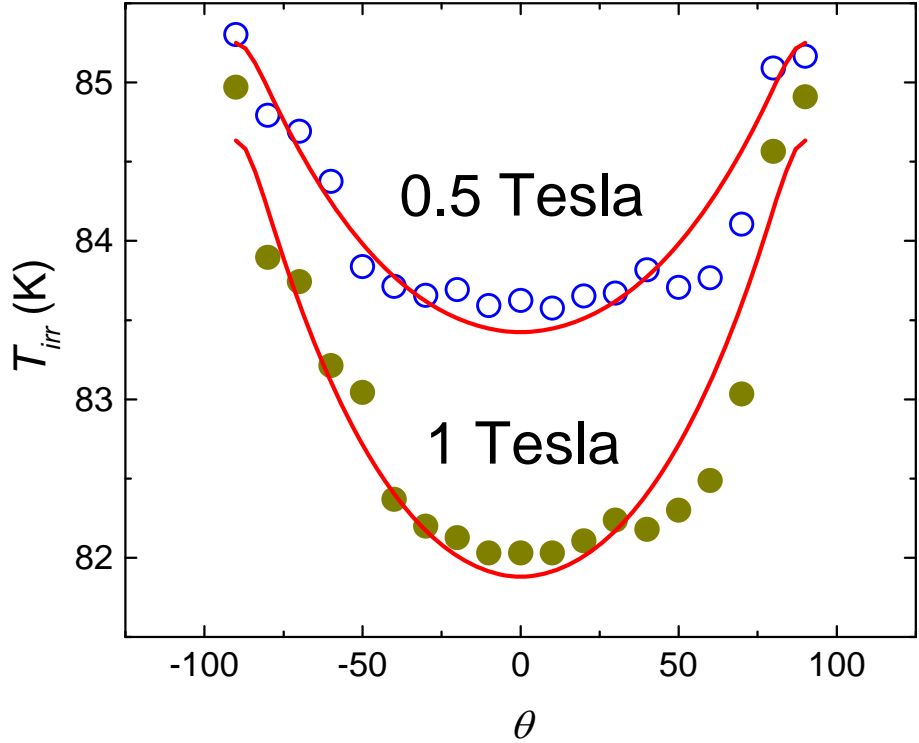

Figure 26. Angular variation of the irreversibility temperature in the unirradiated sample *REF* at two values of the external field: *H*=0.5 and 1 *T*. The solid lines are fits to Eq.(52).

For the *ac* measurements reported below we used a miniature $80\times80\ \mu m^2$ *InSb* Hall-probe, which was positioned in the sample center. The 1 *G ac* magnetic field, always parallel to the *c*-axis, was induced by a small coil surrounding the sample. An external *dc* magnetic field, up to $H_a$=1.5 *T*, could be applied in any direction $\theta$ with respect to the *c*-axis. In our experiments *dc* magnetic field was always turned on at a fixed angle at $T>T_c$ and then the *ac* response was recorded during sample cooling. The irreversibility temperature, $T_{irr}(\theta)$, is defined as the onset of the third harmonic signal in the *ac* response measured by the Hall probe [76, 129, 130]. This procedure was repeated for various *dc* fields and at various angles $\theta$ of the field with respect to the *c*-axis.

Figure 25 presents measurements of $V_3$, the third harmonic in the *ac* response, versus temperature *T*, during field-cooling at 1 *T* for the sample *REF* at various angles between 0 and 90°. Apparently, as the angle $\theta$ increases the whole $V_3$ curve shifts to higher temperatures and becomes narrower. The onset of irreversibility, $T_{irr}(\theta)$, is defined by the criterion $V_3^{onset}=0.05$ in the units of Figure 25.

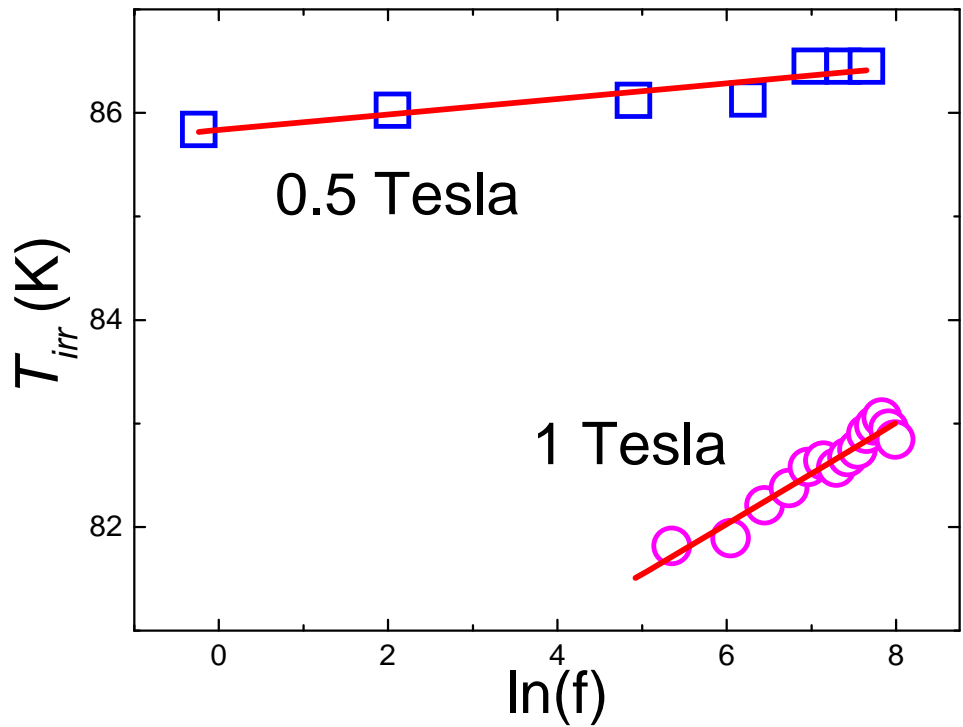

Figure 27. The frequency dependence of $T_{irr}$ in the unirradiated sample *REF* at two values of the external field: *H*=0.5 and 1 *T*. The solid lines are fits to Eq.(53).

Figure 26 exhibits typical $T_{irr}(\theta)$ data for the unirradiated sample *REF*, measured at two values of the external field: 0.5 *T* and 1 *T*. Both curves exhibit a shallow minimum around $\theta=0$ and they reach their maximum values for *H* along the *ab* plane, at angles $\theta=\pm 90°$. We also measured the frequency dependence of $T_{irr}$ for the same values of magnetic field. As shown in Figure 27, the slope $\partial T_{irr}/\partial \ln(f)$ is larger for larger field.

The sample irradiated along the *c-axis* exhibits an additional feature - a *peak* around $\theta=0$. This is clearly shown in Figure 28 where we compare $T_{irr}(\theta)$ at *H=1 T* for the samples *REF* and *UIR*. As discussed below, this peak is a signature of the unidirectional magnetic anisotropy induced by the columnar defects. Intuitively, one would therefore expect two peaks, along $\theta=\pm 45°$, for the third sample, *CIR*, crossed-irradiated at $\theta=\pm 45°$. Instead, we find one strong peak around $\theta=0$, similar to that found in BSCCO crystals. This is demonstrated in Figure 29 where we compare $T_{irr}(\theta)$ at *H=0.5 T* for this sample (*CIR*) and for the unirradiated sample (*REF*). We maintain below, that the peak around $\theta=0°$ is a result of a collective action of the crossed columnar defects, and that its origin is the same as that for unidirectional enhancement of critical current density observed in *BSCCO* crystals [11, 127].

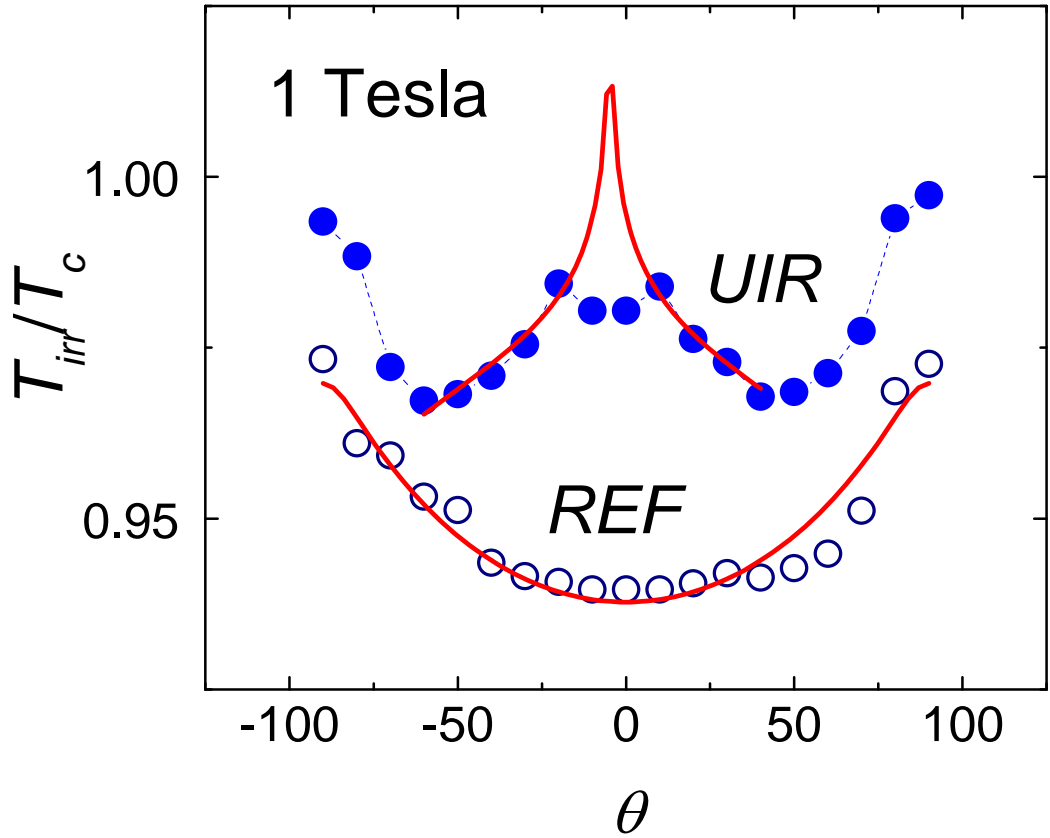

Figure 28. The irreversibility temperature for two samples: *REF* (unirradiated - open circles) and *UIR* (irradiated along the *c*-axis sample, - filled circles) at *H*=1 *T*. Solid lines are fits to Eq.(52) and Eq.(58), respectively.

## B. *Analysis*

The "true" irreversibility temperature $T_0$ is defined as a temperature above which the pinning vanishes. Such disappearance of pinning can be of static (true phase transition), as well as of dynamic origin (gradual freezing, pinning in liquid). In practice, one determines the irreversibility temperature $T_{irr}(\Delta)$ as the temperature above which the critical current density is less than some threshold value $\Delta$. Therefore, by definition, $T_0 = \lim_{\Delta \to 0}(T_{irr}(\Delta))$. The apparent current depends on temperature *T*, magnetic field *B* and the frequency *f* of the exciting field which defines a characteristic time-scale 1/*f* for the experiment. By solving the equation $j(T,B,f)=\Delta$ with respect to $T$ one finds the experimental irreversibility temperature $T_{irr}$ for constant $B$ and $f$. In the following we maintain that in our experiments the measured $T_{irr}$ is a good approximation of $T_0$. In order to estimate $T_{irr}$ we employ a general form for the apparent current density *in the vicinity of the irreversibility line* (*IRL*) [3, 78, 87, 88, 134-137]:

$$j(T,B,f) \propto j_c(0)\frac{(1-T/T_0)^\alpha}{(B/B_0)^\beta}\left(\frac{f}{f_0}\right)^\gamma \qquad (45)$$

where the parameters $B_0$ and $f_0$ are temperature independent. Equation (45) is thus valid only in a narrow temperature interval near the *IRL* and for fields larger than $H_{c1}$). From Eq.(45) we get:

$$T_{irr} = T_0(B)\left\{1-\left[\frac{\Delta}{j_c(0)}\left(\frac{B}{B_0}\right)^\beta\left(\frac{f_0}{f}\right)^\gamma\right]^{1/\alpha}\right\} \qquad (46)$$

Inserting reasonable numerical estimates: $j_c(0) \approx 10^7\ A/cm^2$, $\Delta \approx 100\ A/cm^2$ for our experimental resolution, $B_0 \approx 10^3\ G$, $B \approx 10^4\ G$, $\beta \sim 1$ [3, 78, 87, 88, 137], $\gamma \sim 1$ [87, 88], $f \approx 10^2\ Hz$, and $f_0 \approx 10^7\ Hz$ [87, 88], we get from Eq.(46): $T_{irr} = T_0(B)(1-0.005^{1/\alpha})$. Thus, with 0.5% accuracy we may say that $T_{irr}$, the measured onset of the third harmonic component in the *ac* response, marks some "true" irreversibility crossover line $T_0(B)$. The nature of this line $T_0(B)$ is our main interest, since, as discussed in the introduction, it is directly related to the pinning properties of vortex lattice in type-II superconductors at high temperatures.

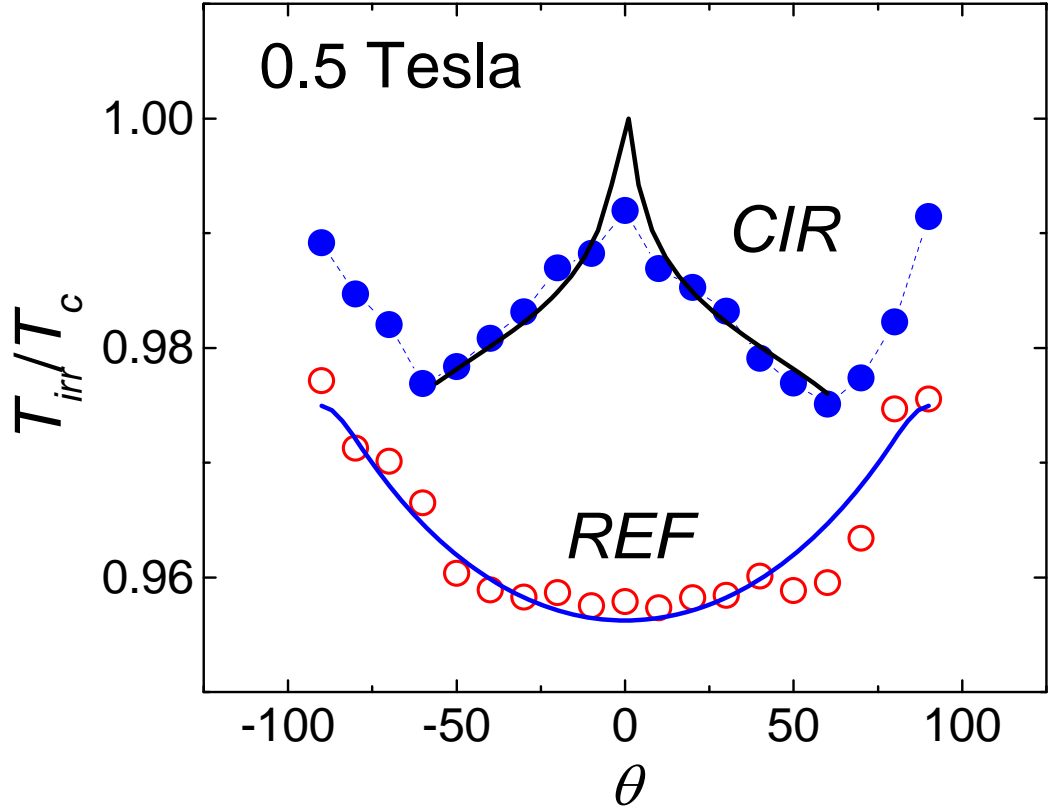

Figure 29. The irreversibility temperature for two samples: *REF* (unirradiated - open circles) and *CIR* (irradiated along $\theta=\pm45°$ - filled circles) at $H=0.5\ T$. Solid lines are fits to Eq.(52) and Eq.(58), respectively.

### 1. Unirradiated $YBa_2Cu_3O_{7-\delta}$ films

We turn now to consider the effect of the intrinsic anisotropy on $T_{irr}(\theta)$. Following the anisotropic scaling approach proposed by Blatter *et al*. [138, 139], we replace $T$ by $\varepsilon T$ and $B$ by $\varepsilon_\theta = \sqrt{\cos^2(\theta)+\varepsilon^2\sin^2(\theta)}$ and $\varepsilon \approx 1/7$ is the anisotropy parameter for $YBa_2Cu_3O_{7-\delta}$. It should be emphasized that we can use this scheme only in the case of *intrinsic* material anisotropy $\varepsilon = \sqrt{m_{ab}/m_c}$, where $m_c$ and $m_{ab}$ denote the effective masses of the electron along the *c*-axis and in the *ab*-plane, respectively. In the case of some *extrinsic* magnetic anisotropy, (columnar defects or twin planes), the critical current depends on the angle not only via the effective magnetic field $B_{eff}$, but also because of this extrinsic anisotropy. See also Chapter III.C for discussion of the induced anisotropy.

As we have already indicated in the Introduction, there are several possible origins for a crossover from irreversible to reversible magnetic behavior in unirradiated samples. We exclude the vortex-glass to vortex fluid transition as a possible origin of the *IRL*, because this transition was shown to occur at temperatures lower than the onset of dissipation [107, 108, 110, 140]. The thermal depinning temperature *increases* with increase of field [2] $T_{dp} \propto \sqrt{B}$ and, therefore, is also excluded. Vortex-lattice melting transition is believed to be responsible for the appearance of reversibility [96, 97, 99, 100]. The implicit angular dependence of $T_m$ was derived by Blatter *et al.* [138] using their scaling approach:

$$T_m(\theta) \approx 2\sqrt{\pi}\varepsilon\varepsilon_0 c_L^2 \sqrt{\Phi_0/B\varepsilon_\theta} \approx \frac{c_L^2 T_c}{\sqrt{\beta_m Gi}}\left(1-\frac{T_m}{T_c}\right)\left(\frac{H_{c2}(0)}{\varepsilon_\theta B}\right)^{1/2} \qquad (47)$$

where $\Phi_0$ is the flux quantum, $\xi$ is the coherence length, $\beta_m \approx 5.6$ is a numerical factor, estimated in [2], $c_L \approx 0.1\$$ is the Lindemann number, $Gi = \left(T_c / \varepsilon H_{c2}(0)\xi^3(0)\right)^2 / 2$ is the Ginzburg number, and $H_{c2}(0)$ is the *linear* extrapolation of the upper critical field from $T_c$ to zero. Solving Eq.(47) with respect to $T_m$ we get

$$T_m(\theta) \approx \frac{T_c}{1+\left(\beta_m Gi / c_L^4 H_{c2}(0)\right)^{1/2}(\varepsilon_\theta B)^{1/2}} \equiv \frac{T_c}{1+C\sqrt{\varepsilon_\theta B}} \qquad (48)$$

Equation (52) predicts that the melting temperature decreases as $B_{\mathit{eff}} = \varepsilon_\theta B$ increases. In agreement with the experimental data of Figure 26 that shows that $T_{irr}$ increases with the angle $\theta$, i.e., decreases with $B_{\mathit{eff}}$. The solid lines in Figure 26 are the fits to Eq.(52). From this fit we get $C \approx 0.0005$. However, a reasonable estimate of $C \approx \sqrt{\beta_m Gi / c_L^4 H_{c2}(0)}$ yields $C \approx 0.01$, where we take $H_{c2}(0) = 5 \cdot 10^6$ G, $c_L = 0.1$,



*Gi*=0.01, and $\beta_m$=5.6 [2]. Also, Yeh et al. showed that the onset of irreversibility occurs above the melting temperature (Ref. [113], Figure 28). In addition, the important effect of the frequency (see Figure 27) is not included in Eq.(52).

We discuss now another possibility for the onset of the irreversibility, namely, pinning in the vortex liquid (for a discussion see Chapter VI in Blatter et al. [2] and references therein). Any fluctuation in the vortex structure in the liquid state has to be averaged over the characteristic time scale for pinning $t_{pin}$. In the absence of viscosity the only fluctuations in the liquid state are thermal fluctuations, which have a characteristic time $t_{th} \ll t_{pin}$. (As shown in [2] $t_{pin}/t_{th} \propto j_0/j_c$, where $j_0$ is the depairing current). Thus, such a liquid is always unpinned. The situation differs for a liquid with finite viscosity. In this case there exists another type of excitation in the vortex structure, i.e., *plastic* deformations with a characteristic time scale $t_{pl}$. The energy barrier, corresponding to plastic deformation, is shown to be [3, 93]

$$U_{pl} \approx \gamma \varepsilon \varepsilon_0 a_0 \approx \gamma \left( \frac{H_{c2}}{4Gi} \right)^{1/2} \frac{(T_c - T)}{\sqrt{B}} \tag{49}$$

where γ is a coefficient of the order of unity. The corresponding characteristic time scale is

$$t_{pl} \sim t_{th} \exp(U_{pl}/T) \tag{50}$$

Thus, depending on the viscosity, $t_{pl}$ can be smaller or larger than $t_{pin}$. In the latter case, after averaging over a time $t_{pin}$, the vortex structure remains distorted and such a liquid shows irreversible magnetic behavior. Thus, on the time scale of $t_{pin}$ the distorted vortex structure is pinned. The crossover between pinned and unpinned liquid occurs at temperature $T_k$ where the characteristic relaxation time for pinning



$t_{pin}(T)$ becomes comparable to that for plastic motion $t_{pl}(T)$. Thus, using Eqs.(49) and (50) we obtain

$$T_k \approx \frac{T_c}{1+\gamma^{-1}\left(4Gi/H_{c2}(0)\right)^{1/2}\ln\left(t_{pin}/t_{th}\right)\sqrt{B}} \tag{51}$$

Finally, using the anisotropic scaling [2] we may rewrite Eq.(51) for $f_{pin} \lesssim f \lesssim f_{th}$ as

$$T_{irr}(\theta) = T_k(\theta) \approx \frac{T_c}{1+\gamma^{-1}\left(4Gi/H_{c2}(0)\right)^{1/2}\ln\left(f_{th}/f\right)\sqrt{\varepsilon_\theta B}} \equiv \frac{T_c}{1+A\sqrt{\varepsilon_\theta B}} \tag{52}$$

with $f_{th} \equiv 1/t_{th}$ and $f_{pin} \equiv 1/t_{pin}$. Note the apparent similarity with the expression for the melting temperature, Eq.(48). The numerical estimate for

$$A = \frac{1}{\gamma}\left(\frac{4Gi}{H_{c2}(0)}\right)^{1/2}\ln\left(\frac{f_{th}}{f}\right) \tag{53}$$

gives: $A \approx 10^{-4}\ln(f_{th}/f)$. This is in agreement with the value found from the fit (solid line in Figure 26) for $H_{c2}(0) = 5 \cdot 10^6$ G, $Gi=0.01$, $f_{th} \sim 10^{10}$ Hz, and $\gamma \approx 4$.

To further confirm that in our $YBa_2Cu_3O_{7-\delta}$ films the most probable physical mechanism for the onset of irreversibility is *a dynamic crossover from unpin to pin vortex liquid* we discuss now the frequency dependence of $T_{irr}$. Equation (52) has a clear prediction for the frequency dependence of $T_{irr}$. To see it directly we may simplify it by using the experimentally determined smallness of value of the fit parameter $A \approx 0.0005$, which allows us to expand Eq.(52) (for not too large fields) as

$$T_k \approx T_c\left[1-\frac{1}{\gamma}\left(4Gi/H_{c2}(0)\right)^{1/2}\ln\left(f_{th}/f\right)\sqrt{\varepsilon_\theta B}\right] \tag{54}$$

which results in a linear dependence of $T_{irr}$ upon $\ln(f)$ and a slope

$$S \equiv \frac{\partial T_{irr}}{\partial \ln(f)} \approx \frac{T_c}{\gamma}\left(\frac{4Gi}{H_{c2}(0)}\right)^{1/2}\sqrt{\varepsilon_\theta B} = T_c A\sqrt{\varepsilon_\theta B}\ln(f/f_{th}) \tag{55}$$



Note that the slope is proportional to $\sqrt{B}$. This is indeed confirmed by the experimental data, as is demonstrated by the solid lines in Figure 27. From this fit we get $S/\sqrt{B} = 0.004$ and we can independently verify the parameter $A$ appearing in Eq.(52) $A = S/\left(T_c \sqrt{\varepsilon_\theta B} \ln(f/f_{th})\right) = 0.0008$, which is in an agreement with the value obtained above.

We note that the approximated expression for the frequency dependence of $T_{irr}$, Eq.(54), is valid in the whole experimentally accessible range of magnetic field since Eq.(52) predicts a maximum in the slope $S$ at $B_{max} = 1/(\varepsilon_\theta A^2) \approx 400\,T$ for the experimental parameters. This value is, of course, beyond the experimental limits, and probably even exceeds $H_{c2}$.

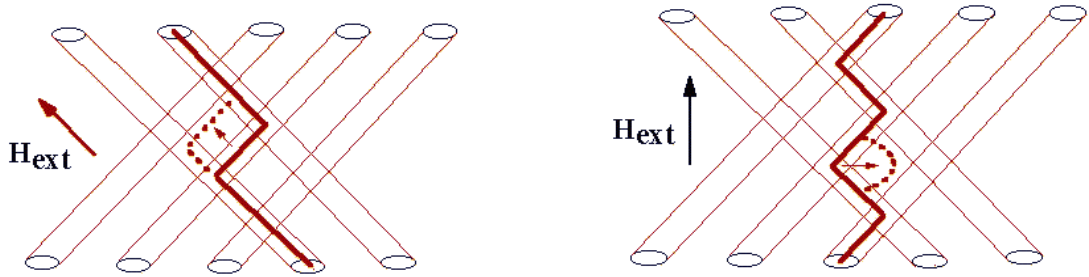

Figure 30. Schematic description of possible depinning modes of a vortex line in the case of crossed columnar defects - (*left*) magnetic field is directed along $\theta = 45°$; (*right*) magnetic field is along $\theta = 0$.

Further support for the onset of irreversibility in a vortex liquid is the *AC* field amplitude dependence of the *IRL*. In both, thermal-activated (TAFF) and flux-flow (FF) regimes the *E-j* curves are linear and the onset of the third harmonic is due to a change in the slope (from $\rho_{FF}$ to $\rho_{TAFF}$). In this case we expect the amplitude dependence for this onset. On the contrary, at the melting transition the onset of irreversibility is sharp and is not expected to depend upon the amplitude of the *ac*



field. In our experiments we find a pronounced amplitude dependence of the *IRL*, thus confirming the above scenario.

## 2. Irradiated $YBa_2Cu_3O_{7-\delta}$ films

For irradiated films the situation is quite different. The models for $T_{irr}(\theta)$ in unirradiated films cannot explain the experimental features exhibited in Figure 28 and Figure 29, in particular the increase in $T_{irr}$ in the vicinity of $\theta=0$. Such a discrepancy can only be due to the angular anisotropy introduced by columnar defects, i.e., the angle-dependent pinning strength. It was shown, both theoretically [Brandt, 1995 #5; Fisher, 1989 #199} and experimentally [123], that for magnetic field oriented along the defects the irreversibility line is shifted upward with respect to the unirradiated system. Thus, our results in Figure 28 suggest that the measured $T_{irr}(\theta)$ is a superposition of the angular variation of $T_{irr}$ in unirradiated film (denoted in this section as $T_{irr}^{REF}$) and the anisotropic enhancement of the pinning strength due to irradiation.

We can estimate the latter contribution by employing the concept of a "trapping angle" $\theta_t$, the angle between the external field and the defects at which vortices start to be partially trapped by columnar tracks. (For a schematic description, see Figure 43 in Blatter *et al.* [2]). As we show in the Appendix (Appendix C)

$$\tan(\theta_t) \approx \sqrt{2\varepsilon_r/\varepsilon_l} \tag{56}$$

where $\varepsilon_r(T)$ is the trapping potential of a columnar defect and $\varepsilon_l$ is the vortex line tension. In the experiment we cool down at a fixed $\theta$, and the onset of irreversibility must occur when $\theta=\theta_t(T)$, provided that the temperature is still larger than $T_{irr}^{REF}(\theta)$.



Otherwise, the onset occurs at $T_{irr}^{REF}$. This defines the condition for the irreversibility temperature $T_{irr}$ for angles $\theta \leq \theta_C \equiv \theta_t \left( T_{irr}^{REF} (\theta_t) \right) \approx 50^o$ in our case

$$\tan(\theta) = \tan(\theta_t (T_{irr})) \approx \sqrt{2\varepsilon_r / \varepsilon_l} \tag{57}$$

At high-temperatures $\varepsilon_r (T) \propto \exp\left( -T / \tilde{T}_{dp}^r \right)$, where $\tilde{T}_{dp}^r$ is the depinning energy [2]. Thus, we can write for $T_{irr}$

$$T_{irr}(\theta) = \begin{cases} T_{irr}^{REF}(\theta) - D \ln\left( C |\tan|\theta_t|| \right), & \theta \leq \theta_c \\ T_{irr}^{REF}(\theta), & \theta > \theta_c \end{cases} \tag{58}$$

where $D$ and $C$ are constants. This expression is in an agreement with our results shown in Figure 28 (solid line). We note some discrepancy, however, in the vicinity of $\theta = 0$, where we find quite weak dependence of $T_{irr}$ on angle. We explain this deviation by considering the influence of relaxation which, in the case of parallel defects, depends on angle. The relaxation rate is maximal, when vortices are aligned along the defects and retains its normal "background" value for perpendicular direction [141, 142]. Vortex, captured by a defect, can nucleate a double kink that slides out resulting in a displacement of a vortex on a neighboring column. In our irradiated samples the defect lattice is very dense (the matching field $B_\Phi = 4$ T, i.e., distance between columns $d \approx 220$ Å) and such double-kink nucleation is an easy process. Thus, the irreversibility temperature should be shifted down around $\theta=0$ as compared to the "ideal", non-relaxed value, Eq.(58). This explains the reduction in $T_{irr}$ in Figure 28.

We may now conclude that in irradiated films, for angles less than the critical angle $\theta_c$, the irreversibility line is determined by the *trapping angle* $\theta_t$. The Bose-glass



transition can probably only be found for small angles within the lock-in angle $\theta_L \approx 10^o$. This conclusion is also indirectly confirmed in [143].

As was pointed out in the Introduction, crossed defects should hinder the relaxation due to forced entanglement of vortices. Thus, the irreversibility temperature is expected to be closer to that predicted by Eq.(58). Figure 29 shows good agreement of the experimental data with Eq.(58) (solid line). To explain why defects crossed at a large angle act collectively and force unidirectional magnetic anisotropy, we follow here the approach outlined in [11, 127], and extend that description to account for arbitrary orientation of the external field with respect to the crossed columnar defects and to the *c*-axis. In Ref. [11, 127] the authors consider possible motion of vortices in a "forest" of crossed defects for field oriented along the *c-axis*. In our case of a dense lattice we may exclude from consideration free kink sliding and consider only depinning from the intersections. We sketch in Figure 30 the two limiting situations: (*a*) the external field is parallel to one subsystem of the columnar defects $\theta=45^o$ and (*b*) the external field is oriented along the *c-axis*, between crossed columns $\theta=0$. In case (*a*), Figure 30 (left), vortices can depin just by nucleation of the single kinks which are sliding from intersection to intersection, or, by nucleation of the super-kinks resulting in a kind of motion, similar to a variable-range hopping. This type of thermally assisted vortex depinning does not extend any additional energy on vortex bending. Another situation arises for field along the *c-axis*, Figure 30 (right). Now vortices can depin *only* via nucleation of multiple half-loops, whose characteristic size depends upon current density. This results in an additional barrier for vortex depinning, which even diverges at zero current [2]. As a result, the relaxation rate is anisotropic, i.e, it is suppressed when the external field is oriented along the mid-



direction between the two subsystems of the crossed columnar defects. This is quite the opposite to the situation of uniformly irradiated samples.

## C. *Summary and conclusions*

In this Chapter we have presented angle-resolved measurements of the irreversibility temperature in unirradiated $YBa_2Cu_3O_{7-\delta}$ film and in two films with columnar defects, induced by 6 *GeV* Pb-ions irradiation, either parallel to the *c-axis* or 'crossed' in $\theta = \pm\ 45°$. We find that in the unirradiated film the transition from irreversible to reversible state occurs *above the melting line* and marks the *crossover from the pinned to unpinned vortex liquid*. In irradiated films, within the critical angle $\theta \approx 50°$, the irreversibility line is determined by the temperature dependent *trapping angle*. For larger angles $T_{irr}$ is determined by the intrinsic anisotropy via the effective field. The formulae for $T_{irr}(\theta)$ for both unirradiated and irradiated films are given. We also discuss the possible influence of anisotropic enhancement in relaxation rate which leads to a smearing of the expected cusp at $\theta=0$ in the $T_{irr}(\theta)$ curve in the uniformly irradiated film. Finally, we demonstrate the collective action of crossed columnar defects, which can lead to suppression of relaxation and enhancement of pinning strength along the mid direction.



# Chapter V.   Local magnetic properties of HTS crystals

## A. *Miniature Hall-probe array measurements*

As discussed in Chapter I "global" magnetic measurements have a serious disadvantage, since they do not allow measurements of the magnetic induction distribution across the sample. All models describing irreversible magnetic properties of superconductors inevitably discuss peculiarities of this distribution [78, 134, 137, 144]. This has forced a search for new, *local*, techniques for the measurement of magnetic induction distribution. One of the well-known methods is based on the magneto-optic effect in which the sample is covered by some magneto-optically active substance (e. g., EuSe or yttrium-iron garnet). Polarized light passed through such material rotates proportionally to the local magnetic field due to the Faraday effect. A detailed description of the magneto-optical technique, together with numerous references, can be found in [5-7, 10, 145].

The magneto-optical technique, however, has some serious limitations, like small working field ranges (few kG), difficulties in calibration and in dynamic measurements. As an alternative, in the last several years another technique has been established, based on miniature Hall-probes.

Miniature, single Hall-probes made of InSb were already used in the early 90's [12, 114, 146, 147]. In addition, single probes of Bi films [14] GaAs/AlGaAs [13, 148, 149] and doped GaAs [15] as well as arrays of doped GaAs sensors [150] were used in static and scanning modes. Recently, the technique was improved substantially by developing arrays of Hall sensors using practically two-dimensional electron gas (2DEG) formed at a hetero-interface of GaAs/AlGaAs [37, 89, 90].



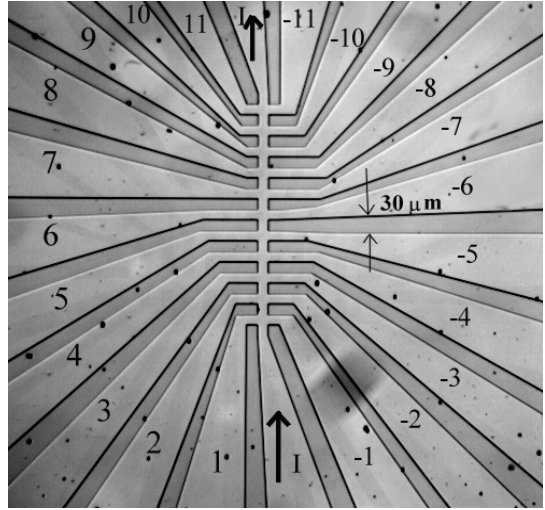

Figure 31. Optical photography of the 11-element Hall-probe array (courtesy of E. Zeldov)

The GaAs 2DEG has a mobility of about $1 \times 10^5 cm^2/V\ sec$ at 80 K and a carrier concentration of about $6.25 \times 10^{11} cm^2$ resulting in sensitivity of about $0.1\Omega/G$. These sensors have the advantage of a quick response to the magnetic field, large field and temperature working ranges, weak temperature dependence, and high sensitivity. The sensors are manufactured using well established photolithographic and etching techniques. The 2DEG is embedded in a GaAs-AlGaAs heterostructure less than 100 nm below the surface. The heterostructures are grown by MBE in a Riber model 32 system under conventional growth conditions. A 1 µm thick buffer layer is first grown on the (001) oriented, semi-insulating GaAs substrate beginning with a short super-lattice followed by undoped GaAs. A 10 µm thick AlGaAs spacer isolates the 2DEG from the ionized impurities. A delta doped region of $10^{12} cm^{-2}$ is used to supply the carriers to the 2DEG which further extends into a 25 µm AlGaAs layer uniformly doped to $10^{18} cm^{-3}$. Nominally 35 % Al is used in these AlGaAs layers and a relatively low Si concentration, in order to ensure a complete freeze-out of the excess carriers. The Al concentration is then ramped down towards the surface to 25 % over 25 nm



also doped to $10^{18}$cm$^{-3}$. A 25 nm GaAs cap layer doped to 5 x $10^{17}$cm$^{-3}$ protects the surface. This structure was chosen in order to facilitate the formation of low resistance Ohmic contacts and to avoid creeping and deterioration due to surface oxidation. Using this procedure Hall sensors with dimensions down to 1μm x 1μm are readily manufactured. Even smaller sensors can be made using more sophisticated e-beam lithography techniques. For a detailed description of the micro Hall-probe technique see [151].

Figure 32 shows a schematic placement of the Hall-probe array on a sample. In practice, the sample is adhered to the surface of the GaAs using a low-melting-temperature wax. The sensors measure the perpendicular to the sample surface component of the magnetic induction.

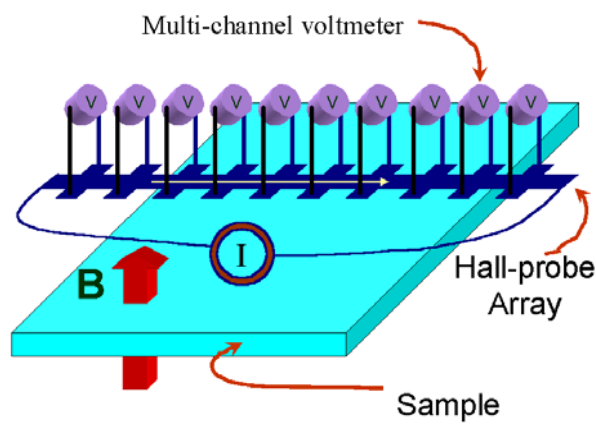

Figure 32. Scheme of the micro Hall-probe array experiment.

Measurements of the magnetic induction using a Hall-probe array may be assisted by the preliminary positioning of the array on the sample surface using a magneto-optical picture, which can readily reveal sample defects.



## B. *Analysis of the local AC magnetic response*

We begin the discussion of the results obtained using micro Hall-probes from the strongest aspect of this technique, i.e., its excellent dynamic characteristics. This means that one can use miniature Hall-sensors in AC mode, which provides ample information about the irreversible magnetic properties of superconductors in the cases when magnetic response is extremely small. This feature has been used to study the so called irreversibility line [84, 92, 93, 98, 113], melting line [2, 3, 17, 99], Bose-glass [2] and vortex-glass [2] transitions and even transport current distributions. In general AC response can be described considering a superconductor as a non-linear conductor with current dependent resistivity [152-155]. We will, however, show that even simple geometrical considerations using the critical state model can give an idea of the physics behind the local AC response.

The following discussion is based on the results of Ref. [156] and describes local AC magnetic response ignoring the magnetic relaxation during the AC cycle (we consider effects of magnetic relaxation in the next section). Here we analyze the *local* AC response, as measured by a miniature Hall probe located on the sample surface at some distance from its edge. Our analysis shows important differences between the AC response at different locations of the Hall probe. Understanding of these differences is important for correct interpretation of the local AC measurements and for the utilization of these measurements in characterizing and mapping of inhomogeneities across the sample.

The analysis is based on the following assumptions:



a) The dimensions of the Hall probe are much smaller than the sample size. As discussed in Chapter V.A this condition is well satisfied for today's Hall probes and usual high-$T_c$ samples.

b) Variations of the magnitude of the persistent current $j$ due to the alternating field $H_{ac}$ can be neglected. This condition is usually satisfied in experiments using AC fields much smaller than the DC field.

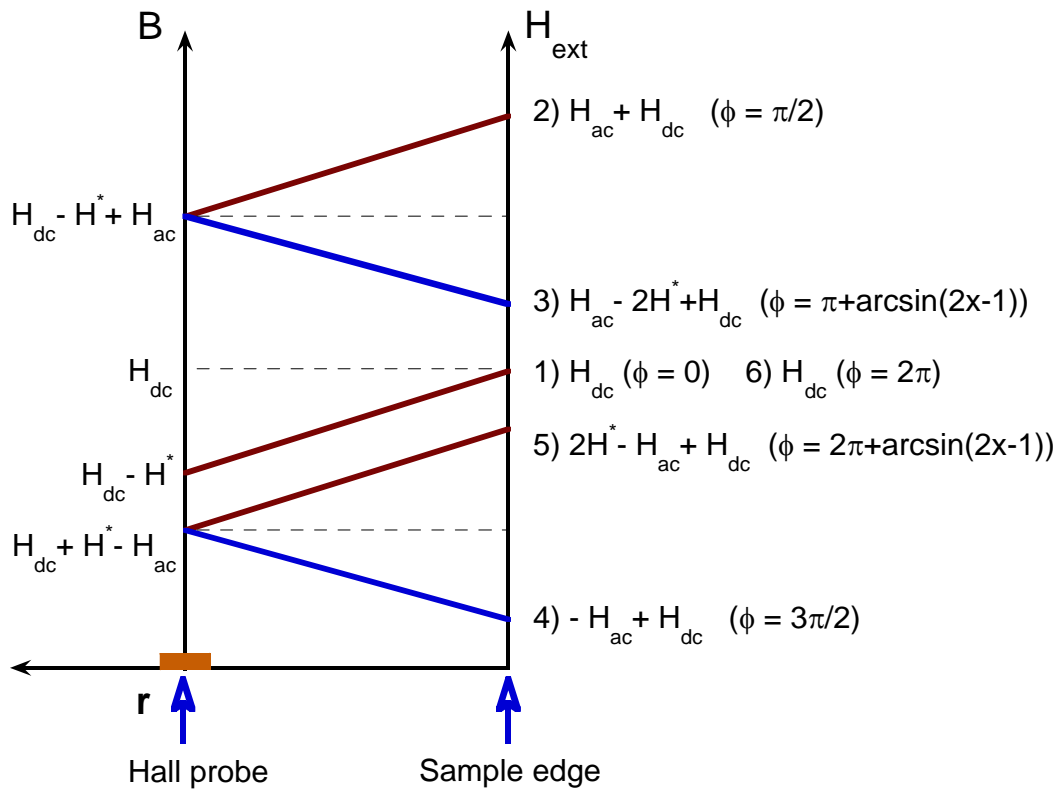

Figure 33. Schematic description of the magnetic induction evolution during one cycle. Straight lines are only for simplicity of the plot.

Assume now that a miniature Hall probe is located at a distance $r$ from the nearest edge of the sample located at $r=0$, as shown in Figure 33. Note that the magnetic induction profiles are not necessarily straight lines [23] and we use straight lines in Figure 33 only for simplicity. We define "a local penetration field" $H^*(r)$ as the first actual AC field at which oscillations of the z-component of the magnetic



induction $B_z$ reach the Hall-probe. Such a definition of $H^*$ is relevant for any geometry, and it takes into account the lower critical field $H_{c1}$. The field $H^*$ is directly related to the persistent current $j$; For a homogeneous infinite slab or cylinder with a zero demagnetization factor $H^*_{inf} = 4\pi \cdot jr/c$. For a homogeneous thin film [22]:

$$H^*_{film} = \frac{4\pi}{c} t(1+\alpha) j_s \operatorname{arccosh}\left(\frac{W}{W-r}\right),$$

where $W$ is a coordinate of the center of the sample, $t$ is the thickness, and $\alpha$ accounts for the effect of demagnetization. The parameter $\alpha$ depends on the position of the Hall probe and on the geometry of the sample, and it determines the actual magnitude of the AC field on the edge of a sample: $h_{ac} = (H + \alpha H^*)\sin(\omega t) = H_{ac}\sin(\omega t)$, where $H$ is the amplitude of the applied AC field.

To illustrate the method of calculation, let us consider the response of the Hall probe for the descending part of $h_{ac}$ (see Figure 33). Down to the field $H = H_{dc} + H_{ac} - 2H^*$ the Hall probe does not feel any change in the external field. After this field is reached, $B_z(h_{ac})$ follows the changes in the external field through a quite complicated relationship [22]. However, it quickly converges to a linear dependence. Assuming a linear relationship between $B_z$ and $h_{ac}$, one can easily calculate the local AC response as a function of time or, more conveniently, as a function of the phase $\varphi = \omega t$ during one period. Figure 33 shows schematic profiles of the field at five different stages during a complete cycle. The dependence of $B_z$ on the phase $\varphi$ during the period between these stages is given in Table 1. The DC part of $B_z$ is omitted since it does not affect the AC response.



Table 1. Magnetic induction $B_z(\varphi)$ at the Hall probe location during one cycle of the AC field. The parameter $x = H^*/H_{ac}$.

| Stage # (Figure 33) | $\varphi$ | $B_z(\varphi)$ |
|---|---|---|
| 1 to 2 | $0 \to \dfrac{\pi}{2}$ | $H_{ac}\sin(\varphi) - H^*$ |
| 2 to 3 | $\dfrac{\pi}{2} \to \pi + \arcsin(2x-1)$ | $H_{ac} - H^*$ |
| 3 to 4 | $\pi + \arcsin(2x-1) \to \dfrac{3\pi}{2}$ | $H_{ac}\sin(\varphi) + H^*$ |
| 4 to 5 | $\dfrac{3\pi}{2} \to 2\pi + \arcsin(2x-1)$ | $H^* - H_{ac}$ |
| 5 to 6 | $2\pi + \arcsin(2x-1) \to 2\pi$ | $H_{ac}\sin(\varphi) - H^*$ |

As an illustration we show in Figure 34 graphs of $B_z(\varphi)$ as calculated from Table 1 for $H^* = 0$ and for $H^* = H_{ac}/2$. When $H^*$ is zero the magnetic induction follows the external alternating field. In contrast, when $H^*$ is not zero (e.g., $x = 1/2$ in Fig. 2) the signal is cut and there is a pronounced phase shift, which is proportional to the ratio $x = H^*/H_{ac}$.



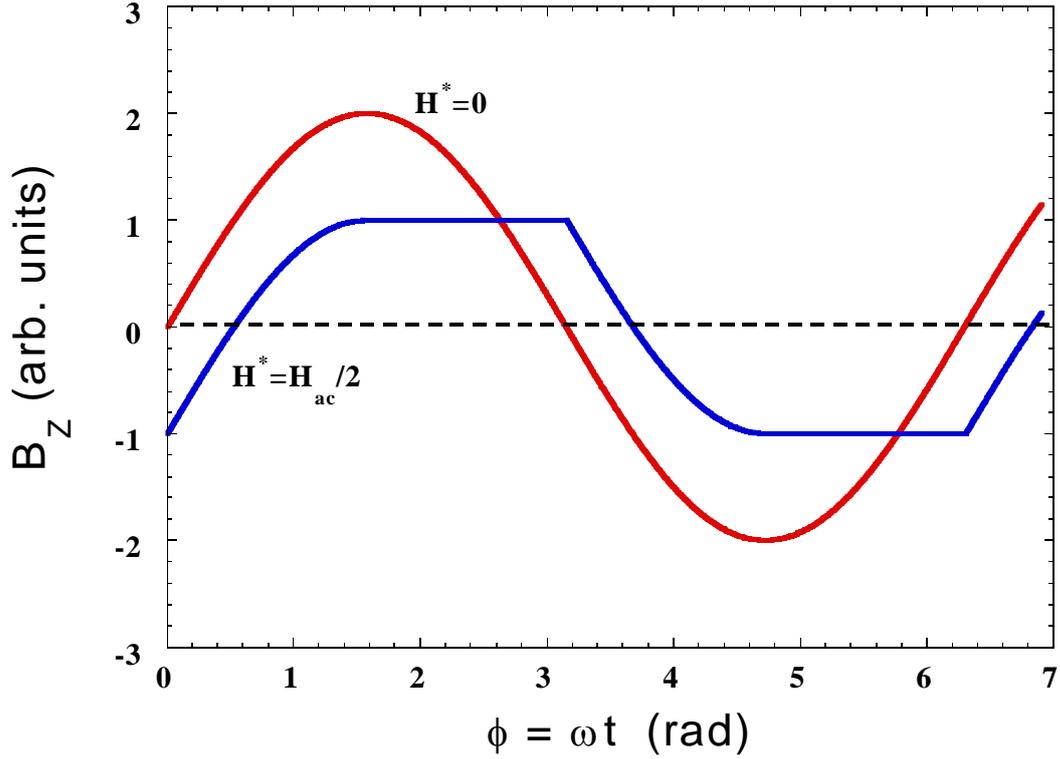

Figure 34. Curves of $B_z(\varphi)$ as calculated from Table 1 for $H^*=0$ and $H^* = H_{ac}/2$.

Wang and Sayer [157] experimentally observed both features, the saturation of the AC signal and a phase shift, although the geometry of their experiment was slightly different. Once the waveform of $B_z$ is known, the real and imaginary parts of the local harmonic susceptibilities, $\chi'_n$ and $\chi''_n$, respectively, can be calculated:

$$\begin{Bmatrix} \chi'_n \\ \chi''_n \end{Bmatrix} = \frac{1}{H_{ac}} \int_0^{2\pi} B(\varphi) \begin{Bmatrix} \sin(n\varphi) \\ \cos(n\varphi) \end{Bmatrix} d\varphi \qquad (59).$$

Note that in the normal state our definition gives $\chi'_1 = \pi$. The global harmonic susceptibilities of the entire sample can be calculated by integration:

$$X_n = \frac{\delta}{\pi} \int_0^k \chi_n(x) dx \qquad (60)$$



where, $\delta = H_{ac}/H_p$ is the parameter of the global response defined in Ref. [158], (inversely) analogous to our local parameter $x$, and $H_p$ is the field of full penetration up to the center of a sample. The upper integration limit $k = \begin{cases} 1/\delta & \text{if } \delta \geq 1 \\ 1 & \text{if } \delta \leq 1 \end{cases}$.

Table 2. Local harmonic susceptibilities $\chi_n$ ($n$ = 1, 3, 5) normalized by $H_{ac}$. Here $x = H^*/H_{ac}$ and $\rho = \sqrt{x(1-x)}$.

| | |
|---|---|
| $\chi'_1$ | $\frac{\pi}{2} - 2\rho(2x-1) - \arcsin(2x-1)$ |
| $\chi''_1$ | $-4\rho^2$ |
| $A_1$ | $\sqrt{\frac{\pi^2}{4} + 2(2x-1)\rho(2\arcsin(2x-1) - \pi) + 4\rho^2 + \arcsin(2x-1)(\arcsin(2x-1) - }$ |
| $\chi'_3$ | $\frac{16}{3}\rho^3(2x-1)$ |
| $\chi''_3$ | $\frac{4}{3}\rho^2(1-8\rho^2)$ |
| $A_3$ | $\frac{4}{3}\rho^2$ <br> The maximum $A_3 = 1/3$ is reached at $x = \rho = 1/2$. |
| $\chi'_5$ | $\frac{16}{15}\rho^3(2x-1)(5-32\rho^2)$ |
| $\chi''_5$ | $\frac{4}{15}\rho^2(8\rho^2(9-32\rho^2) - 3)$ |
| $A_5$ | $\frac{4}{15}\rho^2\sqrt{9-32\rho^2}$ <br> $A_5$ has two maxima: $A_5 = \frac{\sqrt{3}}{20}$ at $x = \frac{1}{4}$ and $x = \frac{3}{4}$, <br> and a minimum $A_5 = \frac{1}{15}$ at $x = \frac{1}{2}$. |

The results for the first, third and fifth harmonics are summarized in Table 2. In this table $A_n = \sqrt{(\chi'_n)^2 + (\chi''_n)^2}$ is the magnitude of the n-th harmonic



susceptibility, and $\rho = \sqrt{x(1-x)}$. Even harmonics do not exist in this model due to the symmetrical functional form of $B_z(\varphi)$.

For the experimental verification of these results we measured the AC response of an $YBa_2Cu_3O_{7-\delta}$ thin film of thickness 3000 Å using a $50 \times 50$ $\mu m^2$ micro Hall probe attached to the surface of the film and located at a distance approximately 500 $\mu m$ from the edge of the sample. The measurements were performed at a fixed temperature ($T$ = 88 K), DC magnetic field ($H_{dc}$ = 200 G) and frequency of the AC field ($f$ = 177 Hz). Figure 35 shows the third and the fifth harmonics (both normalized by $H_{ap}$) as a function of $H_{ap}$. Symbols are experimental points and the solid lines are fits calculated using the expressions given in Table 2. A good agreement of the experiment with the model is evident. Note that the onset of these local responses is obtained at $H_{ac} \cong 0.3$ G, while the onset of the global response is always obtained at $H_{ac} \to 0$.



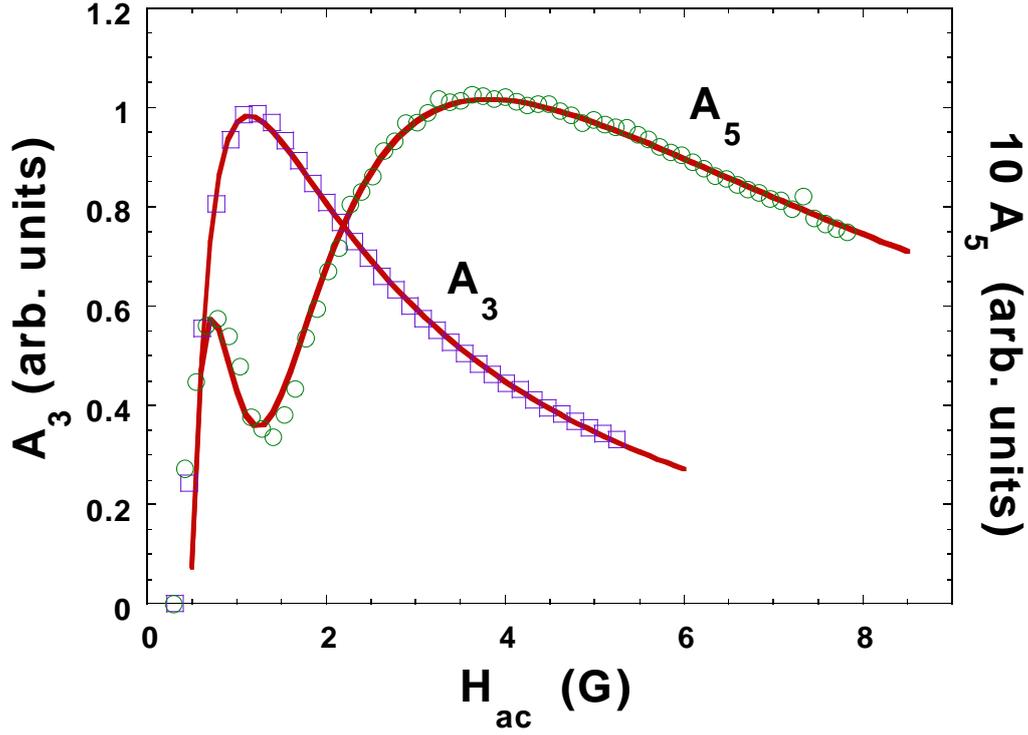

Figure 35. Absolute values of the third ($A_3$) and fifth ($A_5$) harmonic signals measured in $YBa_2Cu_3O_{7-\delta}$ thin film as a function of applied AC field at $T$=88 K and $H_{dc}$ = 200 G. Symbols are experimental data and solid lines are fits.

As was pointed out above, local magnetic AC measurements provide a tool for detecting inhomogeneous regions in the sample. This is illustrated in the following example. Consider a long sample with the edge located at $r$=0 and the sample itself occupies the positive part of the $r$ axis. We denote by $r$=1 the maximum propagation distance of the AC field. Suppose that due to some reasons, such as variations of the oxygen content or defects in the microstructure, the persistent current density varies with the coordinate as $j(r)$. The penetration field then is given by: $H^*(r) = \frac{4\pi}{c} \int_0^r j(\xi) d\xi$. Thus, the shielding current density is simply $j(r) = \frac{c}{4\pi} \frac{dH^*(r)}{dr}$. Therefore, all that we need in order to reconstruct $j(r)$ is just to



measure $H^*(r)$. To obtain $H^*(r)$ at fixed temperature and DC field, one can measure, for instance, $A_3(H_{ac})$ $A_3(H_{ac})$ and detect its peak position $H^{peak}_{ac}$. In general, $H^{peak}_{ac} = \left(1-\alpha + \sqrt{1-\alpha}\right)H^*$, thus measurements of $H^{peak}_{ac}$ yield $H^*(r)$. By locating a single Hall-probe at different distances $r$ from the edge or by using Hall-probes array one can map $H^*(r)$ and hence reconstruct $j(r)$. Instead of measuring the peak position of the third harmonic, the same information can be extracted from the magnitude $A_3(r)$ provided $H_{ac}$ is high enough to penetrate the center of the sample.

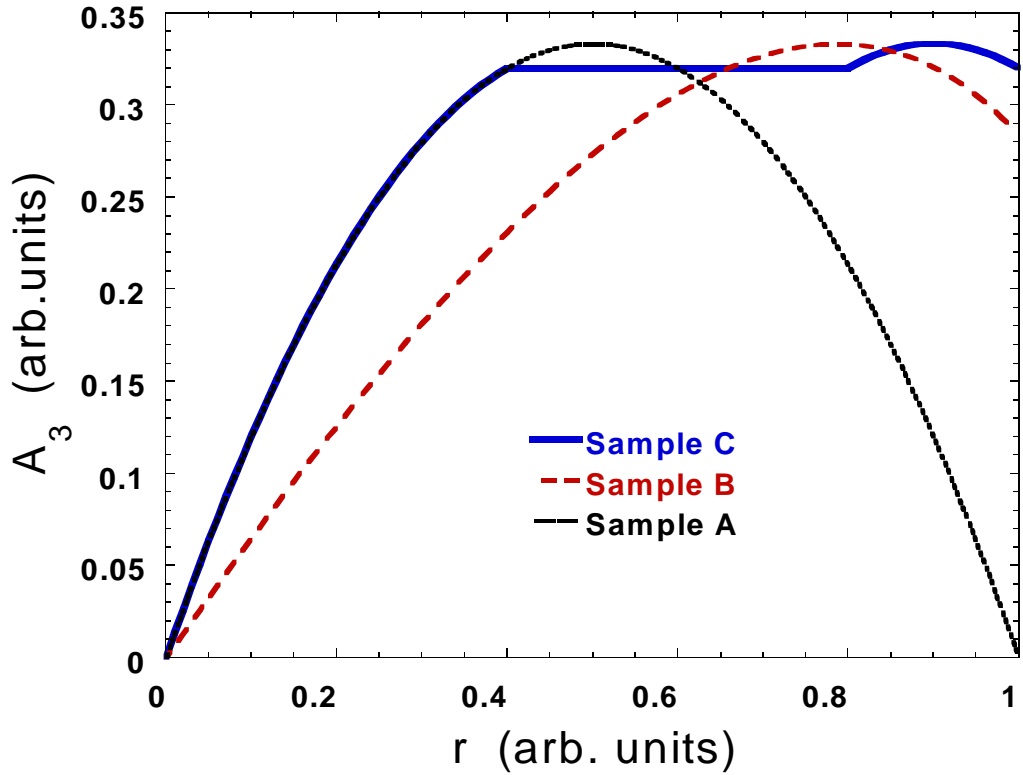

Figure 36. Spatial variations of the magnitude of the third harmonic signal for samples A, B and C described in the text.

As a specific example, consider three model samples: Sample A is homogeneous, i. e., $j$=const across the sample; sample B is inhomogeneous such that the shielding current increases from the edge to the center as $j(r)=1/(2-r)$; and sample



C has a large defect so that for fractional distance from the edge $r = 0.4 - 0.8$, the shielding current $j$ is zero and $j$ is 1 in the rest of material. Figure 36 shows the profiles of the third harmonic amplitude $A_3(r)$ corresponding to samples A, B, and C. Note that for the homogeneous sample A, $A_3(r)$ is **not** constant but rather parabolic. Similarly, in the defect region of sample C, where $j=0$, $A_3(r)$ is not zero but rather a finite constant. The relationship between $A_3(r)$ and $j(r)$ given in Table 2 can be reversed to allow mapping of $j(r)$ from measurements of $A_3(r)$.

Besides local variations of $j$, one may also measure distribution of other parameters such as the irreversibility temperature $T_{irr}$ (see [159] and discussion in the following Section). Measurement of the onset temperature of the third harmonic signal has been established as a reliable method for determining $T_{irr}$ [76, 129, 131, 156]. Our analysis shows that in a homogeneous sample, while the peak position, height and the width of the third harmonic signal vary as a function of the position $r$, the **onset** of this signal (on cooling from above $T_c$) is the same for all positions of the Hall probe. Thus, variations in the onset temperature of the third harmonic signal indicate inhomogeneous distribution of $T_{irr}$ across the sample.

We note that for a detailed interpretation of the experimental data one should take into account such factors as surface barrier, finite area of the Hall probe and finite distance of the Hall probe from the surface of the sample.

## C. *Frequency dependence of the local AC magnetic response*

The discussion in this Section is based on the work [RP13].

In a previous section we have shown that in the framework of the Bean model local AC response is determined by a *single parameter* which is the ratio between the penetration field $H^*$ and the amplitude of the applied alternating field. This implies



that data of the AC response as a function of any experimental variable (e.g., temperature, DC field, amplitude of the AC field, etc.), can be reduced to a universal curve that describes the response as a function of a single parameter. As a particular result, the peak heights of the harmonic susceptibilities should be universal constants for all type II superconductors. This conclusion is borne out in many experiments [160] in which the variable-parameter is either the DC field or the amplitude of the AC field.

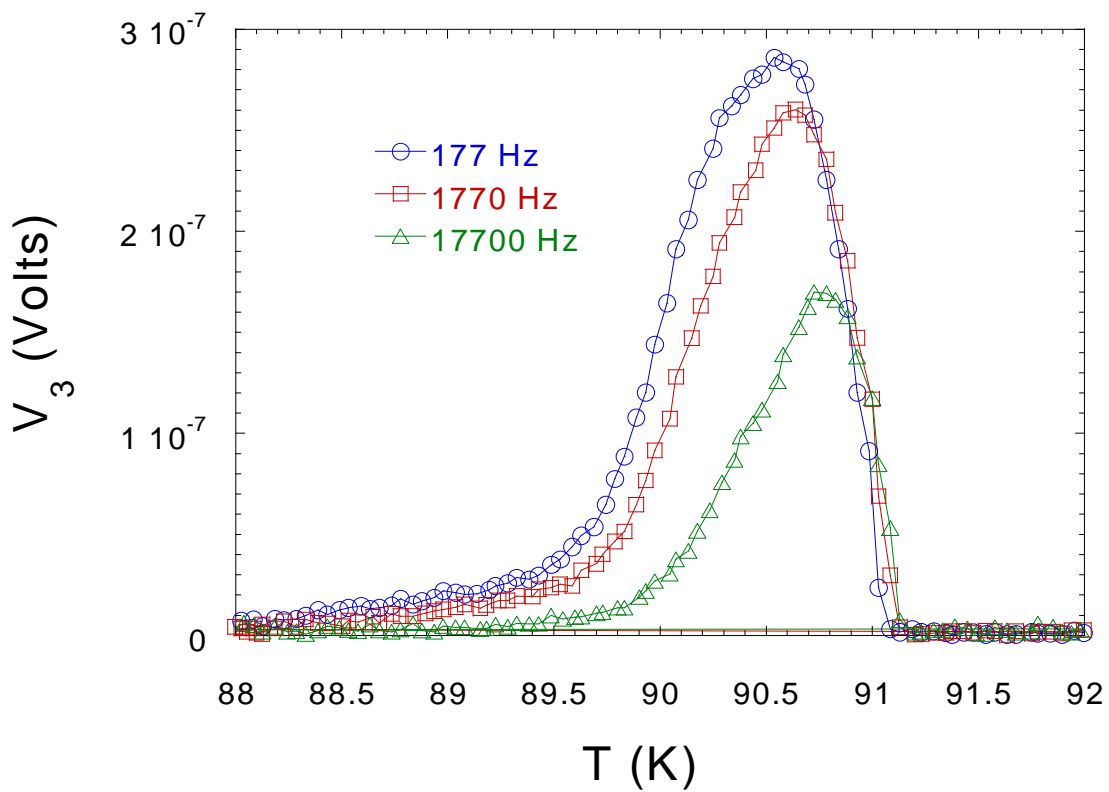

Figure 37. Temperature dependence of the third harmonic response $A_3(T)$ in a $YBa_2Cu_3O_{7-\delta}$ single crystal for the indicated frequencies (after Y. Wolfus *et al.* [129, 161])

However, recent results [129, 161] on $YBa_2Cu_3O_{7-\delta}$ crystals, depicted in Figure 37, exhibit a strong dependence of the peak height and the peak position on frequency. The figure shows a significant reduction in the third harmonic response, $A_3$, as the frequency increases in a relatively narrow range (0.17 - 17.7 kHz). In



addition, the peak of $A_3$ versus temperature becomes narrower and its position shifts toward higher temperatures as the frequency increases. Similar results were reported by Van der Beek *et al.* [117] who measured the third harmonic transmittivity $T_3$ versus temperature in *Bi-Sr-Ca-Cu-O* crystals in the frequency range 0.79 - 2403 Hz. Their data show a monotonous decrease in the peak height of $T_3$ with frequency. It is thus apparent that the frequency cannot be incorporated in the single-parameter description.

Effects of frequency on the AC response were previously analyzed by Gilchrist and Konczykowski [162], modeling the superconductor as one or two inductively coupled loops. This analysis was particularly aimed toward understanding the results of "screening" experiments in which two coils are placed on either side of a planar sheet or film to measure the transmittivity of the sample.

Our approach in this Section is based on the critical state model, taking into account magnetic relaxation effects. Analyzing the local response rather than the global response explains the role of frequency in a most elementary way. We show that magnetic relaxation influences the AC response not only through the frequency dependence of the shielding current density $j$, but also introduces the frequency as an additional parameter to the model. The single parameter approach [149, 150] becomes valid in the limit of high frequencies where magnetic relaxation effects can be neglected.



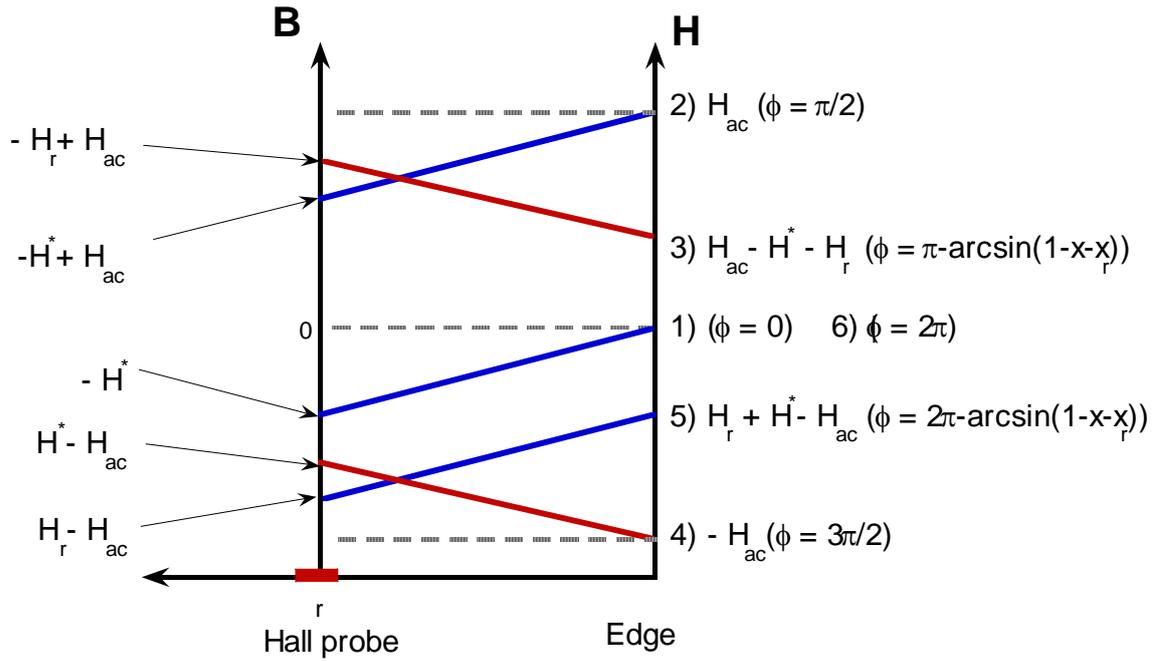

Figure 38. Schematic description of magnetic induction profiles during one cycle of the applied AC field taking into account relaxation.

Here again we use the same method of calculation as in previous Chapter V.B. Here we refer to Figure 38, which describes the evolution of the magnetic induction profiles in a sample during one period in the sample with relaxation. In spite of apparent similarity with Figure 33, these figures differ in details of magnetic induction distribution in the vicinity of the Hall probe. As above, we start from the moment at which the applied AC field is zero (point 1). As the AC field $h_{ac} = H_{ac} \sin(\varphi)$ increases, the magnetic induction at the Hall probe location follows the field up to the point where the applied AC field reaches its maximum value at $\varphi = \dfrac{\pi}{2}$, (point 2). At this moment the magnetic induction at the Hall-probe location is $B_o = H_{ac} - H^*$, where $H^*$ is the "local penetration field". When $h_{ac}$ decreases from point 2, the magnetic induction at the edge of the sample is affected immediately,



however it takes time for the external field to reach a value required to affect the induction at the Hall probe location. During this time, $B_o$ is not affected by the changes in the external AC field, however it increases due to magnetic relaxation effects. At the end of this time period (point 3) the external change in the magnetic field reaches the Hall probe at a phase value $\varphi_r$ corresponding to time $t_r$. In order to calculate $\varphi_r$, let us assume that between stages 2 and 3, $B_o$ has relaxed from $H_{ac} - H^*$ to $H_{ac} - H_r$. Then, according to Figure 38,

$$H_{ac} - H^* - H_r = H_{ac} \sin(\varphi_r) \tag{61}$$

or

$$1 - x - x_r = \sin\left(\frac{\pi}{2} + \omega t_r\right) = \cos(\omega t_r) \tag{62}$$

where $x = \dfrac{H^*}{H_{ac}}$ and $x_r = \dfrac{H_r}{H_{ac}}$.

Once the AC field reaches the Hall probe, we assume that the local induction follows the changes in the external field until the latter changes its sign (point 4 in Figure 38). Thus, in order to calculate the waveform of the local induction during half a cycle, one needs to calculate $t_r$ from Eq.(62), and the time dependence of the local induction during the time period $\{0, t_r\}$. Using similar arguments one can calculate the waveform of the local induction during the second half of the cycle (from point 4 to 5 and back to point 1). From Eq.(62) it is clear that $t_r$ is a function of $x$, $x_r$, and the frequency $\omega$. The parameter $x_r$ is determined by the relaxation law and $x$. Accordingly, the AC response depends on $x$, $\omega$, and the relaxation law. This is the main result of our analysis. It shows that magnetic relaxation affects directly the AC response by introducing the frequency as an additional parameter to the model. The frequency



affects the response also indirectly through the parameter *x* which depends on the shielding current which itself is a function of frequency.

In order to calculate $t_r$ explicitly and the behavior of the local induction in the time period $\{0, t_r\}$, one must engage some analytical form for the relaxation law. At high enough frequencies (small $t_r$) one may assume a linear dependence of $H_r$ upon time: $H_r = H^*\left(1 - \dfrac{t}{t_o}\right)$, where $t_0(T, H_{DC}) > t_r$ is a characteristic time which depends on temperature and the applied DC field.

Table 3. Magnetic induction $B_z(\varphi)$ at the Hall probe location during one cycle of the AC field.

| $h_{ac}$ | $B(\varphi)$ | $\varphi$ |
|---|---|---|
| $0 \to H_{ac}$ | $H_{ac}\sin(\varphi) - H^*$ | $0 \to \pi/2$ |
| $H_{ac} \to H_{ac} - H^* - H_r$ | $H_{ac} - H^*\left(1 - \dfrac{\varphi - \pi/2}{\omega t_o}\right)$ | $\pi/2 \to \pi - \arcsin(1 - x - x_r)$ |
| $H_{ac} - H^* - H_r \to -H_{ac}$ | $H_{ac}\sin(\varphi) + H^*$ | $\pi - \arcsin(1 - x - x_r) \to 3\pi/2$ |
| $-H_{ac} \to -H_{ac} + H^* + H_r$ | $-H_{ac} + H^*\left(1 - \dfrac{\varphi - 3\pi/2}{\omega t_o}\right)$ | $3\pi/2 \to 2\pi - \arcsin(1 - x - x_r)$ |
| $-H_{ac} + H^* + H_r \to 0$ | $H_{ac}\sin(\varphi) - H^*$ | $2\pi - \arcsin(1 - x - x_r) \to 2\pi$ |

Such simplified time dependence serves as an illustration, since it allows us to obtain a relatively simple analytical expression for the local induction. Inserting $x_r = x\left(1 - \dfrac{t_r}{t_o}\right)$ in Eq. (62) results:

$$x + x\left(1 - \frac{t_r}{t_o}\right) \approx \frac{(\omega t_r)^2}{2} \tag{63}$$

from which $t_r$ can be calculated:



$$\omega t_r = \frac{x}{\omega t_0}\left(\sqrt{1+\frac{4\omega^2 t_0^2}{x}}-1\right) \tag{64}$$

For $x_r$ and $\varphi_r$ one obtains:

$$x_r = x\left(1-\frac{t_r}{t_0}\right) \approx x\left(1-\frac{x\left(\sqrt{1+\frac{4\omega^2 t_0^2}{x}}-1\right)}{\omega^2 t_0^2}\right) \tag{65}$$

$$\varphi_r \approx \pi - \arcsin\left(1-2x+\frac{x^2\left(\sqrt{1+\frac{4\omega^2 t_0^2}{x}}-1\right)}{\omega^2 t_0^2}\right) \tag{66}$$

Table 3 gives the functional form of the local magnetic induction $B_z$ during one cycle, and Figure 39 illustrates the waveform of $B_z$ as calculated from this table for $x = 1/2$.

For comparison, the waveform of $B_z$, ignoring magnetic relaxation effects, is also illustrated (dotted curve in Figure 39, see also Figure 34). It is seen that due to relaxation the "cut-off" in the second stage is not constant but varies with time (linearly in our approximation). At this stage the time is related to the phase through: $\omega t = \varphi - \pi/2$. Once the waveform of $B_z$ is known, the real part $\chi'_n$, imaginary part $\chi''_n$ and magnitude $A_n = \sqrt{(\chi'_n)^2 + (\chi''_n)^2}$ of the local harmonic susceptibilities can be calculated from Eq.(59) using Table 3.



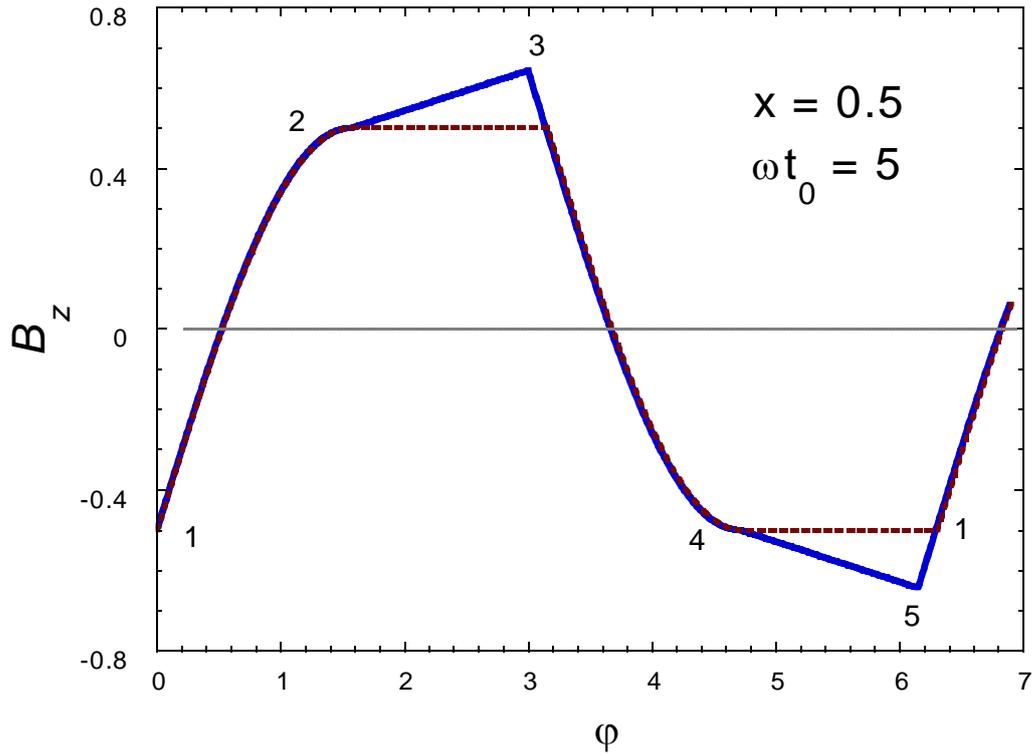

Figure 39. Waveforms of the magnetic induction during one cycle, for $x=H^*/H_{ac}$=0.5, with ($\omega t_0 = 5$) and without relaxation (solid and dotted lines, respectively). Numbers correspond to the stages in Figure 38

Because of the complexity of these expressions we prefer to present the results in graphical form. We use these expressions to show, in Figure 40 and Figure 41, the harmonic susceptibilities $\chi_n$ (n=1,3) as a function of $x$ for two values of frequency corresponding to $\omega t_0$=5 and $\omega t_0$=100. It is seen that the frequency has strong effects on both, the peak height as well as the width of the curves. In order to compare our theoretical results with experiments, we plot in Figure 42 the absolute value of the third harmonic $A_3$ as a function of temperature. Here we have assumed exponential temperature dependence of shielding current [149] and $t_0 \sim (1-T/T_c)^2$.



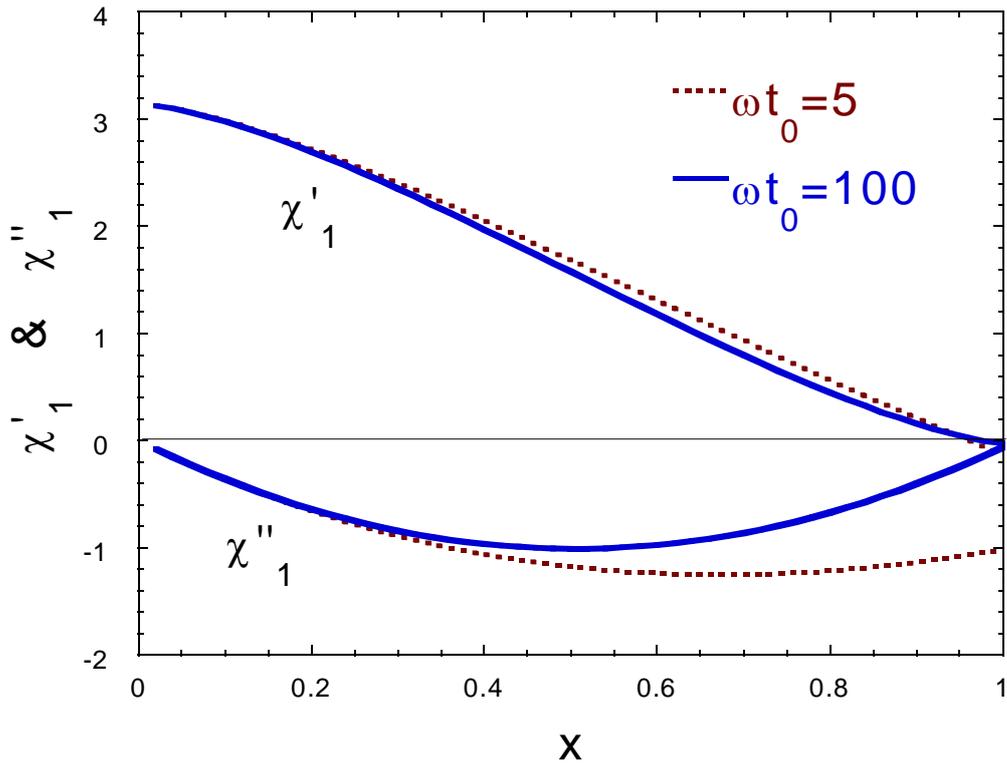

Figure 40. First harmonic susceptibility calculated from Eq.(80) and Table 3

A comparison between Figure 37 and Figure 42 shows that the main experimental observations are satisfactorily described by our analysis; i.e., the peak height of $A_3$ decreases and the location of the peak is shifted towards higher temperature as the frequency increases. Obviously, a quantitative comparison in the whole temperature range between the experimental results and our model can be made knowing the explicit form of the temperature dependence of the shielding current and the exact relaxation law in the material.



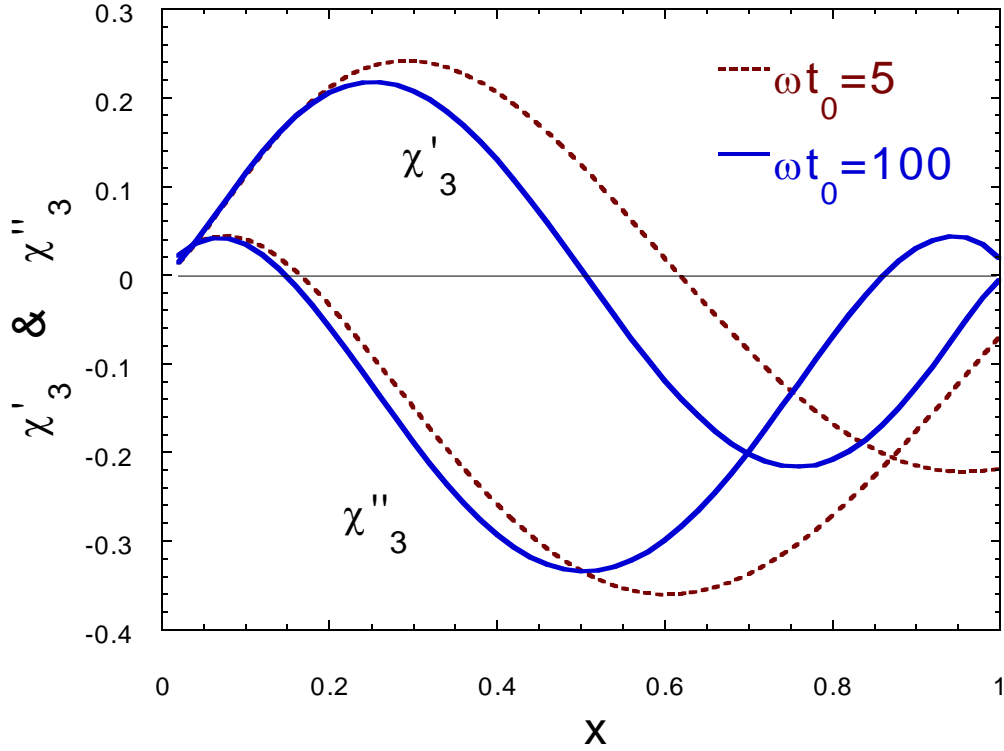

Figure 41. Third harmonic susceptibility calculated from Eq.(59) and Table 3

As a particular application of the above analysis, we outline below a method to obtain the short-time relaxation law $j_s(\omega)$. For example, one can measure the third harmonic susceptibility $\chi_3$ as a function of $H_{ac}$ at various frequencies ω, at a constant DC magnetic field and temperature. The peak height $A_3^p(\omega)$ of each curve is a function of the parameter $\omega t_0$ alone. Thus, by fitting $A_3^p(\omega)$ to the theoretical prediction one can determine $t_0$. The location $H_{ac}^p$ at which the peak in $\chi_3$ occurs is a function of both, $x$ and $\omega t_0$. Using the previously determined value of $t_0$ and the experimental values for the location of the peaks, one can determine the $x$ values, $x^p(\omega)$, corresponding to the peaks. Hence, one can calculate $H^*(\omega) = x^p(\omega) H_{ac}^p(\omega)$, which relates to the shielding current through a numerical factor $\gamma$, determined by the sample geometry [150]. We note that our simplification



of a linear relaxation does not contradict a situation where the latter procedure yields a non-linear $j_s(\omega)$; While the above procedure yields the effective value of $H^*(\omega)$ in a wide frequency range, the time $t_r$ is short and one can make a linear approximation for any relaxation law.

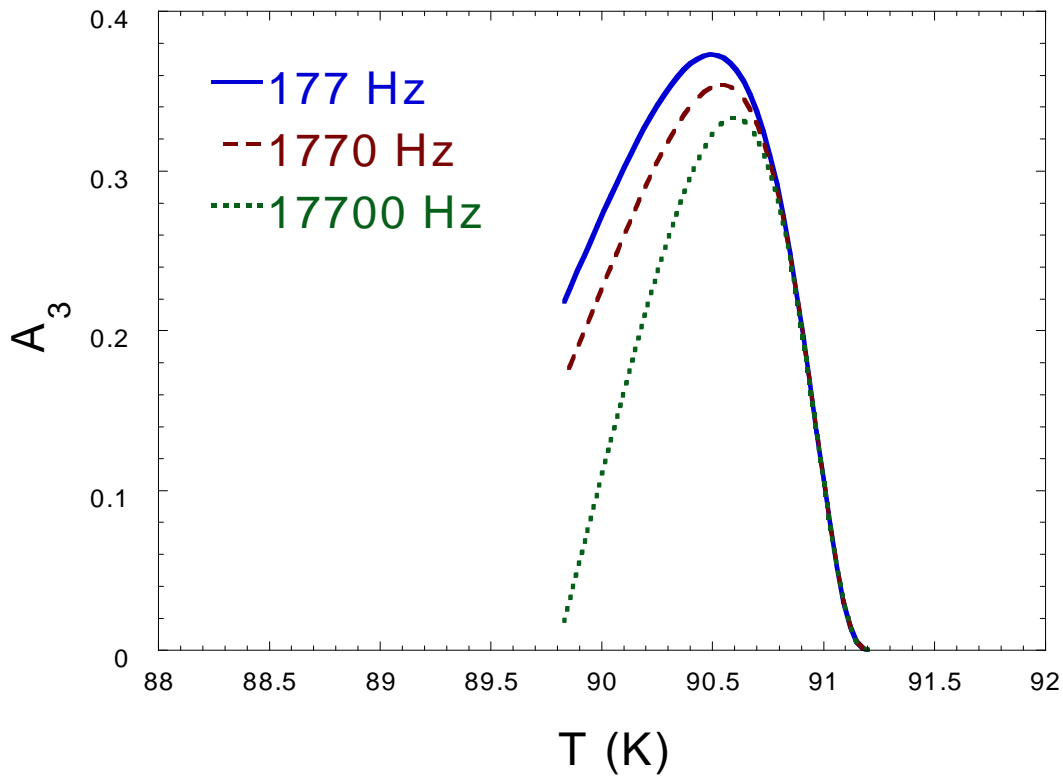

Figure 42. Third harmonic signal $A_3 = \sqrt{(\chi_3')^2 + (\chi_3'')^2}$ as a function of temperature, calculated from Eq.(59) and Table 3 for various frequencies, see text. The qualitative similarity to the experimental data of Figure 37 is apparent.

In summary, local AC magnetic response can be analyzed using a critical state model. Introducing the relaxation makes this model more plausible. The advantage of this method is its clarity. On the other hand the more general treatment of the problem (however, less transparent) is achieved considering a superconductor as a highly non-linear conductor. Then, using Maxwell equations, proper boundary conditions and particular material law $E(j)$, one can describe the static and dynamic response of



superconductors. We deal with experimental determination of *E(j)* from local magnetic measurements in the following sections.



## D. *Local magnetic relaxation in high-$T_c$ superconductors*

Another application of the excellent dynamic performance of miniature Hall-probe arrays is found in local measurements of magnetic flux creep – slow process, so that a good sensitivity is required.

Thermally activated flux creep (or magnetic relaxation) in type-II superconductors is the subject of intensive study (for a review, see [159]). In the critical state Lorentz force $\mathbf{F}_L = (\mathbf{j} \times \mathbf{B})/c$ exerted by the persistent current $\mathbf{j}$ on vortices balances the pinning force. At any finite temperature there is a finite probability for a vortex to overcome a local barrier $U(j)$ and jump to another local minimum. Direction of the Lorentz force determines the direction of the overall displacement of vortices. Thus, with time initially inhomogeneous distribution of the magnetic induction is smoothed out causing a decay of the persistent current density. On the other hand, motion of vortices with mean velocity $\mathbf{v}$ results in appearance of the electric field $\mathbf{E} = (\mathbf{v} \times \mathbf{B})/c$. This in fact means that any type-II superconductor is a (nonlinear) dissipative conductor with finite resistance $\rho = E/j$ [19]. In conventional superconductors magnetic relaxation is very slow [159, 163], whereas in high-temperature superconductors it is usually so fast, that it was called a "giant flux creep" [159].

In principle, one can describe any non-linear conductor using a set of Maxwell equations and the "material law", i. e., $E(j)$ characteristics. Following Anderson and Kim [136, 164] the latter is conventionally written in a Boltzman form $E = E_c \exp(-U(j)/T)$, where $U(j)$ is the average barrier for magnetic relaxation - in general a non-linear function of $j$. Thus, determination of $E(j)$ characteristics in



superconductors has become an experimental and theoretical challenge. Experimentally, most of such study has been conducted using transport measurements, which has several limitations. First, transport current distribution is different from that of persistent "magnetization" current and second, most important, insufficient sensitivity prohibiting measurements of the low electric fields. The alternative approach is to use direct measurement of the magnetic relaxation.

Consider again an infinitely long bar of width $2w$ and thickness $2d$ in magnetic field along the $z$-axis (see Figure 1). Both persistent current and electric field have only one y-component, whereas magnetic induction has two: $B_x$ and $B_z$. Maxwell equation for electric field in this geometry is:

$$\frac{1}{c}\frac{\partial B_z}{\partial t} = -\frac{\partial E}{\partial x} \tag{67}$$

If our reference point is at z=0 and the slab occupies -∞<z<∞ then integrating Eq.(67) equation across the sample yields:

$$E(w) = -\frac{1}{c}\frac{\partial}{\partial t}\left(\int_{\xi=0}^{w} \left(B_z(\xi) - H\right)d\xi\right) \tag{68}$$

As discussed in Appendix B in the case of an *infinite* in *z*-direction slab, the integral on the right-hand side is proportional to the volume magnetization $m = M/V$ and

$$E(w) = -\frac{4\pi w}{c}\frac{\partial m}{\partial t} \tag{69}$$

This equation has been routinely used for the estimation of the electric field from the measurements of the magnetic relaxation. Unfortunately, in finite samples the integral in Eq.(68) is not proportional to the magnetic moment, which instead is described by a more complicated relationship Eq.(99). Also, measurements of the relaxation of the global magnetic moment do not provide information about local dynamics, which may be different in the bulk and on the surface (for example due to



surface barriers). Therefore, one needs to employ local measurements of the magnetic induction and use Eq.(67) for the direct estimation of the electric field in the sample surface.

Important note !

All local techniques provide information about magnetic fields on the sample surface and, in general, this does not reflect what happens in the sample interior. For example, on the top of an infinitely long bar, values of both magnetic induction (see Chapter I.C.2) and electric field [19] are half of their bulk values (in the middle of an infinite bar). In a thick sample, ($d\sim w$), this problem is very significant and may lead to a wrong, quantitative interpretation of the experimental results. Fortunately, in most commonly used samples $d<<w$ and surface values reflect, more or less, the behavior in the sample interior. Naturally, the most favorable samples are thin films although as we saw in Chapter II, current density is inhomogeneous even in this case.

## E. *Measurements of E(j) characteristics*

In this section we demonstrate the application of the Hall-probe array technique described in Chapter V.A for the direct measurements of the local *E(j)* characteristics.

Typical profiles of the local magnetization ($m(x)=B(x)-H$) are shown in Figure 43. Different symbols indicate profiles measured at different values of the external magnetic field *H*.



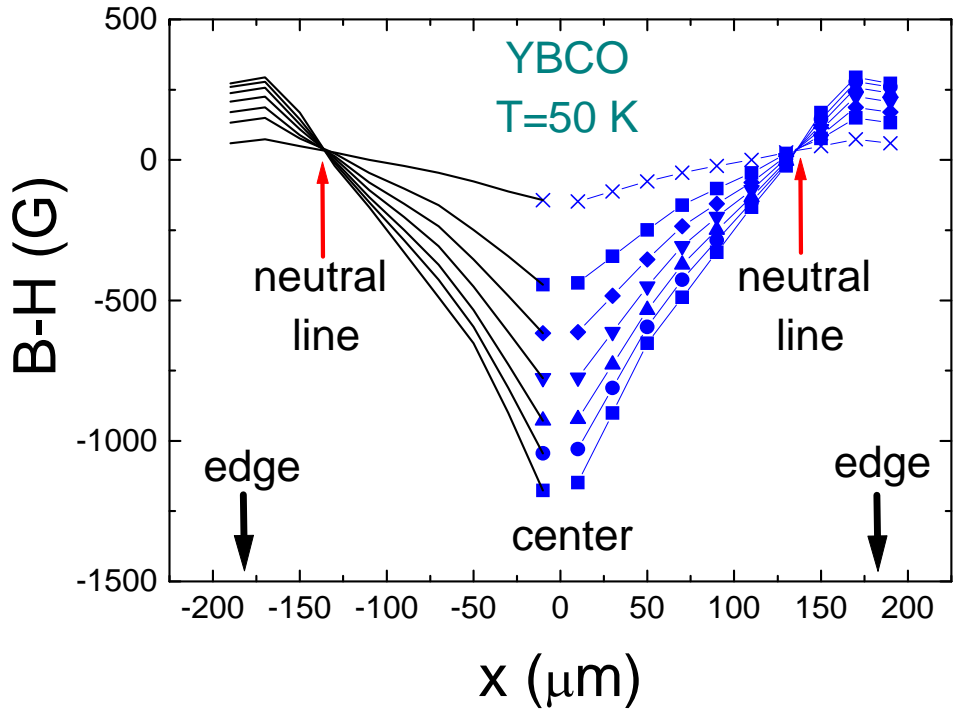

Figure 43. Typical profiles of the local magnetization (B-H) measured in a clean $YBa_2Cu_3O_{7-\delta}$ crystal at $T$=50 K. Different profiles correspond to different values of the external magnetic field.

The distinctive features of the finite geometry are the appearance of a so-called "neutral line" at $B=H$ (or $m=0$) in which all the profiles cross and regions where $B>H$ due to demagnetization.

The next step is to measure the time dependence of the profiles. This enables direct measurement of the local relaxation rate $\partial B_z / \partial t$ needed for estimation of the electric field from Eq.(67). Figure 44 presents local measurements of the magnetic induction relaxation at $H$=860 Oe. Note change in the sign of $\partial B_z / \partial t$ crossing the neutral line. Relaxation is fastest at the sample center and slows down approaching zero on the neutral line.



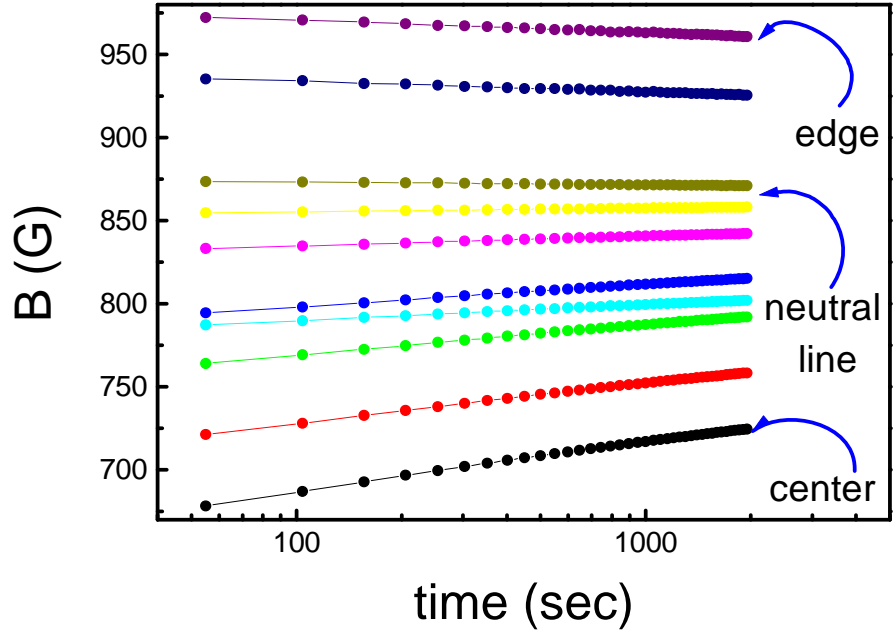

Figure 44. Measurements of the local magnetic relaxation in a $YBa_2Cu_3O_{7-\delta}$ crystal at H=860 Oe.

These measurements are sufficient to calculate the electric field in the sample. Integrating Eq.(67) across the sample we get:

$$E(x,t) = -\frac{1}{c}\int_0^x \frac{\partial B_z(\xi,t)}{\partial t} d\xi \qquad (70)$$

However, in order to build *E(j)* curves we also need to estimate current density, preferably from the same measurements. To achieve that, one may use a direct inversion of the Biot-Savart relationship between current density and magnetic field outside the sample:

$$B_z(x,z) = -\frac{2}{c}\int_{\eta=-d}^{d} d\eta \int_a^b \frac{j(\xi,\eta)(x-\xi)}{(x-\xi)^2+(z-\eta)^2} d\xi \qquad (71)$$



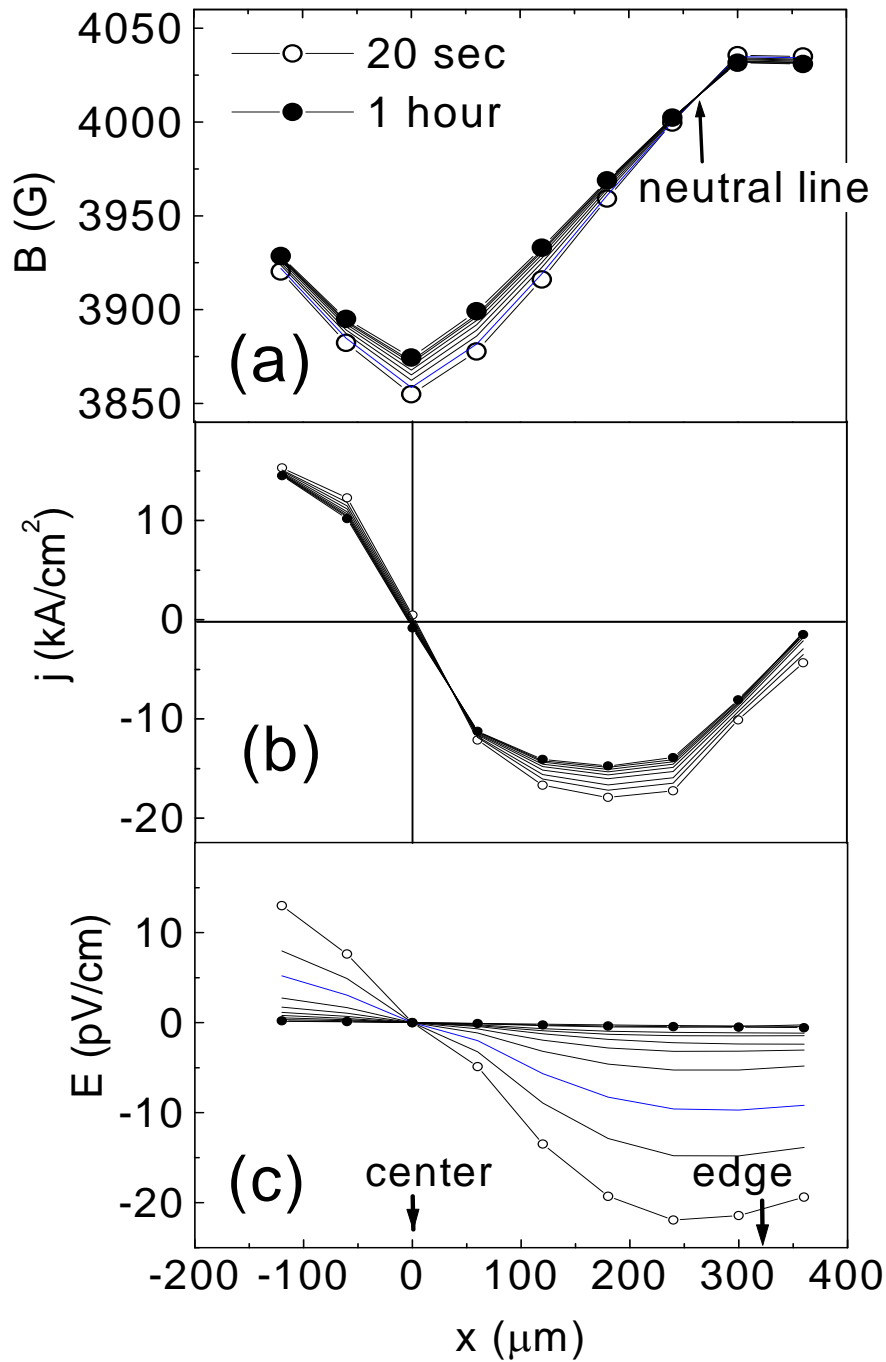

Figure 45. Standard data processing: Building *E(x,t)* and *j(x,t)* from measurements of $B_z(x,t)$ in a $YBa_2Cu_3O_{7-\delta}$ single crystal.

Imaging the sample split into *N* (number of measured points) slices, each bearing constant current density $j_k$, integral equation (71) can be written in a matrix form as



$$B_i = M_{ik} j_k \tag{72}$$

where matrix $M_{ik}$ is calculated analytically for a given sample geometry. Finally, Eq.(72) can be inverted and current density is given by

$$j_k = M_{ik}^{-1} B_i \tag{73}$$

In the more complicated case of 2D measurements of the magnetic induction or when $N$ is the large number, one can use more sophisticated inversion schemes [21, 165].

Figure 45 describes typical data processing for experimental determination of the $E(j)$ curves in superconductors. The uppermost frame presents raw data – measurements of the time dependence of the magnetic induction profile during an hour. This allows estimation of $\partial B_z / \partial t$ at each point. Next, using Eq.(73) we calculate current density distribution during the relaxation process and, finally, the last frame shows profiles of the electric field, calculated using Eq.(70). Note that electric field relaxes much faster than the current density and magnetic induction. Since $E = E_c \exp(-U/T)$ [3] and $U = T \ln(1+t/t_0)$ [166] we obtain time-hyperbolic decay: $E = E_c t_0 / (t_0 + t)$, where $t_0$ is the time for the formation of the critical state. Hence, $1/E$ should be directly proportional to time $t$. This is demonstrated in Figure 46, where we plot $1/E$ as a function of time for different probes.



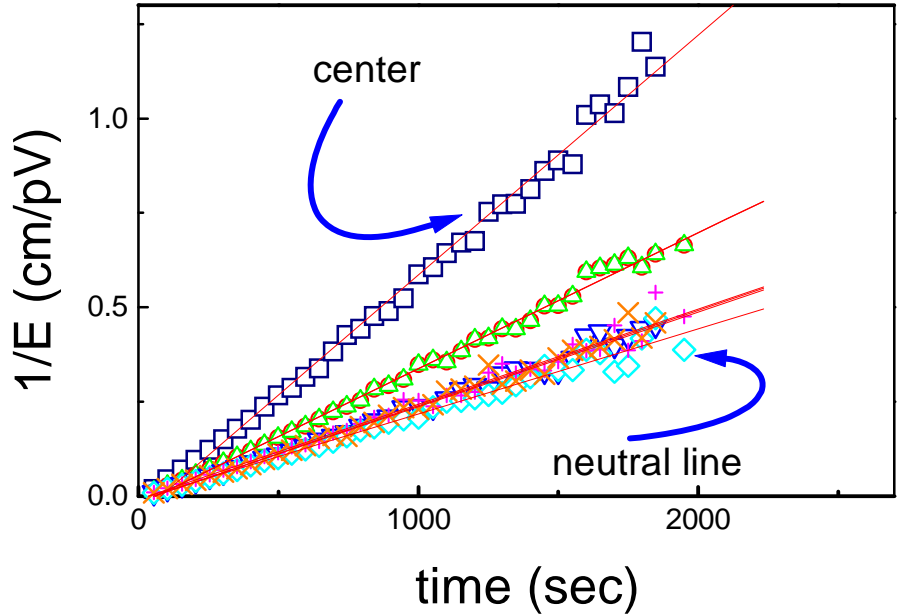

Figure 46. Hyperbolic time dependence of the electric field. Different lines are 1/$E$ at different locations inside the sample.

Having established the technique for the local measurements of the $E(j)$ characteristics we will apply it to the study the "fishtail" magnetization – a puzzling feature observed in clean high-$T_c$ crystals.

## F. $E(j)$ characteristics near the "fishtail" peak in the magnetization curves

As discussed in Appendix B, the width of the magnetization loop is proportional to the persistent (irreversible) current density. The same is true for local magnetization loops, where this width is proportional to the current density and depends on the position of the measured point inside the sample. Most of the existing theories of the critical state predict a decrease of the current density with the increase of the vortex lines density due to interaction between vortices. (The situation is still unclear due to difficulties in the statistical summation of random pinning forces [2, 3, 78, 137]). In high-$T_c$ superconductors the situation is further complicated by the giant



flux creep, since most of the experiments have a macroscopic time window and probe quite a relaxed state.

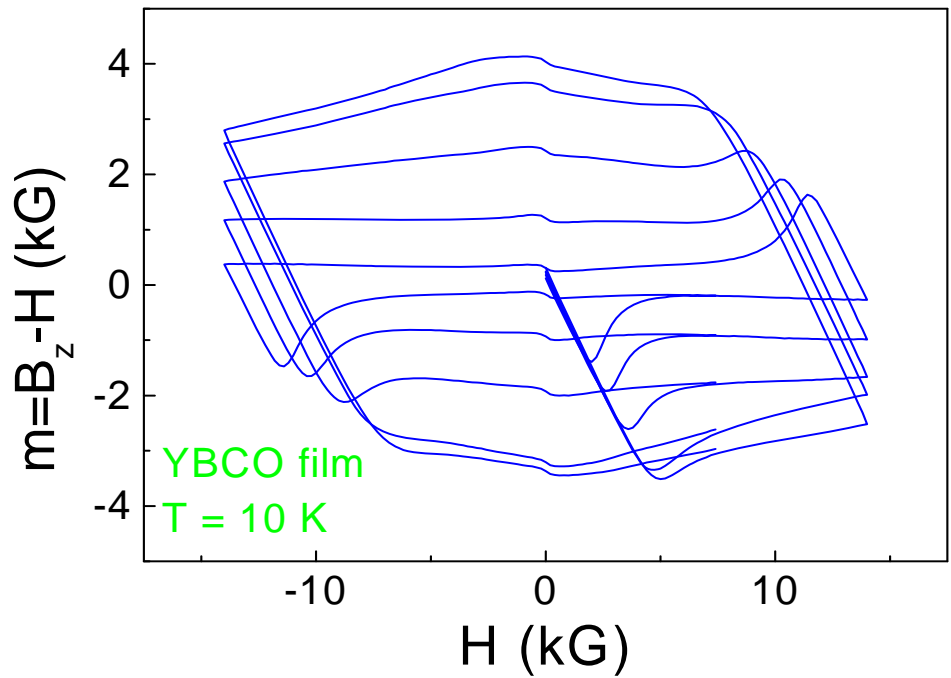

Figure 47. Conventional local magnetization loop in a $YBa_2Cu_3O_{7-\delta}$ thin film at T=10 K.

Magnetization loops observed in high-$T_c$ superconductors may be divided into two groups: 1) "Normal", when width of the loop decreases with increase of the magnetic field indicating a decrease of the persistent current density. Such loops are usually observed in samples which have many intrinsic defects, e.g., thin films and not optimized crystals. Figure 47 demonstrates a typical loop measured in thin $YBa_2Cu_3O_{7-\delta}$ film. 2). "Anomalous" loops, when the width of the magnetization loop first increases up to a field $B_p$ and decreases above this field. Such behavior is sometimes called a "fishtail" magnetization. The example of such behavior is shown in Figure 48.



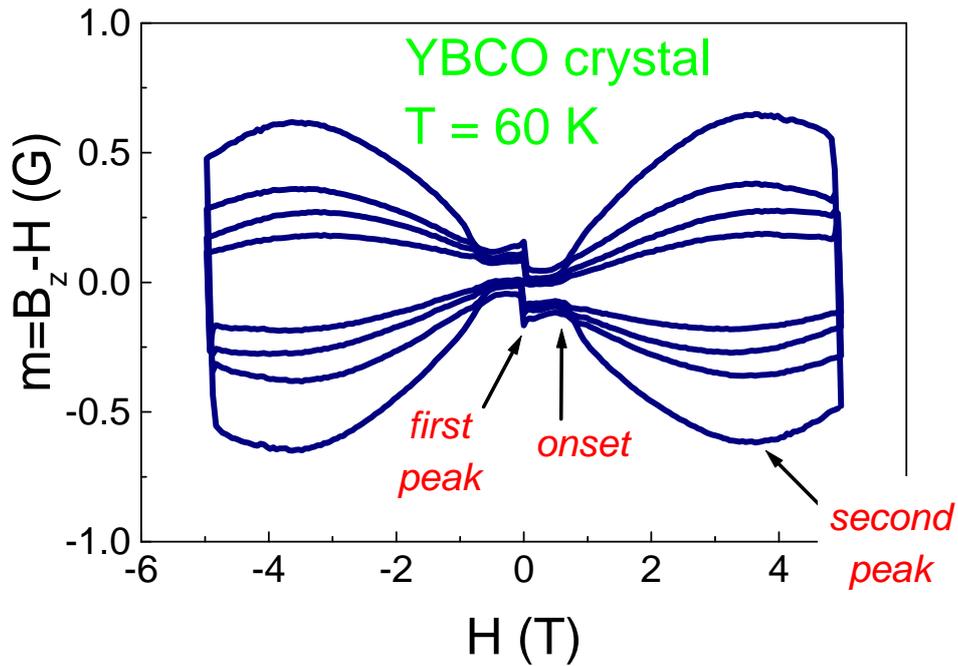

Figure 48. "Fishtail" magnetization in a clean $YBa_2Cu_3O_{7-\delta}$ crystal at T=60 K.

We will put aside the static explanations of the "fishtail" attempting to attribute this anomaly to the similar behavior of the critical current density, due to great deal of uncertainty in those theories [167, 168]. Instead we will concentrate on the dynamic explanation of this feature, which naturally takes into account magnetic flux creep. The most straightforward theory of the fishtail is that based on the collective creep model (see for review [2]). The increase of the correlated volume (flux bundle) results in a monotonous decrease of the critical current density with the increase of the magnetic field and an increase of the barrier for magnetic relaxation.



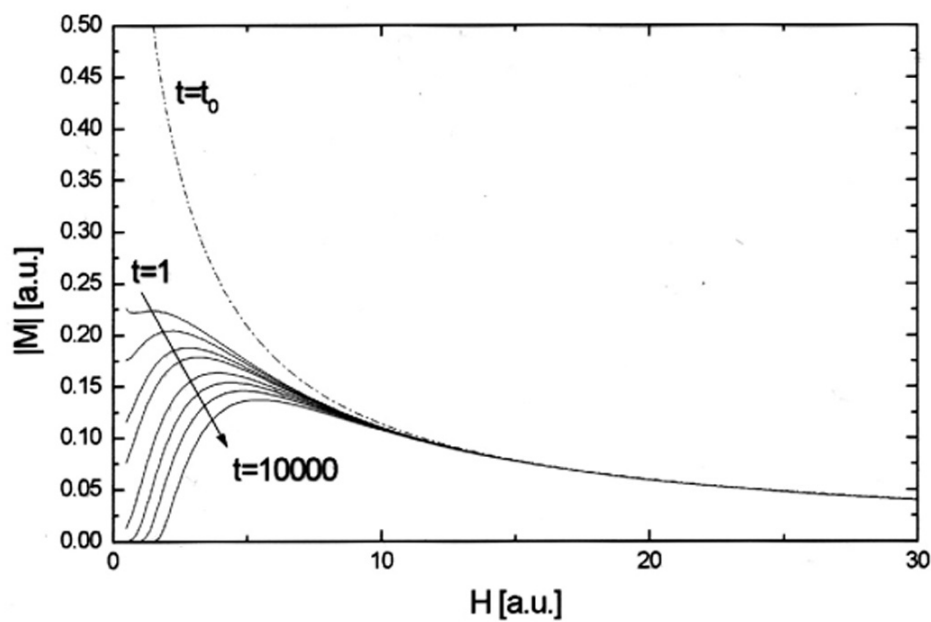

Figure 49. Dynamic development of the "fishtail" in a framework of the collective creep model. Dashed line represents the field dependence of the critical current.

This causes very fast relaxation at low fields and, as a result, non-monotonous field dependence of the persistent current density. This process can be studied numerically using a differential equation of the flux creep: $\partial B / \partial t = -\nabla(v \times B)$ [169]. Critical state is formed at the time $t_0$ and relaxes towards equilibrium homogeneous distribution. As demonstrated in Figure 49, the essential feature of this scenario is that on the increasing ("anomalous") part of the magnetization loop ($B<B_p$) current density must be very far from its critical value $j_c$, whereas on the decreasing part ($B>B_p$) it should closely follow $j_c$. Another very important feature is apparent in Figure 49, namely, a peak in magnetization shifts with time to *larger* fields. Measurements conducted on crystals exhibiting the "fishtail" (see Figure 50) have shown that magnetic relaxation is essential at all values of magnetic field and, in particular, at $B>B_p$ current density



changes up to 80% during measurements. Also, magnetization peak shifts to *lower* fields, contrary to the collective pinning scenario.

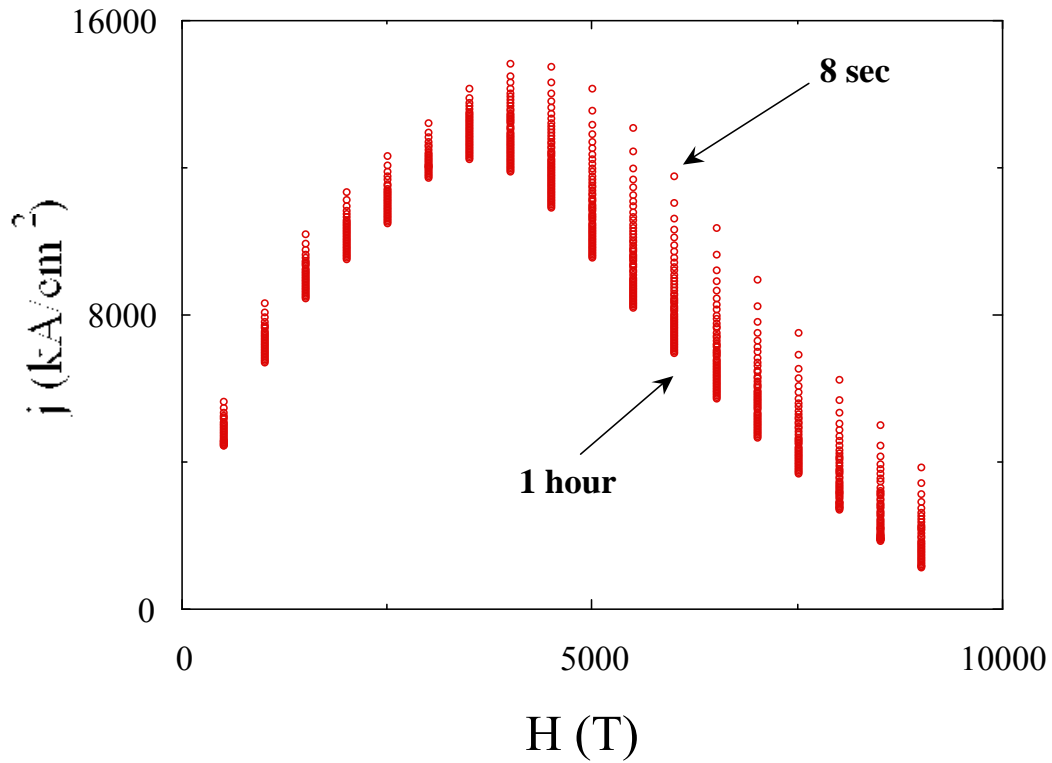

Figure 50. Magnetic relaxation at different values of the external field.

Such behavior indicates that another mechanism, apart from that of collective creep, of flux creep is involved in the process. At low enough current densities the gradient of the magnetic induction (or density of flux lines) is sustained mainly by the elastic deformations of the vortex lattice. This means that displacements of vortices out of the equilibrium positions are small.



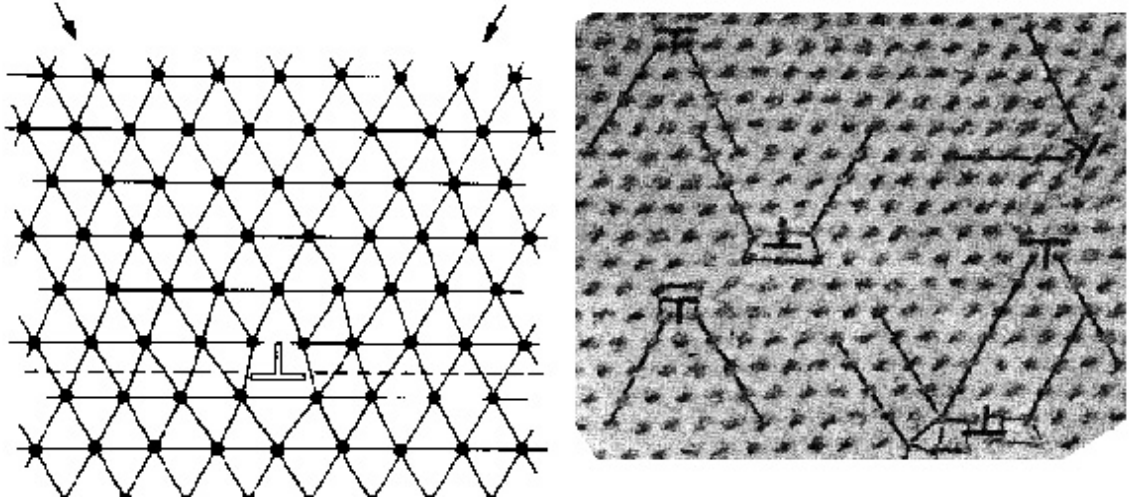

Figure 51. Dislocations in a vortex lattice (from the book of H. Ulmaier [137])

At larger currents, however, the elastic deformations are not sufficient and usually the vortex lattice has many imperfections, such as dislocations and other topological defects. The existence of such defects has been directly proved using the magnetic decoration technique, which is demonstrated in Figure 51. Therefore, there are usually several possibilities for the release of extra energy stored in a deformed (due to macroscopic vortex density gradient) vortex lattice. The first is via elastic accommodations, when the vortex jumps are not accompanied by a change of the local lattice structure. We will call it *elastic* creep. Another possibility is the *plastic* creep, which is realized via the motion of dislocations, jumps of interstitial vortices or (at large currents) channeling of vortices along the easy channels. The barrier for elastic creep is shown to increase with the increase of the magnetic field [2], whereas the plastic barrier (at least for single interstitial vortices and dislocations) decreases with an increase of magnetic field. This is because it is proportional to the inter-vortex distance $a_0 \approx \sqrt{\Phi_0 / B}$. At very large fields, however, it may start to increase due to interaction between dislocations.



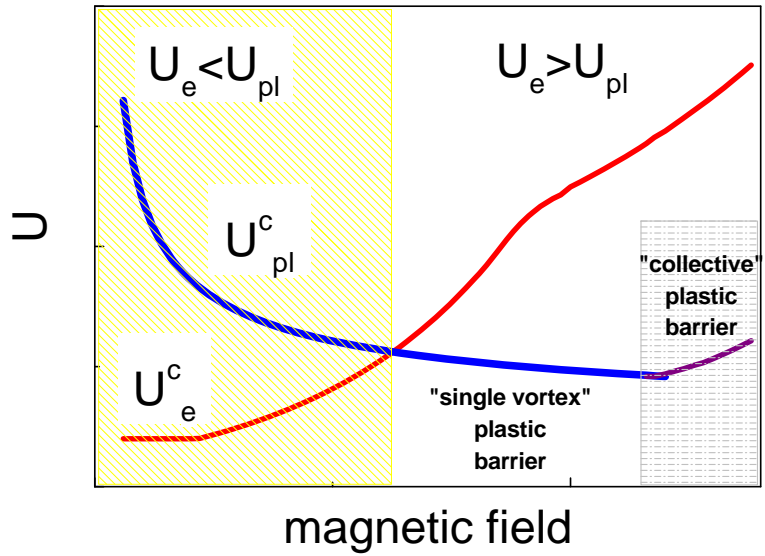

Figure 52. Schematic field dependence of the elastic $U_e$ and plastic $U_{pl}$ barriers for flux creep

Thus, at low fields the elastic barrier is lower than the plastic barrier and the collective creep model describes magnetic relaxation. However, at larger magnetic field plastic creep becomes a more favorable channel for magnetic relaxation. Importantly, the relaxation rate in this case increases with an increase of the magnetic field.

In order to determine which particular mechanism takes place before and after the peak we have measured local magnetic relaxation at different values of the external magnetic field and used these measurements to determine the $E(j)$ characteristics, as described in the previous Section.



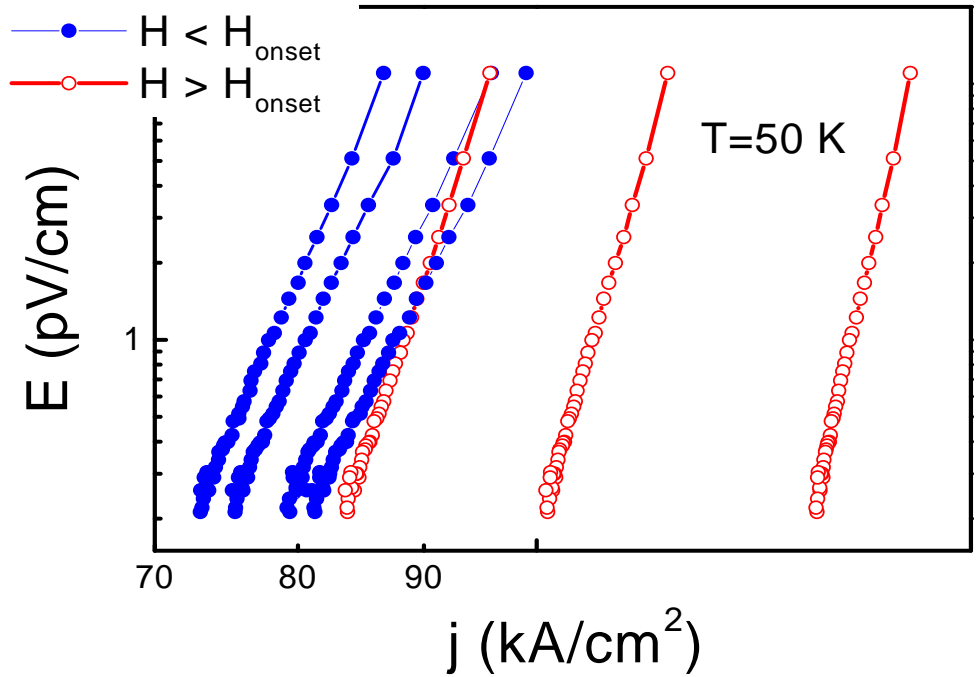

Figure 53. *E(j)* characteristics in the vicinity of the onset of the anomalous increase of the magnetization in $YBa_2Cu_3O_{7-\delta}$ single crystal at T=50 K.

As shown in Figure 53, *E(j)* characteristics look almost linear in a log-log scale, reflecting a high degree of non-linearity. Thus, it is convenient to characterize each curve by a single parameter $n \equiv d\ln(E)/d\ln(j)$ which, as shown below, is related to the physical quantities of the flux creep. Figure 53 exhibits an abrupt change in *n* before and after the onset. Similarly, we observe a dramatic change in *n* in the vicinity of the magnetization peak, Figure 54. Similar features are observed in other high-$T_c$ systems exhibiting the "fishtail".



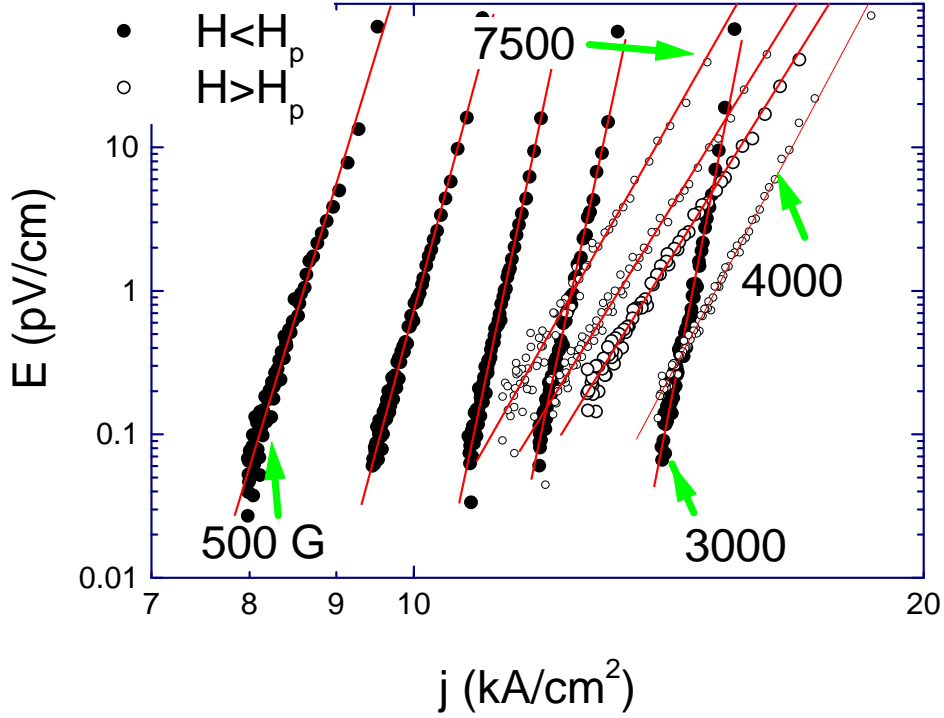

Figure 54. *E(j)* characteristics before and after the anomalous magnetization peak in *YBa$_2$Cu$_3$O$_{7-\delta}$* single crystal at T=50 *K*.

Since $E = E_c \exp(-U(j)/T)$ the logarithmic slope *n* of *E(j)* curve gives:

$$n \equiv \frac{\partial \ln(E)}{\partial \ln(j)} = -\frac{j}{T}\frac{\partial U(j)}{\partial j} \quad (74)$$

In the framework of the collective creep model $U(j) = U_c (j_c/j)^\mu$ and

$$n = \frac{\mu}{T} U(j) \approx \mu \ln(t/t_0) \quad (75)$$

where we have used a logarithmic solution of the flux creep problem [166]. Thus, in this model, field dependence of the parameter *n* is determined by the field dependence of the "collective creep" exponent μ [2]. The last is predicted to be non-monotonous changing from 1/7 of the single-vortex flux creep through 5/2 of the small bundles to 7/9 of the large bundles or even ½ of the CDW creep regime. At the same time, current density must increase monotonously with B as a result of relaxation.



In the framework of the dislocation mediated plastic creep the exact form of $U(j)$ is not authentically known, although recent calculations predict $U(j) = U_c \left(1 - (j/j_c)^\alpha\right)$, where $\alpha \approx 1/2$ (although, values larger than 1 may explain the observed experimental peculiarities, see below) and $U_c \propto \varepsilon \varepsilon_0 a_0 \propto 1/\sqrt{B}$ where $\varepsilon$ is the anisotropy and $\varepsilon_0 = (\Phi_0 / 4\pi\lambda)^2$ is the vortex energy. Differentiating $U$ with respect to $j$ we obtain:

$$n = \frac{\alpha}{T}(U_c - U(j)) \approx \frac{\alpha}{T} U_c - \alpha \ln(t/t_0) \qquad (76)$$

so that the field dependence of $n$ is due to the field dependence of $U_c$.

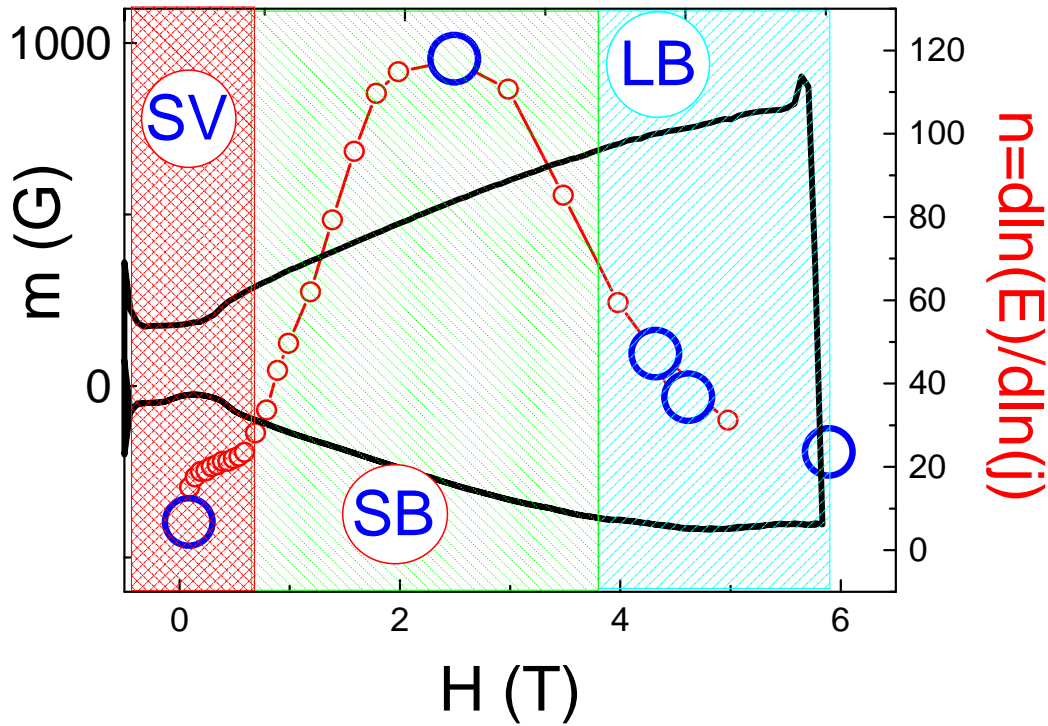

Figure 55. The parameter *n* (see Eq.(75)) *vs.* magnetic field plotted along with the magnetization loop (bold solid line) for a *YBa$_2$Cu$_3$O$_{7-\delta}$* crystal at *T*=50 *K*. Small open symbols represent measured *n* values, whereas large open symbols represent values obtained from the collective creep theory.



Figure 55 exhibits parameter *n* (small connected open circles) plotted along with the magnetization loop for $YBa_2Cu_3O_{7-\delta}$ single crystal at *T*=50 *K*. Persistent current density (proportional to the width of the magnetization loop) increases monotonously, whereas *n* varies non-monotonously. Large open symbols represent values calculated from the collective creep theory. The whole picture is consistent with the collective creep [2].

At larger magnetic field, beyond the peak in magnetization *vs.* H curve, the collective creep theory may not be applied, since persistent current in quite far from its critical value (see Figure 50) and decreases with field.

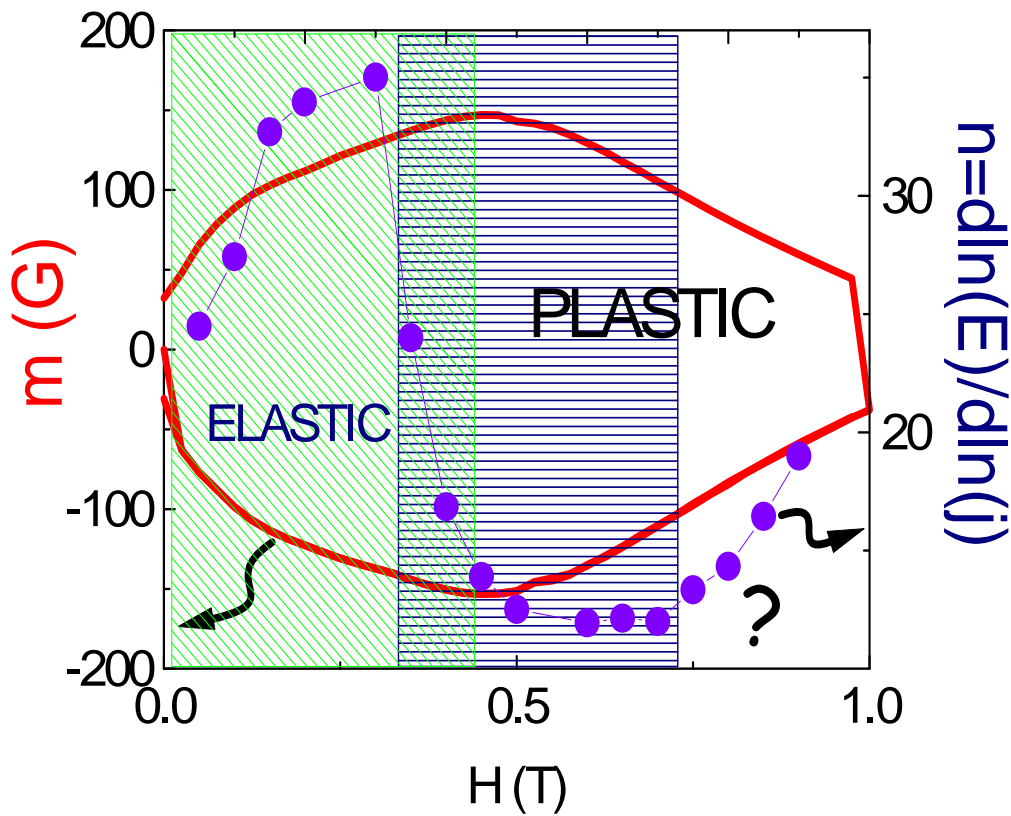

Figure 56. The parameter *n* (solid symbols) *vs.* H plotted along with the magnetization loop (solid bold line) for a $YBa_2Cu_3O_{7-\delta}$ crystal at *T*=85 *K*.



Figure 56 shows the parameter *n* (filled circles) plotted *vs.* magnetic field along with the "fishtail" magnetization loop (bold solid line). Beyond the peak *n* exhibits non-monotonous behavior. The initial decrease is consistent with the plastic creep, Eq.(76). The upturn at large fields may imply a "collective plastic" creep mechanism, namely motion of the interacting dislocations or large topological defects. Another, more trivial explanation of the upturn of *n* at large fields is that the logarithmic term in Eq.(76) decreases with magnetic field due to a decrease of the derivative $\partial U / \partial j$ of the barrier for plastic creep at $j \to 0$. Since $\ln(t/t_0) \propto \ln(|\partial U / \partial j|) \propto \ln(j^{\alpha-1})$ the logarithmic term decreases if $\alpha > 1$, which contradicts to a value of ½ predicted by a simple theory of dislocation motion.

In either case the creep mechanism beyond the second magnetization peak is not elastic.

To determine whether the scenario described is unique to the *YBa$_2$Cu$_3$O$_{7-\delta}$* system or is more general we have performed the same set of measurements on a *Nd$_{1.85}$Ce$_{0.15}$CuO$_{4-\delta}$* (NCCO) single crystal. The shape and the field range of the fishtail in this system is quite different from that in *YBa$_2$Cu$_3$O$_{7-\delta}$*. The NCCO crystal ($T_c \approx 23$ K) exhibits a fishtail with a sharp onset at relatively low fields ($H < 500$ *G*), whereas the YBCO crystal ($T_c \approx 91$ *K*) exhibits a broad peak at fields of order 5000 *G*. The magnetization loops in NCCO are similar to those observed in *Bi$_2$Sr$_2$CaCu$_2$O$_{8+\delta}$* [170, 171].



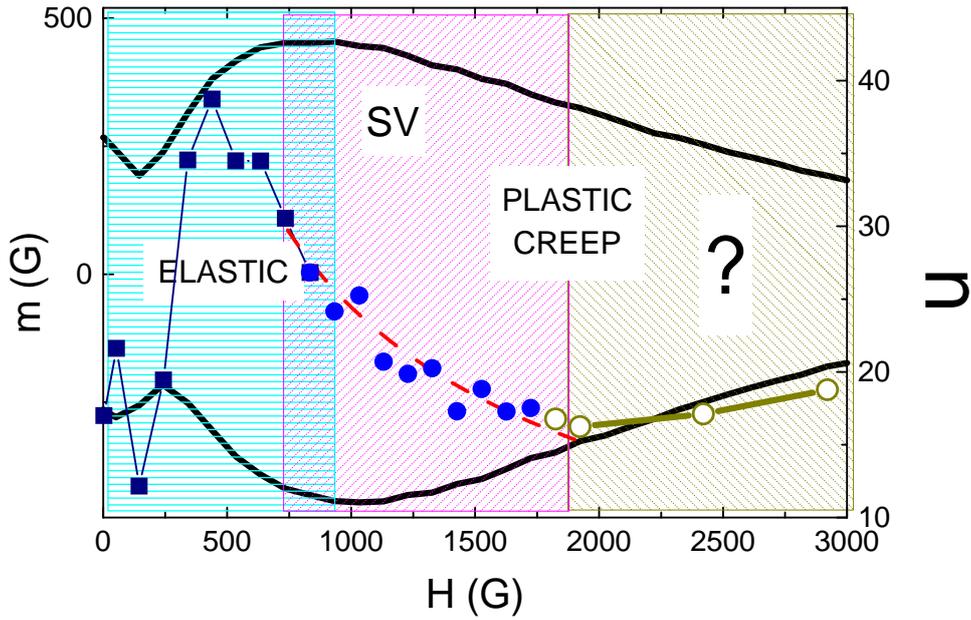

Figure 57. The parameter *n* (solid symbols) *vs.* H plotted along with the magnetization loop (bold solid line) for $Nd_{1.85}Ce_{0.15}CuO_{4-\delta}$ single crystal at *T*=13 K. The dashed line is a fit to $1/\sqrt{B}$ dependence.

Figure 57 exhibits the parameter *n* plotted along with the magnetization loop (bold solid line) at *T*=13 K for a $Nd_{1.85}Ce_{0.15}CuO_{4-\delta}$ single crystal. Due to the lower fields there is an extended range of "non-interacting dislocations" plastic creep. This region is shown by large solid symbols. The solid line is fit to the expected $1/\sqrt{B}$ dependence. At larger fields there is again a region marked by a "?" of a possible "collective plastic" creep or saturating barrier *U*, as discussed above.

## G. *Summary and conclusions*

In conclusion, local dynamics measurements provide a very powerful tool for the direct study of *E-j* curves in type-II superconductors in a regime of flux creep. The dynamic picture observed in $YBa_2Cu_3O_{7-\delta}$ and $Nd_{1.85}Ce_{0.15}CuO_{4-\delta}$ single crystals is consistent with the predictions of the collective creep theory for fields below the field of the "fishtail" peak. The non-trivial prediction of the non-monotonous behavior of



the "collective creep exponent" $\mu$ is experimentally confirmed. At large magnetic field, beyond the peak in magnetization, plastic creep is the dominant mechanism of magnetic relaxation. This region is divided into "single vortex" plastic creep and "collective" plastic creep. In the latter, the barrier for magnetic relaxation increases due to the plastic motion of interacting dislocations or larger topological defects.



# Chapter VI. Self-organization during flux creep

It was shown in previous sections and in Appendix B that the Bean notion of the critical state [134, 136, 144, 164] describes fairly well the magnetic behavior of type-II superconductors, including high-temperature superconductors. Geometrical patterns of the flux distribution in type-II superconductors are, in many aspects, similar to a sand hill formed after some time when, for example, sand is poured on a platelet stage [78, 172-174]. The study of dynamics of such strongly correlated many-particle systems has resulted in a new concept called self-organized criticality (SOC), proposed by Bak and Tang [173-176]. Tang [173, 174] has analyzed a direct application of the SOC idea to type-II superconductors. Numerous studies have later elaborated significantly on this topic [177-182]. However, in real experiments with high-$T_c$ superconductors, persistent current density is usually much lower than the critical one due to flux creep, whereas the concept of SOC is strictly applied only to the critical state [180-182]. Direct application of the notion of self-organized criticality to the problem of flux creep meets a number of serious difficulties. It is clear that simple power laws for noise power spectra, which work in the vicinity of the critical state, must be changed on later stages of relaxation due to the highly non-linear response of a superconductor. Not relating to SOC, the universal behavior of the electric field during flux creep has been analyzed by Gurevich and Brandt [153, 183]. The problem of flux creep has also been analyzed from the point of view of a classical SOC [180-182]; it was found that direct modifications of the relaxation law due to avalanches is minor and can be hardly reliably distinguished from the experimental data. Nevertheless, neither of the existing models give a clear physical description of self-organization in type-II superconductors during flux creep.



However, in this Section we attempt to outline a general picture and show how self-organization fits into this picture.

## A. *Numerical solution of the flux creep equation*

Throughout the following we assume the infinite in a *z*-direction slab (see Figure 1, Appendix B, in the limit $d \to \infty$), so that magnetic field has only one component – $B_z$. As a mathematical tool for our analysis we use a well-known differential equation for flux creep in its simplest form suitable for a creep regime

$$\frac{\partial B}{\partial t} = -\frac{\partial}{\partial x}\left(Bv_0 \exp(-U(B,T,j)/T)\right) \qquad (77)$$

Here *B* is the magnetic induction, $v = v_0 \exp(-U(B,T,j)/T)$ is the mean velocity of vortices in the *x* direction and *U* is the effective barrier for magnetic flux creep. It is important to emphasize that we do not attempt to modify the pre-exponent factor $Bv_0$, as suggested in previous works on SOC (see e. g. [173, 174]). Such modifications result only in a logarithmic correction to the "effective" activation energy and may be omitted in a flux creep regime [180-182]. Noting that in our geometry $4\pi M = \int_V (B-H)dV$, we get for the volume magnetization *m*=*M*/*V* from Eq.(77)

$$\frac{\partial m}{\partial t} = -A\exp(-U/T) \qquad (78)$$

where $A = Hv_0/4\pi w$.

The numeric and analytic solutions of Eq.(78) in different regimes are beyond the scope of this thesis. The interested reader is referred to our paper [169].



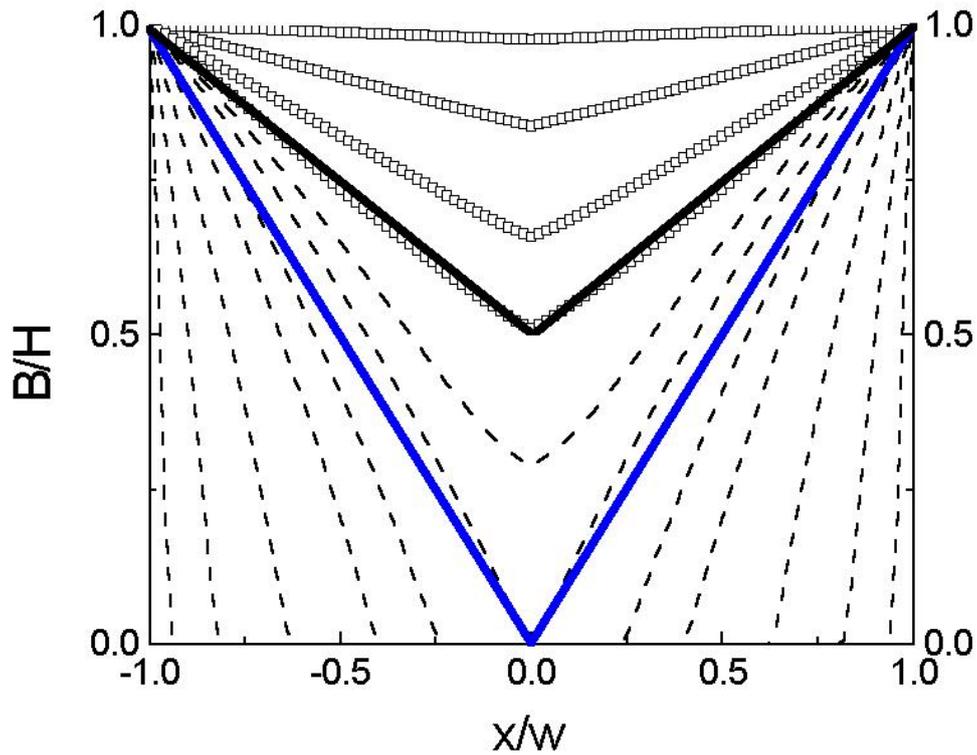

Figure 58  Magnetic induction profiles during flux creep as calculated from Eq.(77).

Here we summarize the main results of [169]. Figure 58 shows distribution of magnetic induction $B(x,t)$ at different times during flux flow and flux creep. In this simulation, the sample was assumed to be zero-field cooled (so at $t=0$ $B=0$) and magnetic field is instantaneously switched on at $t=0$. The barrier $U \propto B^\alpha j^{-\beta}$ describes the collective creep or vortex glass [2, 3]. Figure 59 demonstrates the corresponding evolution of the energy barrier $U(x,t)$ calculated from Eq.(77). The important point to note is that the energy barrier $U(x)$ is nearly $x$ independent, so that its maximal variation $\Delta U$ is of order of $T$ (the upturn in the sample center is an artifact of $j \to 0$ there). This means that the vortex system organizes itself in such a way to maintain a uniform distribution of the barrier $U$.



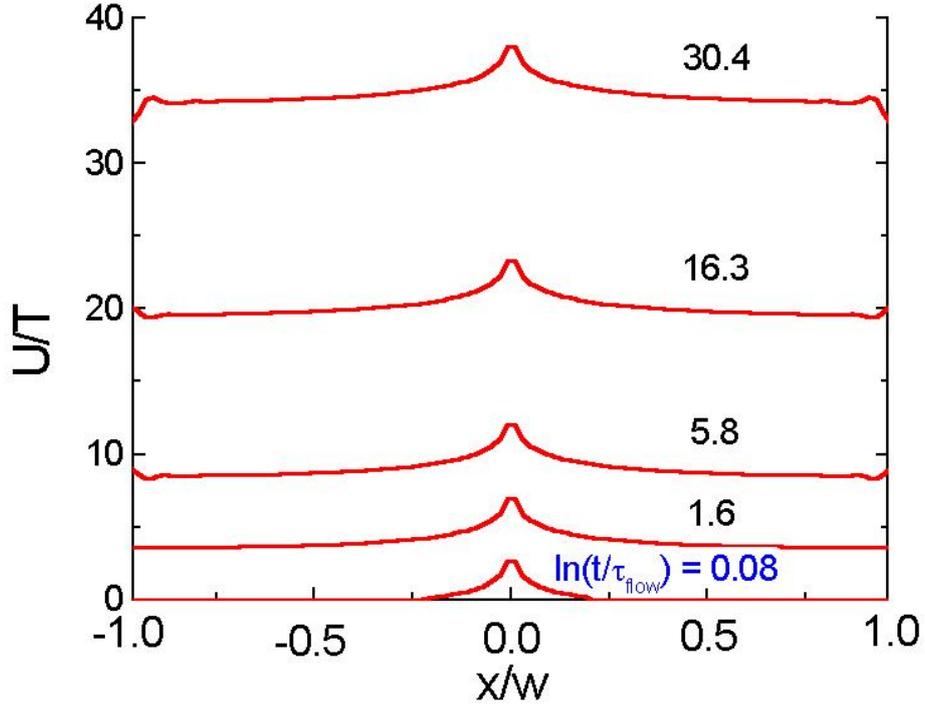

Figure 59. Profiles of the energy barrier for flux creep as calculated from Eq.(77)

## B. *Vortex avalanches and magnetic noise spectra*

We introduce the avalanches in an "integral" way. An avalanche of size *s* causes a change in the total magnetic moment $\delta M \sim s$. This change in the magnetic moment is equivalent in our simple geometry to a change of the current density $\delta j \sim \delta M$. So we may write $\delta j = \gamma s$, where $\gamma$ is a geometrical coefficient determined from the equations given in Appendix B. In particular, for a long slab we get from Eq. (103) $\gamma = 2c/wV$. If the barrier for flux creep is $U(j)$, then variation of current $\delta j$ leads to a variation of the energy barrier

$$\delta U = \left|\frac{\partial U}{\partial j}\right|\delta j = \gamma \left|\frac{\partial U}{\partial j}\right| s \qquad (79)$$

As mentioned above, the largest fluctuation in the energy barrier $\delta U$ is of order of *T* in the creep regime ($\delta U \ll U$), thus the maximal avalanche is:



$$s_m = \frac{T}{\gamma \left|\frac{\partial U}{\partial j}\right|} \qquad (80)$$

The main idea is that in the vicinity of $j_c$ the system of vortices is, indeed, in the self-organized *critical* state, as initially proposed by Tang [173, 174]. During flux creep, it maintains itself in a self-organized state, however *not in the critical* state. The self-organization manifests itself in maintaining almost uniform $U$. Avalanches do not vanish, but there is a constraint on the largest possible avalanche, see Eq. (80). Importantly, $s_m$ depends upon current density and, as we show below, decreases with decrease of current (or with increase of time), so the significance of avalanches vanishes. It is worth noting that Eq. (80) also gives the correct dependence of $s_m$ at criticality on the system size. It is clear that in a finite system the largest possible avalanche must be proportional to the system volume, in agreement with our derivation (we recall that $\gamma = 2c/wV$).

Now we shall derive time dependence of $s_m$. To do that, we shall employ a very useful phenomenological form of the barrier for flux creep, introduced by Griessen [184].

$$U(j) = \frac{U_0}{\alpha}\left(\left(\frac{j_c}{j}\right)^\alpha - 1\right) \qquad (81)$$

Using this formula one may satisfactorily describe all known functional forms of $U(j)$ letting the exponent $\alpha$ attain both negative and positive values. Therefore, $\alpha=-1$ in Eq.(81) describes the Anderson-Kim-like barrier [136, 164]; for $\alpha =-1/2$ the barrier for the plastic creep [185] is obtained; positive $\alpha$ describes collective creep barriers [2]; a logarithmic barrier [186] is obtained in the limit $\alpha=0$. Interestingly, the activation energy written in the form of Eq.(81) results in the famous "interpolation



formula" for flux creep [2] if one applies the logarithmic solution of the creep equation $U(j)=T\ln(t/t_0)$ [166]

$$j(t)=j_c\left(1+\frac{\alpha T}{U_0}\ln\left(\frac{t}{t_0}\right)\right)^{-\frac{1}{\alpha}} \qquad (82)$$

Using this general form of the current dependence of the activation barrier on the current density we obtain from Eq.(80)

$$s_m(j)=\frac{Tj}{\gamma U_0}\left(\frac{j}{j_c}\right)^{\alpha} \qquad (83)$$

and for time dependence

$$s_m(t)=\frac{Tj_c}{\gamma U_0}\left(1+\frac{\alpha T}{U_0}\ln\left(\frac{t}{t_0}\right)\right)^{-\left(1+\frac{1}{\alpha}\right)} \qquad (84)$$

As we see, the upper limit for the avalanche size decreases with a decrease of current density and with an increase of time for all $\alpha>-1$. For $\alpha<-1$ the derivative

$$\frac{\partial^2 U}{\partial j^2}=\frac{(1+\alpha)}{j^2}U_0\left(\frac{j_c}{j}\right)^{\alpha} \qquad (85)$$

This curvature is negative and the largest avalanche does not change with the current (because of the second constraint, that $s_m$ is limited by a system dimension). In this case, a self-organized *critical* state should persist down to very low currents. In practice, however, most of the observed cases obey $\alpha>-1$ form of the barrier and $s_m$ decreases with a decrease of current density (due to flux creep). In the following we estimate the power spectrum density of the magnetic noise in a flux creep regime.

Before starting with calculation of the spectral density of the magnetic flux noise due to flux avalanches, let us stress that the time dependence of $s_m$ is very weak (logarithmic), see Eq.(84). This allows us to treat the process of the flux creep as



quasi-stationary, which means that during the time required for the experimental sampling of the power spectrum, the current density is assumed to remain constant. In more sophisticated experiments [180] one may sweep the external field with a constant rate ensuring that the current density does not change, although it is less than the critical one. In fact, a constant sweep rate fixes a certain time $t/t_0 \sim (\partial H/\partial t)^{-1}$. Thus, decreasing the sweep rate allows the study of the noise spectra on effectively later stages of the relaxation.

Let us now discuss the avalanche size and lifetime distributions and derive the spectral density of the magnetic noise resulting from the avalanches. First we relate the avalanche of size *s* to its lifetime $\tau$ using Eq.(78):

$$\frac{s}{\tau} \approx \left|\frac{\partial M}{\partial t}\right| = Ae^{-U/T} \tag{86}$$

Therefore, $\tau \propto s\exp(U/T)$ and $\tau/\tau_m = s/s_m$. Using simplified distribution of lifetimes derived from computer simulations by Pan and Doniach [181]:

$$\rho(\tau) \propto \exp(-\tau/\tau_m) \tag{87}$$

and we obtain:

$$\rho(s) \propto \exp(-s/s_m) = \exp(-\delta U/T) \tag{88}$$

where we used Eq. (79) and Eq. (80) to obtain $s = s_m \delta U/T$, where $\delta U$ is the fluctuation in *U* related to the avalanche *s*.

To estimate the power spectrum of the flux noise, we assume that the lifetime of each avalanche has same meaning as the waiting time of each avalanche to occur. Thus, avalanches of each size *s* and $\tau$ contribute a Lorentzian spectrum

$$L(\omega,\tau) \propto \frac{\tau}{1+(\omega\tau)^2} \tag{89}$$



Therefore, the total power spectrum of magnetic noise during flux creep is given by

$$S(\omega) = \int_0^\infty \rho(\tau) L(\omega,\tau) d\tau \tag{90}$$

Using Eq. (87) we find

$$S(\omega) \propto \frac{1}{2p^2}\left[\exp\left(\frac{i}{p}\right)Ei\left(\frac{i}{p}\right) + \exp\left(-\frac{i}{p}\right)Ei\left(-\frac{i}{p}\right)\right] \tag{91}$$

or

$$S(\omega) \propto \frac{1}{2p^2}\left[\cos\left(\frac{1}{p}\right)\mathrm{Re}\left(Ei\left(\frac{i}{p}\right)\right) - \sin\left(\frac{1}{p}\right)\mathrm{Im}\left(Ei\left(\frac{i}{p}\right)\right)\right] \tag{92}$$

Here $\rho \equiv \omega \tau_m$ and $Ei(x) = \int_0^\infty \exp(-\eta)/\eta \, d\eta$ is the exponential integral. The spectral density $S(\omega)$ described by Eq.(92) is plotted in Figure 60 (symbols). There is an upper cutoff for the avalanche lifetime - $\tau_m$, hence the lowest frequency is $2\pi/\tau_m$. Thus, only frequency domain $2\pi/\tau_m < \omega$ ($p>1$) is important. In the limit of large $p$, spectral density Eq.(92) has a simple asymptotic

$$S(\omega) \propto \frac{\ln(p) - \gamma}{p^2} \tag{93}$$

where $\gamma \approx 0.5772156649$ is the Euler's constant. This simplified power spectra is shown in Figure 60 by a solid curve.



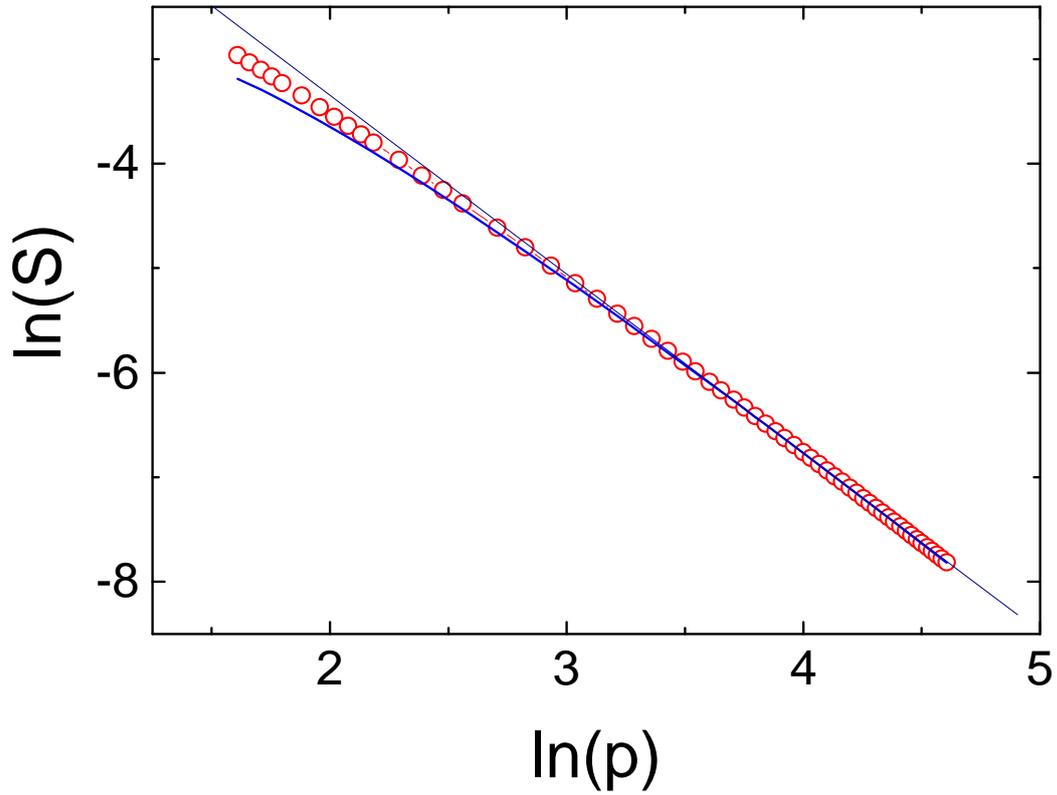

Figure 60. Power spectra calculated using Eq.(92) (symbols) and Eq.(93) (solid curve). The solid straight line is a fit to power law with $\nu \approx 1.71$

For all $p>2$ this approximation is quite reasonable. The usual way to analyze the power spectrum is to present it in a form $S(\omega) \propto 1/\omega^\nu$ and extract the exponent $\nu$ simply as $\nu = -\partial \ln(S)/\partial \ln(\omega)$. In our case the parameter $p=\omega/\tau_m$ is a reduced frequency, so we may formally estimate the exponent $\nu$ as

$$\nu = -\frac{\partial \ln(S)}{\partial \ln(p)} = 2 - \frac{1}{\ln(p) - \gamma} \tag{94}$$

This result is very significant, since it fits quite well the experimentally observed values of $\nu$ which was found to vary between 1 and 2 [180, 187]. (For example, the solid straight line in Figure 60 is a fit with $\nu \approx 1.71$). As seen from Figure 60 it is impossible to distinguish between real $1/\omega^\nu$ dependence and that predicted by Eq.(93)



at large enough frequencies. Remarkably, in many experiments the power spectrum was found to deviate significantly from the $1/\omega^\nu$ behavior, which nevertheless fits very well to Eq.(92).

Using Eq.(92) one can find the temperature, magnetic field and time dependence of the power spectrum, substituting $p = \omega \tau_m(H,T,t) = \omega s_m(H,T,t)\exp(U(H,T,t)/T)$ derived in the previous section. We only emphasize that the power spectrum depends on time. Since the parameter $p$ decreases with the increase of time, the exponent $\nu$ becomes closer to 1 during flux creep At these later stages of relaxation the effect of the avalanches is negligible. Therefore, one of the manifestations of avalanche driven dynamics during flux creep is $1/f^\nu$ noise spectra with $\nu>1$ and changes when sampled at different times during relaxation. This explains very elegant experimental results obtained by Field et al. [180] who measured directly vortex avalanches at different sweep rates. They found that the exponent $\nu$ decreases from $\nu=2$ at a large sweep rate of 20 *G/sec* to $\nu =1.5$ for a sweep rate of 1 *G/sec*. This is in good agreement with our derivations.

### C. *Summary and conclusions*

In conclusion, measurements of the magnetic noise spectra provide new information on the mechanism of flux creep in superconductors. Effects of self-organization highlight themselves in the almost constant across the sample barrier for flux creep. The power spectrum is predicted to be time-dependent due to time dependence of the upper cut-off of the maximal possible avalanche. On the other hand, a superconductor is a perfect model non-linear system to study the effects of self-organization well *below* criticality.



# Chapter VII. Summary and conclusions

The theoretical and experimental studies described in this thesis provide insight into static and dynamic irreversible magnetic properties of high-$T_c$ superconductors. The main results of these studies are summarized below:

The critical current $j$ and magnetic relaxation rate in thin $Y_1Ba_2Cu_3O_{7-\delta}$ films depend strongly on the thickness of the films. Pinning on surfaces perpendicular to the vortices plays an important role, however it cannot explain all of the experimental observation, especially the faster relaxation in the thinner films. Our theoretical analysis predicts inhomogeneous current density distribution throughout the film thickness, even in the absence of surface pinning. The current density is always larger near the surface, and decays over a characteristic length scale, which is between $\tilde{\lambda}$ (of order of the inter-vortex distance $a_0$) and the Campbell length $\lambda_C$. We identify this inhomogeneity in the current density as the origin of the measured decrease in the persistent current and magnetic relaxation rate with the increase of film thickness.

In experiments with $Y_1Ba_2Cu_3O_{7-\delta}$ thin films rotated in an external field, the forward and backward rotation magnetization curves coincide for magnetic fields less or equal to the field of full penetration $H^*$ and not $H_{c1}$, as previously believed. For fields larger than $H^*$, the forward and backward rotation curves are highly asymmetric, exhibiting magnetic irreversibility. We explain these results using the critical state model, taking into account the remagnetization process during the sample rotation.

The angular dependence of the magnetization in $Y_1Ba_2Cu_3O_{7-\delta}$ thin films, irradiated with high-energy Pb ions exhibits a unidirectional anisotropy, so that the persistent current density reaches a maximum when the direction of the external



magnetic field coincides with the direction of the columnar defects produced by the irradiation. The pinning efficiency for different angles between the columnar tracks and the external field are described by different functional forms of the pinning force scaled with the irreversibility field.

Angle-resolved measurements of the irreversibility temperature $T_{irr}$ in $Y_1Ba_2Cu_3O_{7-\delta}$ thin films reveal that in unirradiated films the transition from the irreversible to the reversible state occurs above the melting line and marks a crossover from the pinned to unpinned vortex liquid. In irradiated films with columnar defects, within the critical angle $\theta \approx 50^o$, the irreversibility line is determined by a temperature dependent trapping angle. For larger angles, $T_{irr}$ is determined by the intrinsic anisotropy via an effective field. Formulae for $T_{irr}(\theta)$ for both unirradiated and irradiated films are derived. It is also demonstrated that the collective action of crossed columnar defects leads to the suppression of relaxation and enhancement of the pinning strength.

Analysis of the local AC magnetic response in type-II superconductors, as measured by a miniature Hall probe, shows that the response is a function of the position of the probe even in a homogeneous sample. The local response in an inhomogeneous sample is further influenced by local variations of superconducting properties across the sample. Application of this analysis in characterizing and mapping of inhomogeneities in a sample are illustrated.

Effects of frequency on the local AC magnetic response are analyzed based on the critical state model, taking into account the relaxation of the magnetic induction in the sample interior during the AC field cycle. It is shown that the magnetic relaxation influences the AC response not only through the frequency dependence of the shielding current density $j$, but also introduces the frequency as an additional



parameter to the model. The traditional single-parameter approach is valid only in the limit of high frequencies, where magnetic relaxation effects can be neglected.

Local relaxation measurements provide a powerful tool for the study of flux creep mechanisms. An efficient scheme for direct determination of *E-j* curves in the flux creep regime is developed. Measurements in $Y_1Ba_2Cu_3O_{7-\delta}$ and $Nd_{1.85}Ce_{0.15}CuO_{4-\delta}$ single crystals in the field range corresponding to the anomalous magnetization peak in their magnetization curves show that two different flux creep mechanisms govern the magnetic relaxation on the two sides of the peak. The dynamic behavior before the peak is consistent with the predictions of the collective (elastic) creep theory, whereas above the peak a dislocation mediated plastic creep is dominant. The plastic creep regime can possibly be divided into "single-dislocation" plastic creep and "collective" plastic creep regimes. The latter is described by a motion of interacting dislocations or larger topological defects of the vortex lattice. While in a "single-dislocation" plastic creep regime the activation energy *U* decreases with field, in the "collective" plastic creep region *U* increases with field.

Numerical analysis of the flux creep equation shows that the activation energy $U(x)$ is almost constant across the sample. This is due to self-organization of vortices during the flux creep process. The power spectrum of the magnetic noise due to thermally activated flux jumps is predicted to be time-dependent due to time dependence of the upper cut-off of the maximal possible avalanche. This study extends the model of self-organized criticality to a much wider region of self-organization well below criticality.

The results summarized above are described in detail in a series of publications listed in Appendix D.



# Appendix A. Principles of global magnetic measurements

As we discussed in Chapter I, the main difference between *global* and *local* magnetic measurements is that the former type of experimental techniques provides information on the *total averaged magnetic moment* of the whole specimen, whereas the latter allows spatial resolution of the magnetic induction on the sample surface. In this Appendix we summarize the main principles of the *global* techniques. Discussion of the *local* measurements and techniques are found in Chapter IV.

Some of the global techniques make use of the Faraday law, $V = -\frac{1}{c}\frac{\partial \Phi_m}{\partial t}$, which relates the change of the magnetic flux $\Phi_m$ through the pick-up coils to the induced in it voltage *V*. Other global techniques use the force or torque experienced by a sample when the magnetic field is tilted or the sample is subjected to a gradient magnetic field [188]. The latter type of global techniques (Faraday balance, Gouy method, torque magnetometers) requires certain assumptions about the interaction between the magnetic field and the sample, which are quite delicate in the case of superconductors. On the other hand, these techniques are sensitive, fast and avoid sample motion in a magnetic field which is, unfortunately, a necessary component of techniques based on Eq.(80).

Nevertheless, the global experimental techniques, which employ Eq.(80), are much more convenient, versatile and provide straightforward measurements of the sample magnetic moment. We discuss here the two most famous reincarnations of these techniques – Vibrating Sample Magnetometer (VSM) and Superconducting QUantum Interference Device Magnetometer (SQUID).



In all magnetometers in which the sample is in motion during the measurement, the *raw data* is an output signal having amplitude that is a function of time, position, or both. The magnetic moment of the sample is then computed using some analytical model that fits the raw data. However, in the analysis, assumptions regarding the magnetic behavior of the sample are always made. For example, nearly all analysis methods assume that the magnetic moment of the sample: 1) approximates a magnetic dipole moment; and 2) the sign and value of this moment do not change during the measurement (constant-dipole model). However, superconducting samples do not necessarily behave as a magnetic dipole of constant magnitude and in those cases the analysis algorithms that compute the sample moment can fail, producing unpredictable and misleading results. In other words, the use of a constant-dipole model may lead to erroneous interpretations of a sample's magnetic properties. This pitfall can be largely avoided if one takes the care to determine the actual measurement conditions and then models the magnetic behavior of the sample under the particular circumstances experienced during the measurement. This does require, however, that one construct an appropriate analysis method to extract the relevant information from the output signal, then apply that analysis method to the raw data from the measurement.

As stated above, conventional magnetometers rely either on the change of voltage induced by moving the sample through the pick-up coils as a function of the sample position or as a function of time. The former method is implemented in a state of the art apparatus produced by *Quantum Design Inc.* – MPMS SQUID (Magnetic Properties Measurements System using Superconducting QUantum Interference Device detector), whereas the latter is used, for example, in an *Oxford Instruments Ltd.* VSM (Vibrating Sample Magnetometer).



# A. *The Superconducting Quantum Interference Device Magnetometer*

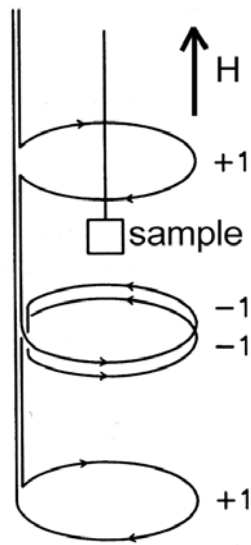

**Figure 61. Second derivative gradiometer configuration**

For a comprehensive discussion of the SQUID magnetometry, see M. McElfresh *et al.* [189]. The MPMS measures the local changes in magnetic flux density produced by a sample as it moves through the MPMS superconducting detection coils. The detection coils are located at the midpoint of a superconducting solenoid that can apply a DC magnetic field up to 5 *Tesla* to the sample. The detection coils are connected to the input of a SQUID located in a magnetic shield some distance below the magnet and detection coils. In this configuration the detection coils are part of the superconducting input circuit of the SQUID, so that changes in the magnetic flux in the detection coils (caused by the sample moving through the coils) produce corresponding changes in the current flowing in the superconducting input current, which is detected by the SQUID.

The detection coil system is wound from a single piece of superconducting wire in the form of three counter-wound coils configured as a second-order (second-derivative) gradiometer. In this geometry, shown in Figure 61, the upper coil is a single turn wound clockwise, the center coil comprises two turns wound counterclockwise, and the bottom coil is a single turn wound clockwise. This, so-called second-order gradiometer configuration rejects noise caused by fluctuations of the large magnetic field of the superconducting magnet, and also reduces noise from nearby magnetic objects in the surrounding environment.



The MPMS determines the magnetic moment of a sample by measuring the output voltage of the SQUID detection system as the sample moves through the coil. Since changes in the output voltage of the SQUID are directly proportional to changes in the total magnetic flux in the SQUID's input circuit, transporting a point-dipole sample along the z-axis of the detection coil system produces the position-dependent output signal, $V(z)$, shown in Figure 62.

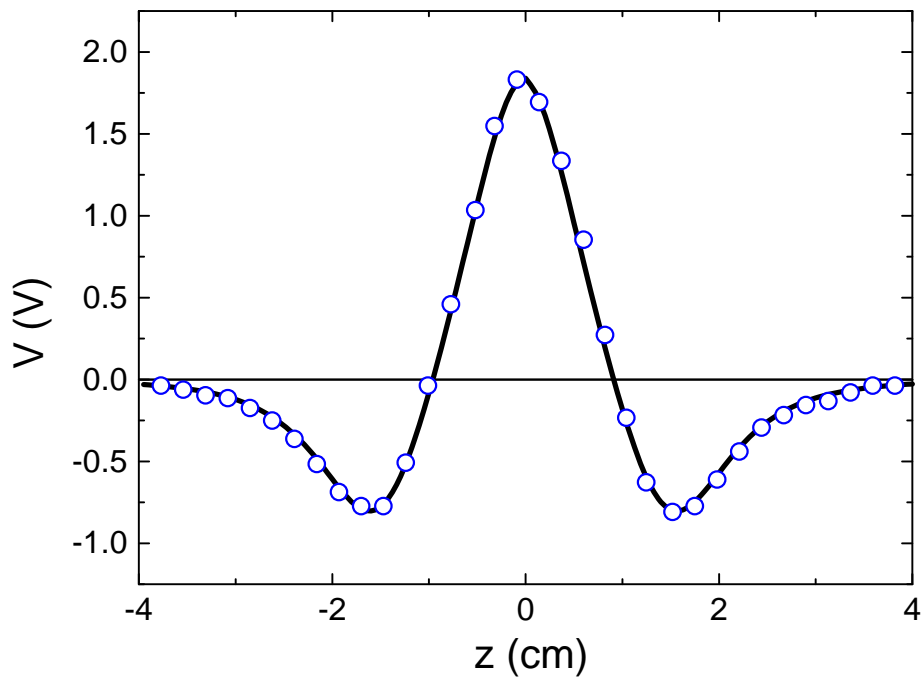

Figure 62. Typical voltage reading from pick-up coils

In practice, a measurement is performed by recording the output voltage of the SQUID sensor at a number of discrete, equally spaced positions as the sample is drawn upward through the detection coils. To ensure that the mechanical motion of the sample transport mechanism does not cause vibration noise in the SQUID detection system, the measurements are performed by recording the SQUID output while the sample is held stationary at a series of points along the scan length. Such a



set of voltage readings from the SQUID detector, with the sample positioned at a series of equally spaced points along the scan length, comprise one complete measurement, which we refer to as a single "scan". After each scan has been completed, a mathematical algorithm is used to compute the magnetic moment of the sample from the raw data.

There are several algorithms utilized by the MPMS software for determining the magnetic moment from the raw data.

The Full Scan algorithm effectively integrates the total area under the voltage-position curve, $V(z)$, by computing the square root of the sum of the squares of the data points, normalized by the total number of data points in the scan. When using the Full Scan algorithm, the magnetic moment $m$ is computed using the equation:

$$m = C\sqrt{\Delta z \sum_{i=1}^{n} V_i^2} \qquad (95)$$

where $\Delta z = L/(n-1)$, $L$ is the scan length, n is the number of data points in the scan, $V_i$ is the voltage reading at the $i$-th point, and C is a calibration factor that includes the current-to-voltage ratio of the SQUID as well as the voltage-to-moment conversion factor for the Full Scan algorithm. The factor $\Delta z$, which is the distance between data points, normalizes the calculation for both changes in the scan length and the number of data points collected. However, measurements made with a different scan length and/or a different number of data points will give slightly different values for the absolute value of the magnetic moment. This, algorithm usually requires a relatively long scan length (5 cm or more) to provide an accurate measurement of the absolute moment of the sample. The Full Scan algorithm also has a significant limitation when used to measure very small magnetic moments. Since noise in the system always adds



to the moment, the algorithm can never give a value of zero when there is any noise in the measurement. Hence, the smallest measurable moment can be severely limited by the background noise from the magnet, and there is no effective way to improve the measurement by additional averaging.

Another algorithm for determining the magnetic moment from the raw data is the linear regression algorithm, which computes a least-squares fit of the theoretical voltage-position curve to the measured data set, using the amplitude of the magnetic moment as its free parameter. Should sample centering be in doubt, MPMS can employ an iterative regression algorithm, which uses an additional free fit parameter – sample center. For the second-derivative detection coils employed in the MPMS, one can easily calculate the theoretical curve, $V(z)$, for the general condition in which the magnetic moment of the sample, $m(z)$, varies with position. For this case, the output signal, $V(z)$, is given by the following expression:

$$V(z) = -\frac{a^2 m(z)}{2} \left[ \frac{1}{\sqrt[3]{(z+d)^2 + a^2}} - \frac{1}{\sqrt[3]{(z+b)^2 + a^2}} - \frac{1}{\sqrt[3]{(z-b)^2 + a^2}} - \frac{1}{\sqrt[3]{(z-d)^2 + a^2}} \right]$$
(96)

where all the coils have the same radius, $a$, and the four coils are positioned at $z = -d$, $z = -b$, $z = b$, and $z = d$. The origin of this coordinate system is at the center of the coil set, and the z axis lies along the axis of the coils. In the ideal case, one assumes that the sample moves in a perfectly uniform magnetic field so the magnetic moment of the sample is constant as the sample moves through the detection coils. In this case, the position-dependent magnetic moment, $m(z)$, in the above equation is just a constant, $m$, and one can easily compute the theoretical curve which is shown in Figure 62 by a solid line. When the regression algorithm is used to compute the magnetic moment of the sample, the theoretical equation given above is fit to the set



of position-voltage readings using a least-squares calculation with the magnetic moment, *m*, as its free parameter (or both moment and sample center). The value of *m* that minimizes the least-squares calculation is then reported as the magnetic moment of the sample.

## B. *The Vibrating Sample Magnetometer*

The vibrating sample magnetometer (VSM) uses essentially the same type of response as the MPMS SQUID discussed above. The difference lies in the acquisition of the signal from the pick-up coils. While MPMS relies on the least square fit to the *V*(*z*) data, the VSM utilizes the *time dependence* of the induced voltage.

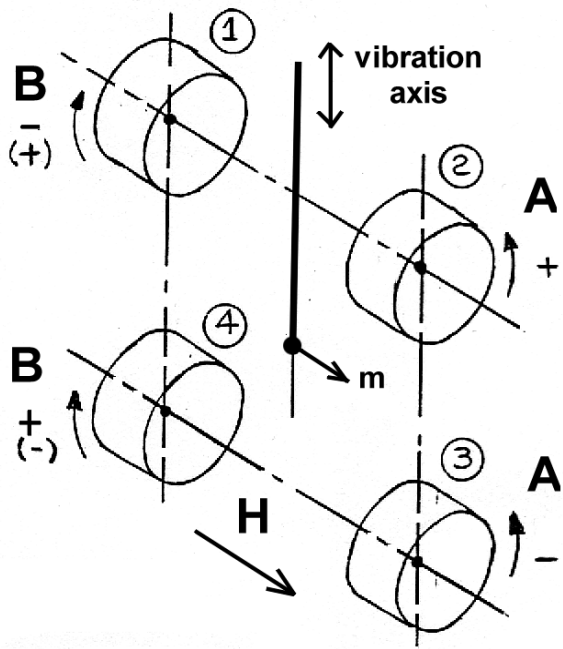

**Figure 63. VSM detection coils**

A VSM operates by vibrating the sample in a controlled manner in a magnetic field. The output from the sample resulting from this vibration is picked up in a sense-coil arrangement. The sample being tested is suspended in the applied field using carbon fiber (tube sections rod), which is used as it is extremely rigid and yet light. Because of the nature and precision of the rod no guidance is necessary, which alleviates the chance of lateral vibration of the sample (which can cause large errors). The VSM sense coil arrangement shown in Figure 63 mounted in the magnet system senses the output from the vibrating sample.



Separate amplification is provided for each coil segment. As most sense coil geometries employ more than one coil, achieving a good balance between these is critical for optimum sensitivity and this is achieved through amplifier fine trimming. Sense coil systems have a sensitivity function against displacement in the *x*, *y* and *z*-axis, (where the *z*-axis is the vibration direction). The output from the sense coils is processed by two amplifier stages. The first is programmable from unit to x $10^7$ in 80 range selections (giving a sensitivity range of 0.1 µV to 10 V). The second is a fine trim gain adjustment with a range from unity to x10. This second stage is used to calibrate the sensitivity of the sample amplifier chain such that a precise relationship may be established between the display reading and a sample during a calibration procedure (normally using a nickel sample). The sample output following the amplification and scaling stage is fed to a synchronous detection system and digitized with each half cycle of the signal sampled by the computer. The vibration frequency chosen gives a precise rejection of mains interference following the averaging of 8 successive samples. This gives a base amplifier time constant of 60 m/s for 50 *Hz* mains. The sample processing also includes noise spike rejection algorithms that further enhance the ultimate usable sensitivity. Larger amplifier time constants are achieved by successive doubling of the total samples averaged. Thus a doubling of the time constant is achieved for each increase in time constant, i.e., 50 Hz mains gives a sequence in the form 60 m/s 120 m/s, 240 m/s. etc. Time constants up to 122.88 s are available for 50 Hz mains.



# Appendix B. Using the critical state model for samples of different geometry

## A. *Relationship between magnetic moment and field in finite samples*

The main physical quantity characterizing the magnetic behavior of a superconductor is the macroscopic shielding current *j* resulting from both reversible Meissner screening and irreversible contribution of pinning. The latter was first described by Bean [134] and Kim et al. [135] and afterwards in many comprehensive reviews, see e.g., [2, 3, 78, 137, 190-192]. The unique feature of superconductors, as compared to other magnetic materials, is that currents contributing to the magnetic moment circulate in loops, which spread over the whole sample. In the absence of pinning the distribution of the reversible Meissner currents is such that it maintains constant magnetic induction inside the sample and is analogous to a "molecular" magnetization in magnetic materials, so that $\nabla \times \mathbf{H} = 0$. This current contributes to the total magnetic moment: $4\pi \mathbf{M}_{rev} = \int (\mathbf{B} - \mathbf{H}(\mathbf{B})) dV$, where **H(B)** is so-called Abrikosov curve derived for type-II superconductors [1].

In the presence of pinning there is a *macroscopic* spatial variation of the density of Abrikosov vortices, which results in appearance of the irreversible persistent current via $\nabla \times \mathbf{H} = \mathbf{j} 4\pi / c$ [134]. Thus, irreversible current is considered as an external current introduced into the system. It contributes to a total magnetic moment via [193]:

$$\mathbf{M} = \frac{1}{2c} \int_V [\mathbf{r} \times \mathbf{j}] d^3 r \qquad (97)$$



where **M** is the irreversible contribution to the total magnetic moment, *V* is the sample volume and *c* is the speed of light. The *macroscopic* nature of the irreversible persistent currents precludes consideration of the relationship $\mathbf{H} = \mathbf{B} - 4\pi\mathbf{m}$ (where **m**=**M**/*V* is volume magnetization), because the magnetic moment depends on the magnetic history and sample size. For example, the Bean model in an infinite slab (*b*>>*w*, see Figure 1) yields *m*=±*jw*/2*c* (also Eq.(103)). Interestingly, this means that if we would cut the sample into several pieces, the total magnetic moment would not be equal to that of the initial sample. Therefore, in a type-II superconducting sample with pinning one cannot use relationships, well-accepted for magnetic materials, such as $B_i=\mu_{i,j}H_j$, $M_i=\chi_{i,j}H_j$ and, most important, $4\pi\mathbf{M} = \int (\mathbf{B}-\mathbf{H}) dV$. To illustrate how one can relate the sample magnetic moment to the magnetic field, let us consider geometry depicted in Figure 1 and assume *b*→∞. In this case, the persistent current has only one component **j**=(0,*j*,0) and we obtain from Eq.(97):

$$M = \frac{1}{c} \int_{\eta=-d}^{d} d\eta \int_{\xi=-w}^{w} j(\xi,\eta) \xi d\xi \qquad (98)$$

where the magnetic moment *M* has only one *z*-component, and is calculated per unit length along the *y*-direction. Note, that we took into account (there is no 2 in a denominator of Eq.(98)) the contribution from the strip's ends (U-turns) [38]. Using the Maxwell equation $4\pi j = -c\nabla \times H$ we obtain:

$$4\pi M = 2 \int_{\eta=-d}^{d} d\eta \int_{\xi=-w}^{w} \left( B_z(\xi,\eta) - B_z(w,\eta) + B_x(\xi,d)\frac{\xi}{2d} \right) d\xi \qquad (99)$$

Thus, in the case of a superconductor the role of the external field *H* is played by the magnetic induction $B_z(w,z)$ and $B_x(x,d)$ along the sample *surface*, perpendicular to a current flow. Moreover, the contribution of the in-plane component of the magnetic induction is inversely proportional to the sample thickness. Only in the



special case of an "infinite geometry", d → ∞, originally considered by Bean [134, 144] the in-plane component of the magnetic field vanishes and the z-component $B_z(w,\eta)$ on the sample edge becomes constant and equal to the external magnetic field *H*. Then Eq.(99) becomes a full analogue of that known in magnetic materials: $4\pi M = \int (B - H) dV$.

To estimate an error using this equation in finite samples (with H being the external magnetic field) we can calculate the integral $4\pi m = \frac{1}{V}\int (B_z(\xi,\eta) - H) dV$ assuming a constant across the sample current density *j*. Note that only the z-component of the magnetic induction (external field *H* is applied in the z-direction) remains under the integral, because the in-plane component of **B** is anti-symmetric with respect to *z* and cancels after integration. In the case of an infinite in the *y*-direction strip, the integration can be done analytically, but this results in a cumbersome expression for *m*, so that we only plot the result in Figure 64.

The solid line in Figure 64 depicts the ratio $m/m_{exact}$, where $m_{exact}$ is calculated using Eq.(103) below and $m = \frac{1}{16\pi dw}\int_{\eta=-d}^{d} d\eta \int_{\xi=-w}^{w} (B_z(\xi,\eta) - H) d\xi$. The dashed line in Figure 64 depicts the magnetization calculated using only magnetic induction in the sample center (at *z*=0) to avoid edge effects, i. e., $m = \frac{1}{8\pi w}\int_{\xi=-w}^{w} (B_z(\xi, z=0) - H) d\xi$. Interestingly, even for very large values of *d/w* (very long in the *z*-direction slab), the magnetic moment calculated integrating the magnetic induction produces a large error. Only in the limit *d/w*→∞ we get exactly $m/m_{exact}$=1 in both cases.



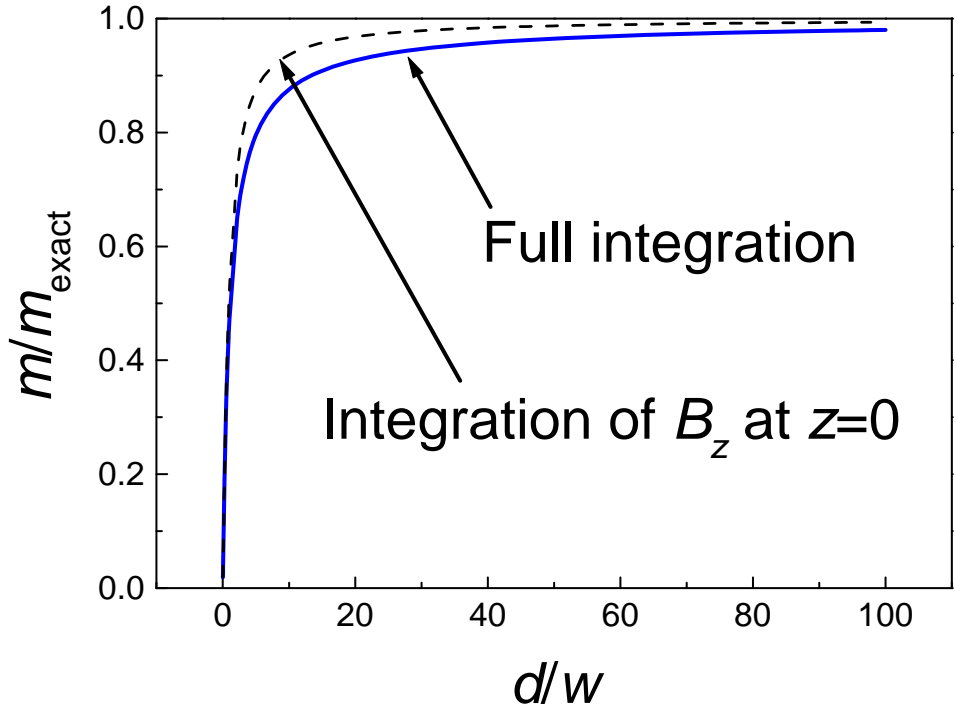

Figure 64. Comparison of the exact result ($m_{exact}$) derived from Eq.(103) and direct integration of the magnetic induction over the volume as described in the text.

In the case of thin films Eq.(80) gives:

$$4\pi M = 2\int_0^w \xi B_x(\xi,d)\,d\xi \qquad (100)$$

Therefore, only the in-plane component of **B** is responsible for the critical state in this case.

Since usual samples are plates and current density is in general not uniform due to its field dependence and anisotropy, the task of evaluating currents out of the measured moment becomes quite complicated. However, some simplifications can be made in order to find general methods of estimating the current density. The most plausible conjecture is a constancy of the current density across the sample. This is well justified at high enough magnetic field $H$ where variation of the magnetic induction across the sample is much less than the applied field and $H$ is directed along



one of the crystallographic symmetry axes. Moreover, magnetic flux creep affects the measured magnetic moment and the persistent current, hence one never measures the critical current $j_c$. As shown in [2, 153, 155, 169, 177], the current density becomes progressively more uniform during flux creep. Thus, under these conditions one may safely assume $j(B) \approx j(H)$.

In addition, a reversible magnetic moment may be separated from the irreversible one using the magnetization loop measurements. If we denote as $M_{dn}$ the descending branch of the magnetization loop (field is ramped down from some high value) and $M_{up}$ the acceding branch (field is ramped up), then the reversible magnetic moment is $M_{rev} = (M_{dn} + M_{up})/2$ and the irreversible magnetic moment is $M_{irr} = (M_{dn} - M_{up})/2$. In the formulae below we deal with the irreversible part of the magnetic moment and denote it simply as $M$.

## B. *Conversion formulae between current density and magnetic moment for different sample geometry*

Using Eq.(97) and assuming the current density $j$ to be constant across the sample one can calculate the magnetic moment for a given shape of a perpendicular to the magnetic field sample cross-section. In the formulae below we use CGS units and denote as $V$ the sample volume. We note that it is often more convenient to use practical units where $j$ is measured in A/cm$^2$, magnetic field in Gauss and length in centimeters. To convert the formulae below (written in CGS) one simply has to put $c$=10.

A simple procedure allows easy calculations of the magnetic moment. Suppose that our sample is a part of an infinite, in z-direction, sample. Apparently



there is only one component of the magnetic induction $B_z$ and the magnetic moment of the whole assembly is a algebraic sum of all samples. On the other hand, as discussed above in the case of "infinite geometry", one can use both Eq.(97) and $4\pi M = \int (B-H) dV$. Thus we can calculate the magnetic moment of an arbitrary shaped sample using a "send-pile" approach [49]. Results of such calculations for the most common cases are presented below.

## 1. Elliptical cross-section

Let us consider a sample of thickness $2d$, which has an elliptical cross-section in the *ab*-plane (ellipse of semi-major axis *b*, semi-minor axis *w*):

$$V = 2\pi wbd$$

$$M = \frac{jw}{2c}\left(1 - \frac{w}{3b}\right)V \qquad (101)$$

In the particular case of a disk of radius $a=b=R$ we have:

$$V = 2\pi R^2 d$$

$$M = \frac{jR}{3c}V \qquad (102)$$

## 2. Rectangular cross-section

In the case of rectangular cross-section of sides $2w$ and $2b$ ($w \leq b$)) we find:

$$V = 8wbd$$

$$M = \frac{jw}{2c}\left(1 - \frac{w}{3b}\right)V \qquad (103)$$

In the particular case of a square of side $2b=2w$ we get:

$$V = 8w^2 d$$



$$M = \frac{jw}{3c}V \tag{104}$$

## 3. Triangular cross-section

If the sample cross-section is a triangle of sides *a*, *b*, *n* with angles defined as $\alpha = (a \wedge n)$, $\beta = (b \wedge n)$ and $\gamma = (a \wedge b)$ we have:

semi-perimeter $s = (a+b+n)/2$

$$V = 2d\sqrt{s(s-a)(s-b)(s-n)} \equiv 2dA$$

$$M = \frac{jA}{3cs}V = \frac{2dba^2 \sin(\beta)}{3c}\left\{\cot\left(\frac{\beta}{2}\right) + \cot\left(\frac{\alpha}{2}\right)\right\} \tag{105}$$

or

$$M = \frac{2}{3}\frac{j}{c}d(s-a)(s-b)(s-n) \tag{106}$$

For a triangle of equal sides *a=b=n* we have

$$V = \frac{\sqrt{3}}{2}a^2 d$$

$$M = \frac{da^3}{6c}j \tag{107}$$

In another particular case of triangle of sides a=b≠n, α=β≠γ

$$M = \frac{dan^2}{3c}j\sin\left(\frac{\alpha}{2}\right) = \frac{dn^3}{12c}j\left(\frac{2a}{n}-1\right) \tag{108}$$

## C. *Variation of the magnetic field across the sample*

Another important quantity is the amplitude of the variation of the magnetic field inside the sample, $\Delta B \equiv |B_z(x=w) - B_z(x=0)|$. Using Biot-Savart law for



calculation of the magnetic field distribution in the superconducting sample of the arbitrary aspect ratio $\eta = d/w$ and considering very long sample $b \to \infty$ we have:

$$B_z = -\frac{2}{c}\int_{v=-d}^{d} dv \int_{\xi=-w}^{w} \frac{j(\xi,v)(x-\xi)}{(x-\xi)^2 + (z-v)^2} d\xi + H \qquad (109)$$

and assuming constant $j$ we get for the variation $\Delta B$:

## 1. In the sample middle plane

Field variation in the sample middle plane at $z=0$ is:

$$\Delta B_{z=0} = \frac{4jw}{c}\left[\eta \ln\left(\frac{(1+\eta^2)^2}{\eta^3\sqrt{4+\eta^2}}\right) + 4\arctan(\eta) - 2\arctan\left(\frac{\eta}{2}\right)\right] \qquad (110)$$

For $\eta \to \infty$ we recover the expected "Maxwell" result: $\Delta B_{z=0} = \frac{4\pi}{c} jw$. In the opposite limit of a thin film $(\eta \to 0)$ we get:

$$\Delta B_{z=0} = \frac{jw}{c}\left[3 - \ln(2\eta^3)\right] \qquad (111)$$

Note that the simplified Eq.(111) gives an accurate estimation of $\Delta B$ for $\eta \leq 0.1$, which is applicable to most of the high-$T_c$ superconducting samples.

For comparison, the variation of magnetic field resulting from transport current $I$ flowing uniformly in a strip of $2w \times 2d$ cross-section is:

$$\Delta B_{z=0}^{transp} = \frac{I}{2dc}\left[\eta \ln\left(1 + \frac{1}{\eta^2}\right) + 4\arctan(\eta)\right]$$

## 2. On the sample top surface

This case of $z=d$ is interesting for local magnetic measurements discussed later. Field variation is:



$$\Delta B_{z=d} = \frac{4jw}{c}\left[\eta\ln\left(\frac{(1+4\eta^2)^2}{16\eta^3\sqrt{1+\eta^2}}\right) + 2\arctan(2\eta) - \arctan(\eta)\right] \quad (112)$$

For $\eta \to \infty$ we get *half* of the usual "Maxwell" result $\Delta B_{z=d} = \frac{2\pi}{c}jw$ (similarly to that of electric field calculated by Brandt [19]) and for thin film Eq. (112) results in

$$\Delta B_{z=d} = \frac{jw}{c}\left[3 - \ln(16\eta^3)\right] \quad (113)$$

One should to note that the usual notion of demagnetization is inapplicable in a superconductor with pinning. One may apply it only in the Meissner state when the superconductor is perfectly diamagnetic and $B=0$. For discussion, see [3, 78]. In order to estimate the magnetic field on the edge of a superconductor in the critical state one should use the above formulae.

For comparison, the variation of magnetic field resulting from transport current $I$ flowing uniformly in a strip of $2w \times 2d$ cross-section is:

$$\Delta B_{z=d}^{transp} = \frac{I}{2dc}\left[\eta\ln\left(1+\frac{1}{\eta^2}\right) + 2\arctan(\eta)\right]$$

### D. *Field of full penetration*

Another important quantity is the full penetration field $H^*(z)$ - the first external field at which vortices reach the sample center ($x=0$) for given $z$ during the field rise after zero-field cooling (as above we assume $B_{c1}=0$). Following this definition we must to find the maximal external field $H$ at which $B_z(x=0,z)$, given by Eq.(109), is still zero. Let us consider two places inside the - center ($z=0$) and top ($z=d$).



## 1. At the sample middle plane

At $z=0$ the field of full penetration is:

$$H^*_{z=0} = \frac{4dj}{\eta c}\left(2\arctan(\eta) + \eta \ln\left(1 + \frac{1}{\eta^2}\right)\right) \tag{114}$$

and for an infinite slab $(\eta \to \infty)$ we get a "Maxwell" result, $\Delta B_{z=0} = \frac{4\pi}{c} jw$ equal to that found in Section 1. For thin film $(\eta \to 0)$ we obtain from Eq.(114):

$$H^*_{z=0} \approx \frac{8jd}{c}(1 - \ln(\eta)) \tag{115}$$

Note that in this case critical state is created over the film thickness, which is manifested in $d$ appearing in a combination $jd$ in Eq.(115).

## 2. On the sample top

On the sample top at $z=d$ calculations give:

$$H^*_{z=d} = \frac{4jd}{\eta c}\left[\arctan(2\eta) + \eta \ln\left(1 + \frac{1}{4\eta^2}\right)\right] \tag{116}$$

For an infinite slab we get half of the "Bean" result (see also [19]) $\Delta B_{z=d} = \frac{2\pi}{c} jw$ and for a thin film we obtain from Eq.(116):

$$H^*_{z=d} \approx \frac{8jd}{c}(1 - \ln(2\eta)) \tag{117}$$

Note that numerical estimations of this field by Brandt [20] yield almost correctly Eq.(117), whereas his interpolation formula does not yield an exact result, Eq.(116).



## E. *Neutral line*

Finally, one may be interested in a neutral line, which is usually defined as the line (in the *x-y* plane in our geometry) along which $B_z=H$. However, in a thick sample this line becomes a 3D surface. Since the analytic calculations are very difficult, we carry them out numerically for the case of infinite in the *y*-direction sample. We calculate the position of the neutral line on the sample top surface (i. e. at $z=d$) and in the sample middle plane (i. e. at $z=0$).

Figure 65 shows the position of the neutral line on the sample top (squares) and in the sample middle plane (circles) as a function of the aspect ratio η. The solid lines are the fits to approximate expressions: $X_{nl} = 0.7071 + 0.2486\ \tanh(\eta/0.02329)$ and $X_{nl} = 0.7071 + 0.23798\ \tanh(\eta/0.03767)$ for top and mid-plane values, respectively. In both cases the value at $\eta \to 0$ approaches the exact result for thin films: $X_{nl} = 1/\sqrt{2} \approx 0.7071$.



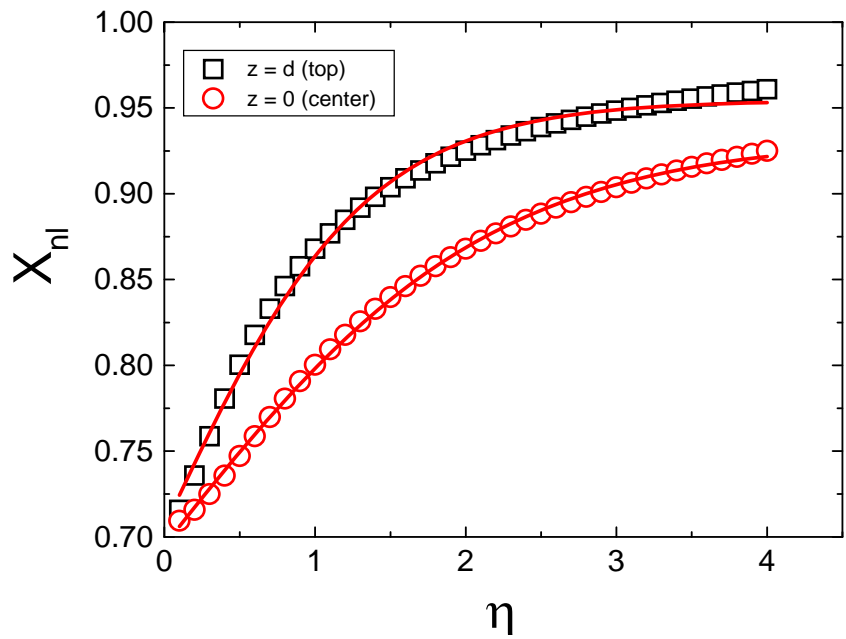

Figure 65. Position of the neutral line as a function of aspect ratio $\eta=d/w$. Squares correspond to the sample top, whereas circles to the sample middle plane. Solid lines are fits as described in the text.



# Appendix C. Trapping angle

We describe here the derivation of our Eq.(56), which differs from similar Eq.(9.17) of Blatter *et al.* [2]. We derive it using exactly the same approach (and notions) as in [2] but, in view of the experimental situation, avoid assumption of small angles, which allows Blatter *et al.* to approximate $\tan(\theta) \approx \sin(\theta) \approx \theta$. In order to estimate the trapping angle one must optimize the energy change due to the vortex trapping by columnar defects. This energy is written as [2]

$$\varepsilon(r,\theta) = \varepsilon_l \left\{ r + \left[ d^2 + \left( \frac{d}{\tan(\theta)} - r \right)^2 - \frac{d}{\sin(\theta)} \right] \right\} - r\varepsilon_r$$

(118)

where $r(\theta)$ is the length of the vortex segment, trapped by a defect; $d$ is the distance between the columns; $\varepsilon_l$ is the line tension; and $\varepsilon_r$ is the trapping potential of the defects. The variation of Eq.(118) with respect to $r$ at a fixed angle $\theta$ defines the angular dependence of $r(\theta)$. The trapping angle $\theta_t$ can be found by solving the equation $r(\theta_t)=0$. This results in

$$\tan(\theta_t) = \frac{\sqrt{\varepsilon_r(2\varepsilon_l - \varepsilon_r)}}{\varepsilon_l - \varepsilon_r}$$

(119)

which, at sufficiently small $\varepsilon_r$, can be approximated as

$$\tan(\theta_t) = \sqrt{\frac{2\varepsilon_r}{\varepsilon_l}} + O(\varepsilon_r^{3/2})$$

(120)

Apparently, at very small angles we recover the original result of Ref. [2]. In this thesis, for the sake of simplicity, we use expression Eq.(120) instead of the full expression Eq.(119). However, as noted above we cannot limit ourselves to small angles and, generally speaking, the trapping angle may be quite large ($\theta \approx$40-50º in our case). The error due to use of Eq.(120) can be estimated as follows: at $\theta \approx$40º Eq.(119) gives $\varepsilon_r/\varepsilon_l \approx$0.24, whereas Eq.(120) gives $\varepsilon_r/\varepsilon_l \approx$0.35, which is suitable for our implication of Eq.(120), since we consider an exponential decrease of $\varepsilon_r$. In addition, as shown in [2] in a system with anisotropy $\varepsilon$, the trapping angle is enlarged by a factor of $1/\varepsilon$.

# Appendix D. List of publications – *Ruslan Prozorov (RP)*

**Institute of Solid State Physics, Chernogolovka, Russia**
*(Magneto-optic study of high-temperature superconductors)*

RP1. V. K. Vlasko-Vlasov, M. V. Indenbom, V. I. Nikitenko, Y. A. Osip'yan, A. A. Polyanskii, R. Prozorov, *"Actual structure and magnetic properties of high-temperature superconductors"*, Superconductivity: Physics, Chemistry, Technology (English translation) **3** (6), S50 (1990).

RP2. V. K. Vlasko-Vlasov, L. A. Dorosinskii, M. V. Indenbom, V. I. Nikitenko, A. A. Polyanskii, R. Prozorov, *"Visual representation of the Meissner expulsion of the magnetic flux in YBaCuO single crystals"*, Superconductivity: Physics, Chemistry, Technology (English translation) **5** (11), 1917 (1992).

RP3. V. K. Vlasko-Vlasov, M. V. Indenbom, V. I. Nikitenko, A. A. Polyanskii, R. Prozorov, I. V. Grekhov, L. A. Delimova, I. A. Liniichuk, A. V. Antonov, M. Y. Gusev, *"Effect of the shape of a superconductor on the flux structure"*, Superconductivity: Physics, Chemistry, Technology (English translation) **5** (9), 1582 (1992).

RP4. L. A. Dorosinskii, M. V. Indenbom, V. I. Nikitenko, A. A. Polyanskii, R. Prozorov, V. K. Vlasko-Vlasov, *"Magnetooptic observation of the Meissner effect in YBaCuO single crystals"*, Physica C **206**, 306 (1993).

RP5. R. Prozorov, A. A. Polyanskii, V. I. Nikitenko, I. V. Grehov, L. A. Delimova, I. A. Liniychuk, *"The distribution of superconducting currents under remagnetization of the thin YBCO film in the normal field"*, Superconductivity: Physics, Chemistry, Technology (English translation) **6** (4), 563 (1993).

RP6. V. K. Vlasko-Vlasov, L. A. Dorosinskii, M. V. Indenbom, V. I. Nikitenko, A. A. Polyanskii, R. Prozorov, *"The distribution of normal and planar components of magnetic induction in HTSC thin films"*, Superconductivity: Physics, Chemistry, Technology (English translation) **6** (4), 555 (1993).

RP7. V. K. Vlasko-Vlasov, L. A. Dorosinskii, V. I. Nikitenko, A. A. Polyanskii, R. Prozorov, *"A study of the surface current in HTSC single crystals and thin films"*, Superconductivity: Physics, Chemistry, Technology (English translation) **6** (4), 571 (1993).

RP8. R. Prozorov, *"Study of the magnetic flux distribution in high-temperature superconductors YBaCuO using magneto-optical method"*, research thesis (Master of Science, specialization: Physics of Metals), Institute for Solid State Physics (Chernogolovka, Russia) and Moscow Institute of Steel and Alloys (Moscow, Russia), February 1992.

RP9.  M. V. Indenbom, T. Schuster, M. R. Koblishka, A. Forkl, H. Kronmüller, L. A. Dorosinskii, V. K. Vlasko-Vlasov, A. A. Polyanskii, R. Prozorov and V. I. Nikitenko, *"Study of flux distributions in high-Tc single crystals and thin films using magneto-optic technique"*, 2$^{nd}$ Israeli International Conference on High-T$_c$ Superconductivity, 4-7 Jan. 1993, Eilat, Israel, Physica C **259 - 262** (1993).

## Department of Physics, Bar-Ilan University

*(Magnetic study of high-temperature superconductors)*

RP10.  R. Prozorov, A. Tsameret, Y. Yeshurun, G. Koren, M. Konczykowski and S. Bouffard, *"Angular dependence of the magnetic properties of thin YBaCuO films irradiated with Pb and Xe ions"*, Physica C **234**, 311 (1994).

RP11.  R. Prozorov, E. R. Yacoby, I. Felner and Y. Yeshurun, *"Magnetic properties of YNi$_2$B$_2$C superconductor"*, Physica C **233**, 367 (1994).

RP12.  R. Prozorov, A. Shaulov, Y. Wolfus and Y. Yeshurun, *"Local AC magnetic response in type-II superconductors"*, J. Appl. Phys. **76**, 7621 (1994).

RP13.  R. Prozorov, A. Shaulov, Y. Wolfus and Y. Yeshurun, *"Frequency dependence of the local AC magnetic response in type-II superconductors"*, Phys. Rev. B **52**, 12541 (1995).

RP14.  Y. Abulafia, A. Shaulov, Y. Wolfus, R. Prozorov, L. Burlachkov, Y. Yeshurun, D. Majer, E. Zeldov, V.M. Vinokur, *"Local magnetic relaxation in high-temperature superconductors"*, Phys. Rev. Lett. **75**, 2404 (1995).

RP15.  R. Prozorov, A. Poddar, E. Sheriff, A. Shaulov and Y. Yeshurun, *"Irreversible magnetization in thin YBCO films rotated in external magnetic field"*, Physica C **264**, 27 (1996).

RP16.  Y. Abulafia, A. Shaulov, Y. Wolfus, R. Prozorov, L. Burlachkov, Y. Yeshurun, D. Majer, E. Zeldov, H. Wühl, V. B. Geshkenbein and V. M. Vinokur, *"Plastic vortex-creep in YBa$_2$Cu$_3$O$_{7-\delta}$ crystals"*, Phys. Rev. Lett. **77**, 1596 (1996).

RP17.  R. Prozorov, M. Konczykowski, B. Schmidt, Y. Yeshurun, A. Shaulov, C. Villard and G. Koren, *"Origin of the irreversibility line in thin YBa$_2$Cu$_3$O$_{7-\delta}$ films with and without columnar defects"*, Phys. Rev. B **54**, 15530 (1996).

RP18.  R. Prozorov, A. Tsameret, Y. Yeshurun, G. Koren, M. Konczykowski and S. Bouffard, *"Frequency dependent irreversibility line and unidirectional magnetic anisotropy in thin YBa$_2$Cu$_3$O$_{7-\delta}$ films irradiated with heavy-ions"*, International M$^2$S Conference, 5-9 July 1994, Grenoble, France, Physica C **235-240**, 3063 (1994).

RP19.  Y. Wolfus, R. Prozorov, Y. Yeshurun, A. Shaulov, Y. Abulafia, A. Tsameret and K. Runge, *"Frequency dependence of the irreversibility temperature in Y-Ba-Cu-O"*, International M$^2$S-IV Conference, 5-9 July 1994, Grenoble, France, Physica C **235-240**, 2719 (1994).